\documentclass{aa}

\usepackage{txfonts}

\DeclareRobustCommand{\VAN}[3]{#2}
\let\VANthebibliography\thebibliography
\def\thebibliography{\DeclareRobustCommand{\VAN}[3]{##3}\VANthebibliography}

\usepackage{graphicx}	
\usepackage{amssymb}	

\usepackage{xcolor}
\usepackage{overpic}
\usepackage{subfigure}
\usepackage{booktabs}
\usepackage{multirow}
\usepackage{caption}
\usepackage[flushleft]{threeparttable}

\usepackage[version=4]{mhchem}

\usepackage[colorlinks=true,allcolors=blue,urlcolor=blue]{hyperref}
\usepackage{hyperref}

\newcommand{\stagger}{\texttt{Stagger}}
\newcommand{\garstec}{\texttt{GARSTEC}}
\newcommand{\gastag}{\texttt{GASTAG}}
\newcommand{\mean}[1]{\ensuremath{\langle #1 \rangle}}

\hypersetup{draft}

\begin{document} 

\title{\gastag{} evolutionary tracks and isochrones from coupled 1D and 3D models: Systematic temperature offsets in red giants}

\titlerunning{\texttt{gastag} tracks and isochrones}

\author{Yixiao Zhou\inst{\ref{rocs},\ref{uc}}
\and
Jiaqi (Martin) Ying\inst{\ref{dartmouth}}
\and
Yaguang Li\inst{\ref{UH}}
\and
Luca Casagrande\inst{\ref{ANU}}
}

\institute{
\label{rocs}Rosseland Centre for Solar Physics, Institute of Theoretical Astrophysics, University of Oslo, P.O. Box 1029, Blindern, NO-0315 Oslo, Norway
\and
\label{uc}School of Physical and Chemical Sciences --- Te Kura Mat{\=u}, University of Canterbury, Private Bag 4800, Christchurch 8140, Aotearoa, New Zealand\\
\email{yixiao.zhou@qq.com}
\and
\label{dartmouth}Department of Physics and Astronomy, Dartmouth College, 6127 Wilder Laboratory, Hanover, NH 03755, USA
\and
\label{UH}Institute for Astronomy, University of Hawai‘i, 2680 Woodlawn Drive, Honolulu, HI 96822, USA
\and
\label{ANU}Research School of Astronomy and Astrophysics, The Australian National University, Canberra, ACT 2611, Australia
}

\abstract{
Models of stellar structure and evolution describe the global and internal properties of stars throughout their lifetimes and are indispensable for studies of individual stars, stellar clusters, and Galactic evolution. However, most 1D evolutionary calculations rely on a simplified treatment of convection, resulting in inaccurate near-surface structures and nonnegligible uncertainties in the predicted fundamental parameters of low-mass stars.
In a series of previous studies, a novel approach was developed to couple 1D stellar interior models with 3D model atmospheres throughout the evolutionary calculation. This 1D-3D coupling method makes predicted stellar properties effectively independent of the mixing-length parameter, while providing oscillation frequencies that agree more closely with asteroseismic observations.
To expand this framework to ensemble studies of stars and age determinations of clusters, we present the \gastag{} stellar evolutionary tracks and isochrones constructed using the 1D-3D coupling approach. Comparing effective temperatures from the APOGEE-\textit{Kepler} catalog with \gastag{} predictions, we find the theoretical temperatures are cooler by about 70~K near solar metallicity. Our isochrones are compared with observed color-magnitude diagrams of star clusters spanning from $\rm [Fe/H] = 0.3$ to $-1.9$. In all cases, the synthesized and observed diagrams agree excellently in the main-sequence, turn-off, and subgiant regions, while isochrones predict systematically cooler red giant branches. Taking these independent findings together reveals that the temperature mismatch is most likely due to deficiencies in stellar models.
Because \gastag{} is constructed using a method that substantially reduces uncertainties associated with surface boundary conditions and the mixing-length parameter, the difference between modeling and observation can be more confidently attributed to other ingredients in the models, such as $\alpha$-element abundances or uncertainties in low-temperature opacities.
}


\date{Received / Accepted}

\maketitle



\section{Introduction}

  Through the ages of stars we can reveal the history of our Galaxy and the Universe. Ages derived for large stellar samples, when combined with kinematic and chemical abundance information, provide a time-resolved view of the Milky Way's assembly history \citep{2022Natur.603..599X}, constraining its chemical evolution \citep{2019MNRAS.489.1742F,2020A&A...640A..81N} and enabling the dating of past merger events \citep{2020NatAs...4..382C,2022MNRAS.514.2527B}. The ages of the oldest stars and clusters set a lower limit on the age of the Universe \citep{2023AJ....166...18Y,2025A&A...703A.232L}. On a smaller scale, precise stellar ages can be used as an additional dimension to probe the evolution of planets \citep{2020AJ....160..108B}.
  
  As one of the most important stellar fundamental properties, accurate ages remain elusive. While masses and radii of solar-type stars can typically be determined with systematic uncertainties of a few percent \citep{2022ApJ...927...31T}, uncertainties in ages exceed 10\% even for stars with high-quality asteroseismic data \citep{2015MNRAS.452.2127S}. The reason for this is that stellar ages are more model-dependent than other fundamental properties. 
  The major methodologies used to determine the ages of low-mass stars include stellar evolution modeling and isochrone fitting, asteroseismology, and gyrochronology (see \citealt{2010ARA&A..48..581S} for a comprehensive review). Asteroseismology gives precise radii, masses, and evolutionary stages, from which ages are determined using grids of stellar models \citep{2013ARA&A..51..353C}. Gyrochronology relies on empirical relations between stellar rotation and age, which require calibration using benchmark stars with asteroseismic ages or well-dated young star clusters \citep{2007ApJ...669.1167B}.
  Consequently, models of stellar evolution are the fundamental framework underlying all three age-determination approaches.
  
  The relatively large uncertainties in stellar ages partly stem from deficiencies in models of stellar structure and evolution. \citet{2023AJ....166...18Y,2025ApJ...987...52Y} quantified contributions to the age error budgets for globular clusters. Apart from uncertainties in observational inputs -- such as the distance, reddening, and chemical composition of the cluster -- the treatment of helium diffusion and the choice of mixing-length parameter stand out. Each corresponds to a major problem in stellar physics: the diffusion and mixing of elements in stellar interiors and the modeling of near-surface convection, respectively.

  Convection is often described by the mixing length theory (MLT; \citealt{1958ZA.....46..108B}) in 1D stellar evolution modeling. For low-mass stars that possess a convective envelope, MLT is well known to fail in the near-surface convective region where the temperature gradient far exceeds the adiabatic temperature gradient. The MLT predicts incorrect near-surface stratifications that lead to significant offsets between theoretical and observed oscillation frequencies, known as the asteroseismic surface effect \citep{1988Natur.336..634C,1996Sci...272.1286C}. In addition, the mixing-length parameter $\alpha_{\rm MLT}$, which is the major free parameter in MLT governing the efficiency of convective heat transport, is tightly correlated with the model's effective temperature. In standard stellar evolution calculations, a 20\% change in $\alpha_{\rm MLT}$ could shift the effective temperatures of the main-sequence and red giant models by more than 200 K \citep{2025MNRAS.540.3400Z}.
  
  On the other hand, recent years have seen substantial progress in 3D radiative-hydrodynamical simulations of stellar surface convection. These simulations have been extensively tested against observations and proven superior to their 1D counterparts in all aspects \citep{2009LRSP....6....2N,2013A&A...554A.118P}. Convection occurs naturally in these simulations without the need for adjustable parameters and operates in a way that is fundamentally different from the physical picture of MLT. To overcome the challenges in modeling near-surface convective layers while simultaneously providing realistic outer boundary conditions for 1D stellar evolution calculations, a novel method of coupling stellar evolution with 3D hydrodynamical simulations of surface convection was developed and validated in a series of studies \citep{2018MNRAS.481L..35J,2020MNRAS.491.1160M,2025MNRAS.540.3400Z}. In this 1D-3D coupling approach, the near-surface region of the 1D model, where MLT is particularly problematic, is replaced by the horizontal- and time-averaged 3D (mean 3D or \mean{\rm 3D}) models. The stellar structural calculation extends to the near-adiabatic convective layer below the stellar surface (0.14\% of the solar radius below the surface in the solar case), while the structure above is derived from interpolated \mean{\rm 3D} models rather than from the MLT.

  The method has been thoroughly described and validated in previous work. \citet{2018MNRAS.481L..35J} and \citet{2020MNRAS.491.1160M} compared oscillation frequencies predicted by the 1D-3D coupled models with the observed frequencies for the Sun and two main-sequence stars with asteroseismic data from \textit{Kepler}. They demonstrate that the use of coupled models reduces the asteroseismic surface effect in all cases.
  \citet{2025MNRAS.540.3400Z} further compared evolutionary tracks computed with the 1D-3D coupling method with standard calculations employing different atmospheric boundary conditions. At solar metallicity, the predicted effective temperatures are similar to those obtained using a gray atmosphere throughout all evolutionary phases, with temperature differences at the same surface gravity generally below 20 K. The method was also validated against eclipsing binaries with precisely determined fundamental parameters.
At different evolutionary stages and metallicities, the 1D-3D coupling approach produces stellar models that satisfy most observational constraints.

  Another key advantage is that stellar evolution calculations using the 1D-3D coupling method are insensitive to the choice of $\alpha_{\rm MLT}$. This is because the super-adiabatic near-surface layers, whose structure is highly sensitive to $\alpha_{\rm MLT}$ in standard MLT-based models, are instead provided by 3D simulations. Although convective energy transport in the deeper interior of the 1D model is still described by MLT, a 20\% change in $\alpha_{\rm MLT}$ results in temperature changes of less than 30 K when using the 1D-3D coupling method. This contrasts with shifts exceeding 200 K in the standard treatment (Fig.~4 of \citealt{2025MNRAS.540.3400Z}).
  Since the temperature change associated with the mixing-length parameter is smaller than typical observational uncertainties, $\alpha_{\rm MLT}$ can in practice be treated as an invariant parameter fixed by solar calibration.
  
  Recently, \citet{2026ApJ...996...83L} performed a systematic validation of the 1D-3D coupling method using 18 stars from the \textit{Kepler} LEGACY sample \citep{2017ApJ...835..172L,2017ApJ...835..173S}. Consistent with earlier studies, they find that using 1D-3D coupled models reduces the discrepancies between theoretical and measured frequencies in all cases, and the remaining frequency offsets exhibit similar shapes. The inferred stellar masses, radii, and ages are in good agreement with previous determinations, even though $\alpha_{\rm MLT}$ was not treated as a free parameter in the modeling.

  Coupling stellar interior models with 3D simulations during evolution not only produces models with more realistic near-surface structures but also effectively eliminates a major uncertainty in stellar evolution calculations. To expand its scope of application to the determination of fundamental parameters of stars and the ages of clusters, we present the \gastag{} stellar evolution tracks and isochrones based on this novel method (Sect.~\ref{sec:GASTAG}). Global stellar parameters predicted by the new evolutionary tracks are compared with measurements from the APOGEE-\textit{Kepler} joint spectroscopic and asteroseismic catalog (Sect.~\ref{sec:track-APOGEE}). In Sect.~\ref{sec:CMD}, we validate synthetic color-magnitude diagrams (CMDs) generated from our isochrones against detailed observations of star clusters at different ages and metallicities.

\section{The 1D-3D coupling method}

\begin{figure}
\includegraphics[width=\columnwidth]{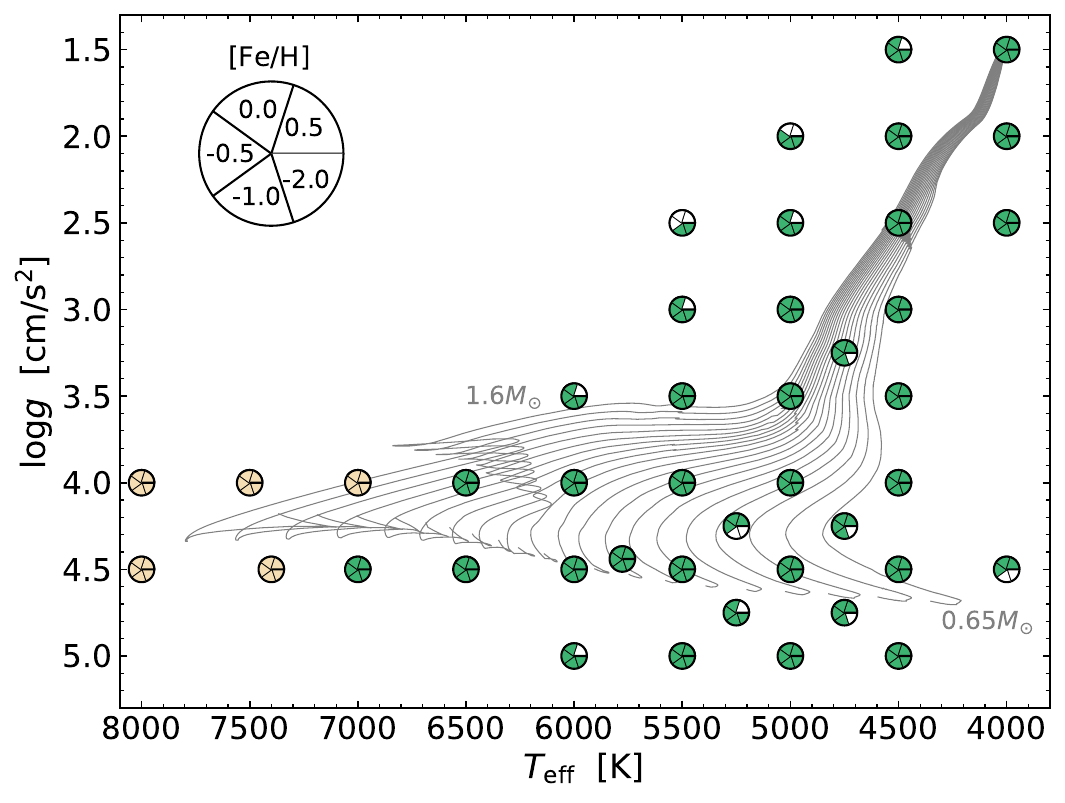}
\caption{Kiel diagram showing the global parameters of the \stagger{}-grid models used in the construction of the \gastag{} evolutionary tracks. The green-shaded portions indicate the 3D model atmospheres used in this work, while the light yellow-shaded portions denote 1D model atmospheres supplied solely to enable the complete evolution of the slightly higher-mass tracks. Note that the mean effective temperatures of the 3D models are close to, but generally not exact multiples of, 500 K. The
\gastag{} evolutionary tracks with an initial metallicity of $\rm [Fe/H]_i = 0$ are shown in the background.
\label{fig:grid}}
\end{figure}

  Models of stellar interior were computed using the Garching Stellar Evolution Code (\garstec{}, \citealt{2008Ap&SS.316...99W}). \garstec{} solves the equation of mass and energy conservation, hydrostatic equilibrium, and heat transport at each vertical grid point with the \citet{1965ApJ...142..841H} method. Structural models were evolved based on changes of chemical composition throughout the stellar interior, due to nuclear reactions and element diffusion. The equilibrium quantities were kept fixed when solving the time-dependent equations in an evolutionary timestep. 
  The code is equipped with several realistic equations of state (EOSs) and Rosseland mean opacities for stellar interior conditions. The available EOSs include the \citet{1988ApJ...331..815M} EOS, the OPAL EOS \citep{1996ApJ...456..902R}, and FreeEOS \citep{2004Irwin...feos1,2012ascl.soft11002I}. Opacity tables are based on the OPAL opacity \citep{1993ApJ...412..752I,1996ApJ...464..943I} smoothly merged with the \citet{2005ApJ...623..585F} data at lower temperatures ($T < 10^{4.1}$ K). 
  Opacity tables with varying $\alpha$-element abundance are available, with $\rm [\alpha/Fe]$ ranging from -0.2 to 0.6 dex in steps of 0.1.
  Nuclear reaction networks treated in the original \garstec{} code consist of the $p$-$p$ chain, CNO cycles, and helium burning reactions, which include \ce{^{4}He}, \ce{^{12}C}, \ce{^{16}O}, \ce{^{20}Ne}, \ce{^{24}Mg}, \ce{^{28}Si}, and \ce{^{56}Ni}. The networks have been expanded substantially by \citet{2013A&A...559A...4C} and \citet{2024A&A...687A.260R}. The nuclear reaction rates were taken from the NACRE database \citep{1999NuPhA.656....3A} and the JINA REACLIB library \citep{2010ApJS..189..240C}.
  Convection was treated with the mixing-length theory in the formulation of \citet{kippenhahn2012stellar}. We note that the implementation of \citet{1986A&A...160..116K} turbulent convection model is in active development \citep{2024A&A...689A.292B}. Atomic diffusion for elements H, He, C, N, O, Ne, Mg, Si, and Fe was calculated following the method of \citet{1994ApJ...421..828T}.
  
  The outer boundary conditions of stellar interior models are provided by the \stagger{}-grid \citep{2013A&A...557A..26M,2024A&A...688A.212R}, a grid of more than 250 3D model atmospheres computed using the \stagger{} code \citep{2018MNRAS.475.3369C,2024ApJ...970...24S}. 
  All 3D models were constructed with a customized version of the \citet{1988ApJ...331..815M} EOS \citep{2013ApJ...769...18T}. The bolometric flux and effective temperature for each simulation snapshot were obtained from frequency-dependent radiative transfer calculations.
  The simulation domain was discretized on a 3D Cartesian mesh (i.e.,~plane-parallel approximation) with a constant gravitational acceleration applied throughout. Vertically, the simulations span the lower atmosphere, the photosphere, and the super-adiabatic convective layers below the surface. The bottom boundary of the simulation domain extends down to Rosseland mean optical depth $\tau_{\rm Ross} \sim 10^6 - 10^7$, where the temperature gradient is close to adiabatic. The horizontal extent of simulations was determined based on the typical size of granules at corresponding stellar parameters. Magnetic fields were not taken into account in the calculation of these 3D models. 
  
  The \stagger{}-grid covers a wide range of stellar parameters for low mass stars, with $T_{\rm eff}$ ranging from $3500$ to $7000$ K in steps of approximately 500 K, $\log (g/{\rm [cm/s^2]})$ from $1.5$ to $5$ in steps of 0.5 dex, and $\rm [Fe/H] = -4,\; -3,\; -2,\; -1,\; -0.5,\; 0,\; 0.5$ dex, corresponding to F, G, K-type dwarfs and giants from extremely metal-poor to metal-rich. All $\rm [Fe/H] \geq -0.5$ models adopted the \citet{2009ARA&A..47..481A} metal mixture scaled based on their metallicity, while a 0.4 dex abundance enhancement for $\alpha$-elements was applied for metal-poor ($\rm [Fe/H] \leq -1$) models.
  In this work, we utilized all 3D models from $\rm [Fe/H] = -2$ to $0.5$, as illustrated in Fig.~\ref{fig:grid}. Since evolution calculations with the 1D-3D coupling method at lower metallicities are more uncertain due to interpolation and convergence issues, we refrained from systematic calculation of stellar tracks and isochrones below $\rm [Fe/H] = -2$. This extremely metal-poor regime will be investigated further in future studies.
  3D simulations were averaged in the horizontal plane and time, yielding mean 3D structures ready for interpolation over $T_{\rm eff}$, $\log g$ and [Fe/H]. The effective temperature of the mean 3D model is the time-averaged value from all simulation snapshots. 
  The relationship between pressure and density shows a similar structural pattern for all \mean{\rm 3D} models, with the ``density inflection region'' being a key feature. The region is located just below the stellar surface and has a relatively small density gradient. Based on this property, physical quantities were scaled by their value at the density inflection point, where the gradient of density with respect to pressure is minimal, yielding comparable scaled quantities across all models for a more stable and robust interpolation \citep{2017MNRAS.472.3264J}.
  For the entire scaled \mean{\rm 3D} structure, 2D interpolations were first performed in the $(T_{\rm eff}, \log g)$ plane, followed by cubic monotonic interpolations \citep{1990A&A...239..443S} in [M/H]\footnote{${\rm [M/H]} = \log_{10} (Z/X) - \log_{10} (Z/X)_{\odot}$, where $X, Z$ are the mass fractions of hydrogen and metal, respectively.}. Quantities at the density inflection point were interpolated in the same manner as the scaled stratification to reconstruct the mean 3D model at given stellar parameters.  
  We refer readers to \citet{2017MNRAS.472.3264J} and \citet{2025MNRAS.540.3400Z} for details of the interpolation technique, and \citet{2013A&A...557A..26M} and \citet{2024A&A...688A.212R} for comprehensive descriptions of the construction of 3D models and \stagger{}-grid.
 
  The hottest models in the \stagger{}-grid have effective temperatures of approximately 7000 K, which effectively impose an upper limit on the stellar mass for which a complete evolutionary calculation can be performed. To enable the evolution of stars with slightly higher masses, we supplemented the grid with a few additional 1D model atmospheres at $\log (g/{\rm [cm/s^2]}) = 4.0$ and $4.5$ that correspond to A and F-type stars (Fig.~\ref{fig:grid}). These 1D model atmospheres were computed with the \texttt{ATMO} code, a 1D atmosphere code that employs the same EOS and opacity table as the \stagger{} code (cf.~\citealt{2013A&A...557A..26M} Appendix A for a description of the code).

  A prominent feature of the 1D-3D coupling method is that the outer boundary conditions for stellar structure calculations are provided by \mean{\rm 3D} models at every evolutionary time step. The matching point, which is equivalent to the outer boundary of the stellar interior model, is located in the convective region below the surface, where the temperature gradient is close to adiabatic.
  At each evolutionary time step, the temperature at the matching point from the 1D model, $T_{\rm m,1D}$, together with the gravitational acceleration and metallicity at the same location, was used to derive the effective temperature by interpolating the function $T_{\rm eff}(T_{\rm m}, \log g, {\rm [Fe/H]})$ defined by mean 3D models. The resulting stellar fundamental parameters were used to interpolate the corresponding $\mean{\rm 3D}$ structure. 
  The thermal (gas plus radiation) pressures of the 1D and \mean{\rm 3D} models at the matching point are required to be identical to ensure a smooth transition. The second boundary condition requires that the luminosity evaluated from the Stefan-Boltzmann law, $L = 4 \pi R^2 \sigma T_{\rm eff}^4$, equals the luminosity given by \garstec{}. Here, $R$ is the photospheric radius determined by integrating the \mean{\rm 3D} model on top of the 1D model, and $\sigma$ is the Stefan-Boltzmann constant. The numerical solver in \garstec{} iteratively adjusts the interior structure and repeats the interpolation procedure until the pressure and luminosity conditions are fulfilled.

\section{Evolutionary tracks and isochrones} \label{sec:GASTAG}
  
  In this section, we describe the input physics and ingredients used in the stellar evolution calculations. We then present the construction of the \gastag{} evolutionary tracks and isochrones, followed by the calculation of bolometric corrections.
  
\subsection{Evolution calculations}

\begin{table*}
\centering
\caption{Summary of input physics and parameters adopted in the \garstec{} and \stagger{} models.
\label{tb:physics}}
{\begin{tabular*}{\textwidth}{@{\extracolsep{\fill}}lcc}
\toprule[2pt]
	Physics 			& Parameter/Choice	& Note/Reference		
  \\
\midrule[1pt]
	\multicolumn{3}{c}{\garstec{}} 
\\
\midrule[1pt]
  	EOS				& \texttt{FreeEOS}	& \citet{2012ascl.soft11002I}
  \\
    	Opacity			& OPAL				& \citet{1996ApJ...464..943I}
  \\
    Solar mixture 	& {\citet{2009ARA&A..47..481A}} &
  \\
  	Helium enrichment law	& $Y_{\rm i} = 1.105 Z_{\rm i} + 0.246$	& $Y_{\rm p} = 0.246$ from {\citet{2020A&A...641A...6P}}
  \\
  	Reaction network	& $p$, \ce{^{3}He}, \ce{^{4}He}, \ce{^{12}C}, \ce{^{13}C}, \ce{^{14}N}, \ce{^{15}N}, \ce{^{16}O},  	& Hydrogen and helium networks
  \\
  					& \ce{^{17}O}, \ce{^{20}Ne}, \ce{^{24}Mg}, \ce{^{28}Si}, \ce{^{56}Ni} & 
  \\
  	Convection		& MLT &	The \citet{kippenhahn2012stellar} formulation
  \\
  	$\alpha_{\rm MLT}$	& 2.76		& Fixed by solar calibration
  \\
  	Overshoot 		& Exponential, $f_{\rm ov}$ increases with	& {\citet{1996A&A...313..497F}}
  \\
  					& stellar mass as a ramp function			&
  \\
  	Atomic diffusion	& Yes				& {\citet{1994ApJ...421..828T}}
  \\
  	Turbulent diffusion & Yes			& The {\citet{2017ApJ...840...99D}} formulation
  \\
  	Mass loss		& No 				&
  \\
\midrule[1pt]
  	\multicolumn{3}{c}{\stagger{}} 
\\
\midrule[1pt]
  EOS				& Customized Mihalas EOS		& {\citet{2013ApJ...769...18T}}
  \\
  Opacity			& Continuum opacities compiled internally;	& {\citet{2010A&A...517A..49H}}
  \\
  					& line opacities from MARCS	& {\citet{2008A&A...486..951G}}
  \\
  Solar mixture 		& {\citet{2009ARA&A..47..481A}} &
  \\
  Convection 		& Ab initio			&
  \\
  Radiative transfer	& Opacity binning; long characteristics,	& {\citet{2018MNRAS.475.3369C}}
  \\
  					& nine different angles including the vertical	&
  \\
\bottomrule[2pt]
\end{tabular*}}
\end{table*}

\begin{table}
\centering
\caption{Initial metallicities, helium mass fractions, and corresponding available mass ranges for the \gastag{} evolutionary tracks.
\label{tb:mass-range}}
{\begin{tabular*}{\columnwidth}{@{\extracolsep{\fill}}cccc}
\toprule[2pt]
	$\rm [Fe/H]_i$		& $\rm [\alpha/Fe]$	& $Y_{\rm i}$	& Mass range ($M_{\odot}$)
  \\
\midrule[1pt]
  	0.4				& 0					& 0.2806			& 0.75--1.90
  \\
    	0.25				& 0					& 0.2711			& 0.72--1.80
  \\
    0 				& 0 					& 0.2605			& 0.65--1.60
  \\
  	-0.25			& 0					& 0.2543			& 0.65--1.45
  \\
  	-0.5				& 0					& 0.2507			& 0.65--1.28
  \\
  	-0.75			& 0.2 				& 0.2497			& 0.63--1.25
  \\
  	-1				& 0.4 				& 0.2490			& 0.60--1.20
  \\
  	-1.25			& 0.4				& 0.2477			& 0.60--1.14
  \\
  	-1.5 			& 0.4				& 0.2469			& 0.60--1.05
  \\
  	-1.75			& 0.4				& 0.2465			& 0.60--1.05
  \\
  	-1.9				& 0.4				& 0.2464			& 0.60--1.00
  \\
\bottomrule[2pt]
\end{tabular*}}
\end{table}


  We used the \texttt{FreeEOS} equation of state\footnote{Available at \url{http://freeeos.sourceforge.net/documentation.html}} \citep{2004Irwin...feos1,2012ascl.soft11002I} and OPAL opacity tables \citep{1996ApJ...464..943I} when calculating the stellar interior models. Both EOS and opacity were calculated according to the \citet{2009ARA&A..47..481A} metal mixture. 
  We note that our interior and atmosphere models adopted different EOSs. Although the thermodynamical quantities given by \texttt{FreeEOS} and \citet{1988ApJ...331..815M} are in good agreement in the parameter space that corresponds to the near-surface region where the matching takes place (see Fig.~A.1 of \citealt{2023A&A...677A..98Z}), this slight inconsistency was mended by reconstructing the density profile above the matching point using \texttt{FreeEOS} based on pressures and temperatures from the \mean{\rm 3D} model (\citealt{2020MNRAS.491.1160M} Sect.~2). Different opacity sources in 1D and 3D models are of little concern, because the energy transport is dominated by convection at the matching location -- radiative flux is negligible compared to the total energy flux. 
  The default basic hydrogen and helium reaction network (Table \ref{tb:physics}; see also Sect.~3.2.1 in \citealt{2008Ap&SS.316...99W}) was employed in this work, which is sufficient for the evolution of low-mass stars until the red clump phase. 
  
  Since a key feature of the 1D-3D coupled method is its insensitivity to the mixing-length parameter, the solar-calibrated value, $\alpha_{\rm MLT} = 2.76$, was used throughout the study. We refer readers to Sect.~3.1 of \citet{2025MNRAS.540.3400Z} for a detailed description of our solar calibration procedure, as well as a discussion of how the role of $\alpha_{\rm MLT}$ in our approach differs from that in standard stellar evolution calculations.
  Convective overshoot leads to element mixing beyond the convective boundary defined by the Schwarzschild criterion. Overshooting beyond the convective core extends the main-sequence lifetime of stars with $M \gtrsim 1.2 M_{\odot}$, and changes the morphology of the Henyey hook in isochrones \citep{2010ApJ...718.1378M,2018ApJ...863...65C}. At boundaries of each convection zone, \garstec{} models overshoot mixing through a diffusion coefficient that decays exponentially with the distance to the Schwarzschild convective boundary,
\begin{equation}
D_{\rm ov} = D_0 \exp\left( \frac{-2 z}{f_{\rm ov} H_P} \right).
\end{equation} 
Here, $H_P$ is the pressure scale height evaluated at the convective boundary, $z$ is the distance from the boundary. The symbol $D_0$ denotes the reference diffusion coefficient within half pressure scale height inside the convective boundary (or half the thickness of the convection zone, whichever is smaller) given by the MLT. The decay length scale of the diffusion coefficient is controlled by the overshooting parameter $f_{\rm ov}$. To avoid the unphysical scenario in which a tiny convective core has an extended overshooting region, $f_{\rm ov}$ was restricted as
\begin{equation} \label{eq:limit-fov}
f_{\rm ov} =  \frac{1}{2}\left\{ 
\tanh\left[ 5 \left(4\frac{R_{\rm CZ}}{H_P} - 1\right) \right] + 1 \right\}
f_{\rm ov,in},
\end{equation}
where $R_{\rm CZ}$ and $f_{\rm ov,in}$ are the thickness of the convection zone and the input value of the overshooting parameter, respectively. The hyperbolic function ensures that $f_{\rm ov}$ approaches zero for small convection zones, while converging to the input value when the size of the convection zone exceeds half of the pressure scale height. It is worth noting that condition \eqref{eq:limit-fov} is a relatively weak constraint on $f_{\rm ov,in}$. Stronger reductions of the input overshooting parameter that takes effect when $R_{\rm CZ} < H_P$ or $R_{\rm CZ} < \alpha_{\rm MLT} H_P$ have been adopted in other studies \citep{2010ApJ...718.1378M,2016A&A...589A..93D}. Using Eq.~\eqref{eq:limit-fov} in the evolution code together with a constant $f_{\rm ov,in}$, such as 0.01, results in a long-lived convective core in the main-sequence phase for stars with $M \leq 1.1M_{\odot}$, which is likely unphysical as concluded in \citet{2023MNRAS.525.1416W}. It hence should be avoided.
  Previous investigations using constraints from binaries \citep{2018ApJ...859..100C}, asteroseismology \citep{2021A&A...650A..58M}, and hydrodynamical simulations \citep{2021A&A...646A.133H} indicate that the overshooting parameter for the convective core increases with stellar mass. In constructing the \gastag{} tracks, we simplified the relation between $f_{\rm ov}$ and $M$ estimated in \citet{2018ApJ...859..100C} to a ramp function starting at $1.1 M_{\odot}$ to set the overshooting parameter as
\begin{equation}
f_{\rm ov,in} = 
\begin{cases} 
0 & (M/M_{\odot}  \leq 1.1 )
\\ 
0.02(M/M_{\odot} - 1.1) & (1.1 < M/M_{\odot} < 2)
\\
0.018 & (M/M_{\odot} \geq 2)
\end{cases}.
\end{equation}
For a given stellar mass, the same value of $f_{\rm ov,in}$ was adopted for overshooting at all convective boundaries in the stellar interior model\footnote{Overshooting above the convective envelope is in the regime of \mean{\rm 3D} models and is therefore irrelevant here.}, since overshooting beneath the convective envelope has little influence on the evolutionary tracks.

  Considering atomic diffusion in stellar evolution calculations leads to significant or complete depletion of helium and heavier elements in the surface convective envelope for F-type and warmer stars, which contradicts the observed chemical composition of A and F dwarfs \citep{1999A&A...351..247V,2010A&A...523A..71G,2019MNRAS.489.1850V}. To counter the strong sedimentation of elements in thin convective envelopes, we included turbulent diffusion as an extra mixing mechanism below the bottom boundary of the surface convection zone following the formulation of \citet[see \citealt{2025MNRAS.540.3400Z} Sect.~3.2 for details]{2017ApJ...840...99D}.
  
  Rotation, mass-loss, and thermohaline mixing were not accounted for in our evolution calculations. As the 1D-3D coupling method was confined to $\log(g/{\rm [cm/s^2]}) > 1.5$ due to the coverage of 3D models, the effect of mass-loss is negligible. Table \ref{tb:physics} summarizes input physics adopted in 1D \garstec{} and 3D \stagger{} models.
  
  With the aforementioned input physics and settings, stellar evolution tracks were computed from initial metallicities $\rm [Fe/H]_{\rm i} = -1.9$ to $0.4$. The step is 0.25 dex from $\rm [Fe/H]_{\rm i} = -1.75$ to $0.25$. To maintain consistency with the 3D models, opacity tables without $\alpha$-enhancement were used for tracks with $\rm [Fe/H]_{\rm i} \geq -0.5$. For tracks with $\rm [Fe/H]_{\rm i} = -0.75$, $\rm [\alpha / Fe] = 0.2$ was adopted, and for those with $\rm [Fe/H]_{\rm i} \leq -1$,  $\rm [\alpha / Fe] = 0.4$ was used (Table \ref{tb:mass-range}). The corresponding range of the initial metal mass fraction is $Z_{\rm i} = 3.41 \times 10^{-4}$ to 0.0313. The initial mass fraction of helium was determined using a linear helium enrichment law, $Y_{\rm i} = (\Delta Y / \Delta Z) Z_{\rm i} + Y_{\rm p}$.
  The primordial helium mass fraction, $Y_{\rm p} = 0.246$, was converted from the baryon fraction given by \citet{2020A&A...641A...6P}. The gradient $\Delta Y / \Delta Z$ was fixed by solving the linear equation with the initial helium and metal mass fractions obtained from the solar calibration, which yields $\Delta Y / \Delta Z = 1.105$. Models with different helium mass fractions will be presented in future work.
  Because the scope of application for the 1D-3D coupling method is dictated by the coverage of the \stagger{}-grid, the mass range of evolutionary tracks, shown in Table \ref{tb:mass-range}, depends on initial metallicity. 
  For each track, we provide all structural models at every evolutionary time step from the zero-age-main-sequence, defined as the point where hydrogen burning contributes to 98\% of the total luminosity, to at least $\log(g/{\rm [cm/s^2]}) = 2$ at the red giant branch (RGB). In addition, each structural model is accompanied by theoretical oscillation frequencies of radial modes up to the acoustic cutoff frequency, computed using the Aarhus adiabatic oscillation package (\textsc{adipls}; \citealt{2008Ap&SS.316..113C}). Previous work by \citet{2020MNRAS.491.1160M,2021MNRAS.500.4277J,2025MNRAS.540.3400Z,2026ApJ...996...83L} have verified that theoretical frequencies predicted from the 1D-3D coupled models agree better with the observed frequencies for both main-sequence stars and giants. 
  Models for high-luminosity red giants and later evolutionary stages are not available.

\subsection{Isochrones construction}

We constructed isochrones using the equivalent evolutionary point (EEP) formalism described in \citet{dotter2016}, which maps stellar evolution tracks onto a uniform basis defined by identifiable evolutionary stages. This approach enables interpolation between tracks of different initial masses while preserving the morphology of rapid evolutionary phases.

Primary EEPs correspond to key evolutionary landmarks along each track, while secondary EEPs were inserted between adjacent primary points to provide a smooth and uniform sampling of the evolutionary sequence. Transforming all tracks to a common EEP grid ensures that interpolation is performed between homologous evolutionary stages rather than between arbitrary timesteps. For each EEP, the stellar age was treated as a monotonic function of initial mass. Isochrones were constructed by interpolating in mass at fixed EEP to determine the model with the age matching the desired isochrone age. Repeating this procedure across all EEPs produces a continuous sequence of stellar models that share a common age and span the full mass range covered by the tracks. This method ensures smooth morphology across all evolutionary phases, including regions of rapid structural change.

For each isochrone point, we recorded the fundamental stellar parameters required for subsequent analysis. These include the initial and current stellar mass, radius, luminosity, effective temperature, surface gravity, and surface mass fractions of hydrogen and helium.

\subsection{Bolometric corrections}

\begin{figure*}
\includegraphics[width=0.49\textwidth]{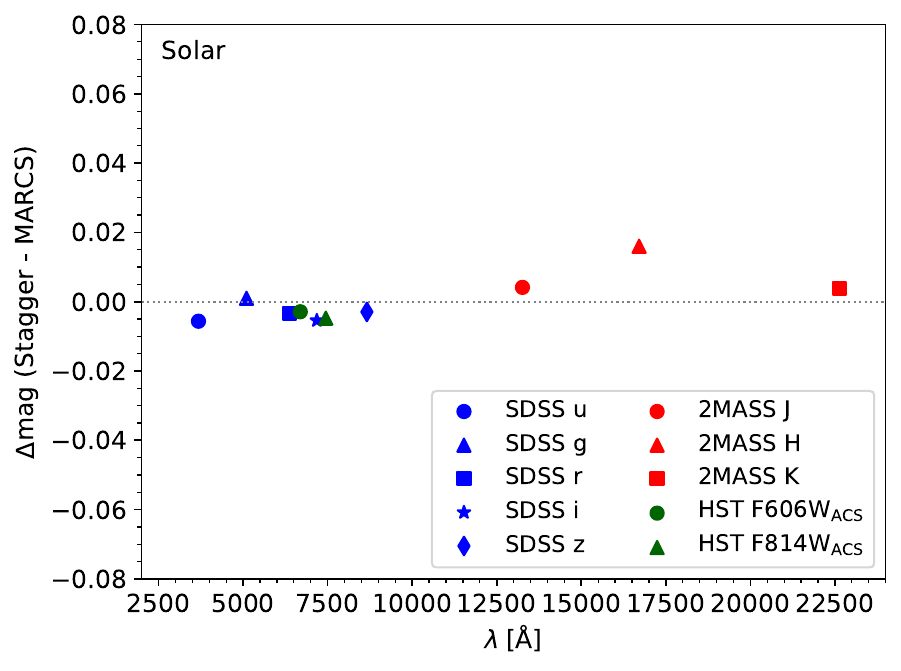}
\includegraphics[width=0.49\textwidth]{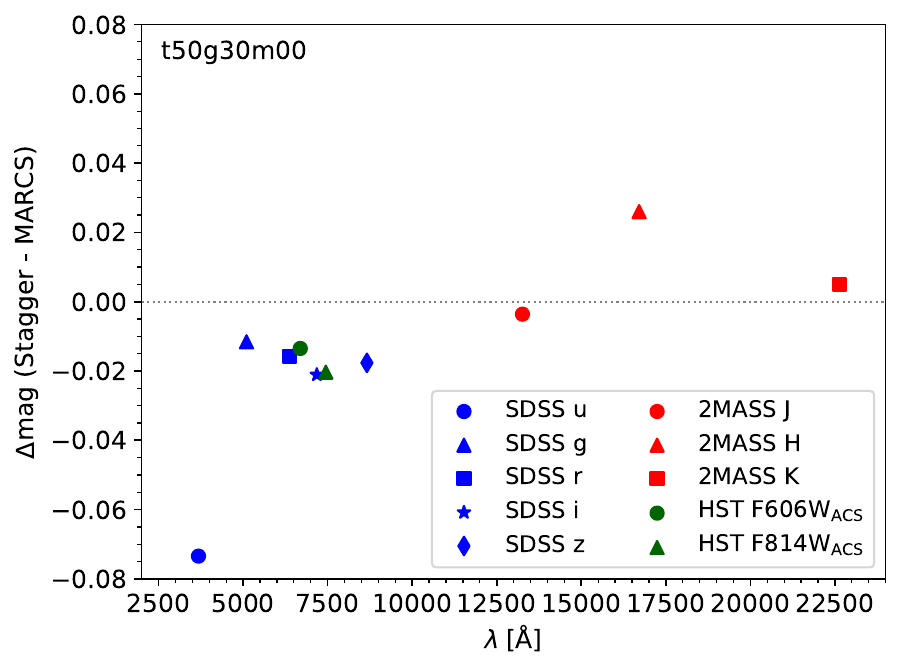}
\caption{Differences in magnitude between predictions from the \stagger{} and MARCS model atmospheres in the SDSS $ugriz$, the 2MASS $JHK_S$, and the HST ACS F606W, F814W filters. \textit{Left panel}: Results for the solar models. \textit{Right panel}: Differences for red giant models with $T_{\rm eff} = 5000$ K, $\log (g / {\rm [cm/s^2]}) = 3$, and solar metallicity. Each filter is plotted at the wavelength corresponding to the maximum of its spectral response function.
\label{fig:diff-mag}}
\end{figure*}

  Bolometric correction is the difference between absolute bolometric magnitude and the magnitude measured at a certain band $\zeta$ of a photometric system 
\begin{equation}
BC_{\zeta} = \mathcal{M}_{\rm bol} - \mathcal{M}_{\zeta}.
\end{equation}
It converts fundamental stellar properties, such as effective temperatures and surface gravities, given by theoretical isochrones to magnitudes at different photometric pass bands, which can be directly compared with observed CMD of star clusters. The absolute bolometric magnitude reflects the stellar luminosity relative to the solar value,
\begin{equation}
\mathcal{M}_{\rm bol} = 
-2.5\log\left( \frac{L}{L_{\odot}} \right) + \mathcal{M}_{\rm bol,\odot},
\end{equation}
whereas the magnitude measured at bandpass $\zeta$ is expressed as \citep{2014MNRAS.444..392C}
\begin{equation}
\mathcal{M}_{\zeta} = -2.5\log
\left[ \left(\frac{R_{\odot}}{d_{10}}\right)^2 \mathcal{K}_{\zeta} \right] + 
zp_{\zeta}.
\end{equation}
Here the subscript ``$\odot$'' denotes the solar value, $d_{10} = 10$ parsec, and $zp_{\zeta}$ is related to the definition of the zero magnitude for $\zeta$.
The term $\mathcal{K}_{\zeta}$ is proportional to $\int F_{\lambda} \mathcal{T}_{\zeta} \; d\lambda$, i.e.,~the wavelength integral of the spectral energy distribution (SED) $F_{\lambda}$ weighted by the spectral response function $\mathcal{T}_{\zeta}$. Its exact form depends on the adopted magnitude system. Therefore, the key step in deriving bolometric correction is the computation of SED. A comprehensive introduction of synthetic photometry is provided by \citet{2014MNRAS.444..392C}.

  Bolometric corrections used in \gastag{} were computed from the Synthetic Stellar Photometry Package\footnote{Available at \url{https://github.com/casaluca/bolometric-corrections}; see also \citet{2014MNRAS.444..392C} Appendix A.} presented in \citet{2014MNRAS.444..392C,2018MNRAS.475.5023C,2018MNRAS.479L.102C}, which takes $T_{\rm eff}$, $\log g$, and $\rm [Fe/H]$ of theoretical isochrones as input and outputs bolometric corrections covering over 100 widely adopted photometric bandpasses. The package employs the MARCS grid of synthetic spectra computed with the opacity sampling method \citep{2008A&A...486..951G,2008PhST..133a4003P} and assuming a microturbulence of 2 km/s. 
  
  We note that since \gastag{} isochrones are constructed with the near-surface and photospheric structure of the averaged \stagger{}-grid models, the most self-consistent approach is to compute SEDs based on \stagger{} model atmospheres for bolometric corrections. In view of this, we compared bolometric corrections derived based on SEDs from \stagger{} and MARCS model atmospheres for $ugriz$ filters of the Sloan Digital Sky Survey (SDSS) photometry, $JHK_S$ filters of 2MASS, as well as F606W and F814W filters of the Hubble Space Telescope Advanced Camera for Surveys (ACS), whose wavelengths range from near-ultraviolet to near-infrared. The test was carried out for the solar models and red giant models with $T_{\rm eff} = 5000$ K (approximately 5000 K for the \stagger{} model), $\log (g/{\rm [cm/s^2]}) = 3$, $\rm [Fe/H] = 0$. 
  In the 3D case, the SEDs were computed using the 3D radiative transfer code \textsc{scate} \citep{2011A&A...529A.158H}, which takes \stagger{} model atmospheres as input and employs identical microphysics for radiative transfer calculations (see \citealt{2021MNRAS.503...13Z} Sect.~5.2 for further details). Magnitudes at different filters predicted by 3D models are compared with MARCS results in Fig.~\ref{fig:diff-mag}. The discrepancies are generally below 0.02 mag for both the solar and red giant models, except for the SDSS $u$ filter. 
  Given that (1) agreement within 0.01-0.02 magnitude in synthetic photometry is considered excellent in light of the observational uncertainties and the accuracy of the photometric zero-point \citep[Sect.~4]{2014MNRAS.444..392C}, and (2) near-ultraviolet SEDs are less reliable due to uncertainties in (or missing) opacities, our test indicates that using SEDs calculated from the \stagger{} models does not lead to a qualitative improvement in the resulting bolometric corrections. This is in agreement with \citet{2018A&A...611A..11C}.

\section{Comparing \gastag{} evolutionary tracks with fundamental parameters from the APOGEE-\textit{Kepler} catalog} \label{sec:track-APOGEE}

\begin{figure*}
\includegraphics[width=0.99\textwidth]{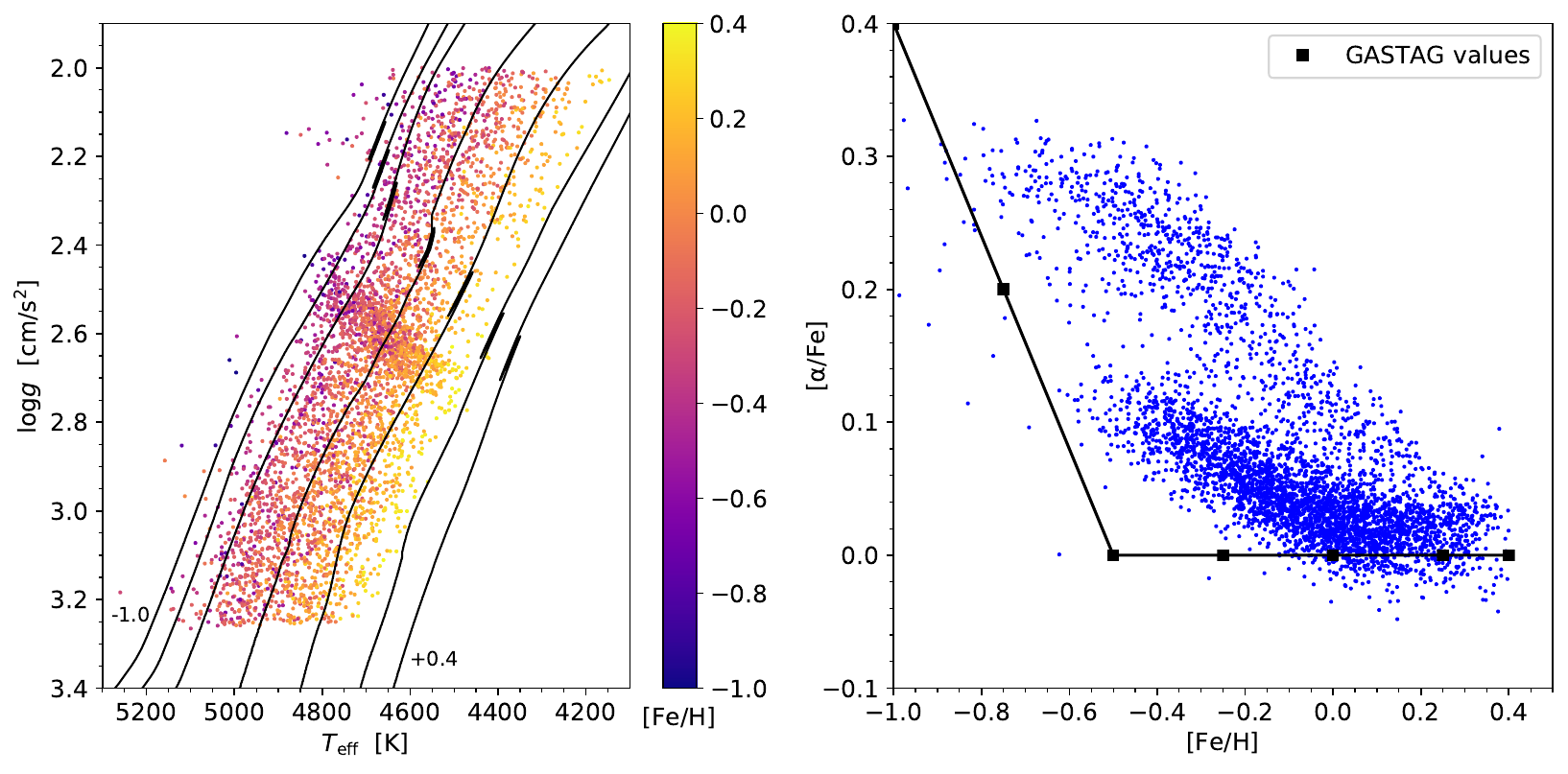}
\caption{\textit{Left panel:} Kiel diagram of our selected stars from the APOKASC3 catalog, color-coded with their observed iron abundance. The black lines indicate the \gastag{} $1.1 M_{\odot}$ evolutionary tracks with various initial iron abundances $\rm [Fe/H]_i = -1, -0.75, -0.5, -0.25, 0, 0.25$, and $0.4$. 
\textit{Right panel:} Our selected sample in the $\rm [\alpha / Fe]-[Fe/H]$ plane. The two $\alpha$ sequences are clearly seen. The black squares represent $\alpha$-element-to-iron ratios adopted in the \gastag{} tracks, which are consistent with the 3D models in the \stagger{}-grid.
\label{fig:sample}}
\end{figure*}
\begin{figure}
\includegraphics[width=\columnwidth]{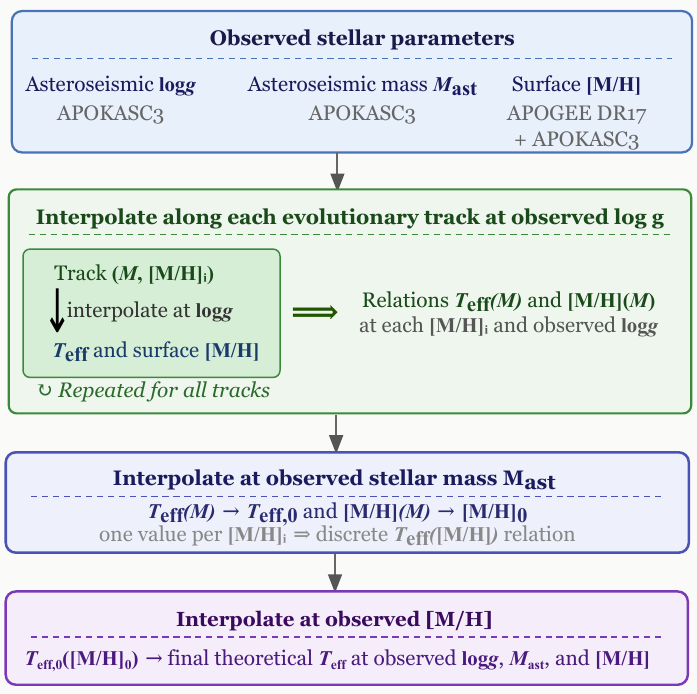}
\caption{Flowchart illustrating how theoretical effective temperatures at observed stellar parameters were obtained by interpolating evolutionary tracks. The overall surface metallicity of a star, [M/H], is calculated from its APOGEE DR17 iron abundance and overall $\alpha$-element abundance provided by APOKASC3 (label \texttt{APOKASC3\_ALPHA\_M}). This assumes the \citet{2009ARA&A..47..481A} solar metal mixture and that all $\alpha$ elements are either enhanced or depleted by the same amount relative to iron.
\label{chart:Teff}}
\end{figure}
\begin{figure*}
\includegraphics[width=0.49\textwidth]{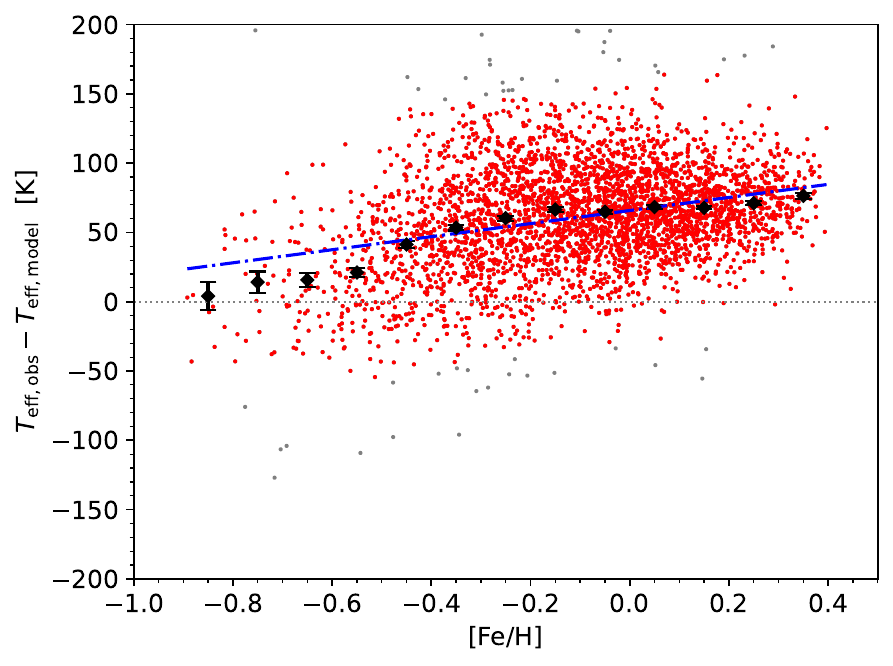}
\includegraphics[width=0.49\textwidth]{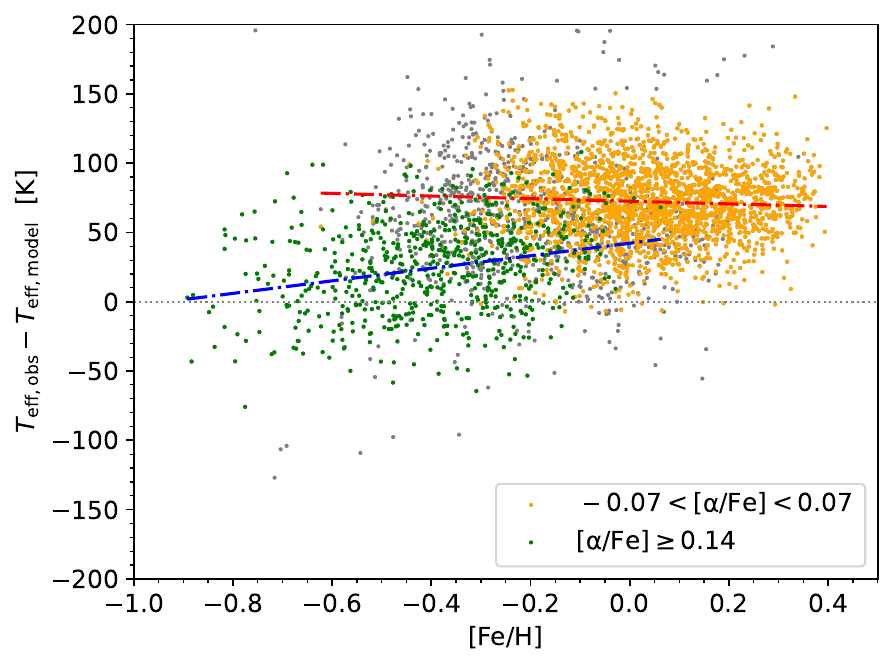}
\caption{\textit{Left panel:} Differences between spectroscopically measured effective temperatures and predictions from \gastag{} evolutionary tracks as a function of observed iron abundance for all red giants selected from the APOKASC3 catalog. The dash-dotted blue line represents a linear fit to the temperature differences, performed without outliers (gray dots), identified based on the median of the fitting residual (see text). The solid diamonds with error bars denote the mean and standard error of the mean for non-outlier stars in each [Fe/H] bin, with a width of 0.1 dex.
\textit{Right panel:} Stars in the $\alpha$-solar (orange dots) and $\alpha$-rich (green dots) group, selected from our sample according to their $\alpha$-element-to-iron ratio. The linear fits for both groups are indicated with the dash-dotted lines using the same methodology as for the full sample.
\label{fig:dTeff}}
\end{figure*}
\begin{figure}
\includegraphics[width=\columnwidth]{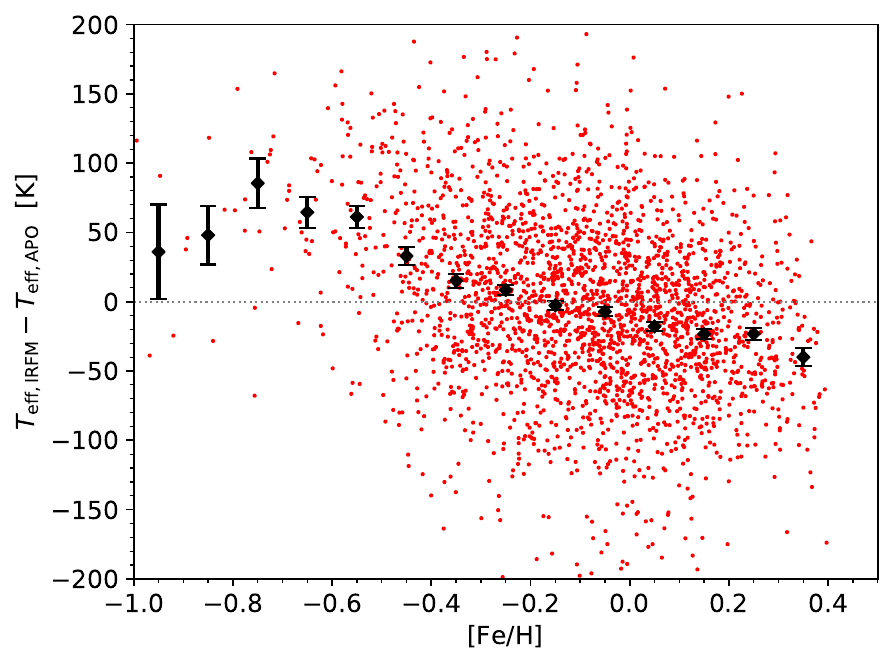}
\caption{Differences in effective temperatures measured from the IRFM and the APOGEE values for the red giant sample from \citet[see their Fig.~1]{2024ApJ...974...77L}. The solid diamonds with error bars indicate the mean and standard error of the mean of all stars within a given [Fe/H] bin of width 0.1 dex.
\label{fig:dTeff-IRFM-APO}}
\end{figure}
\begin{figure*}
\includegraphics[width=0.49\textwidth]{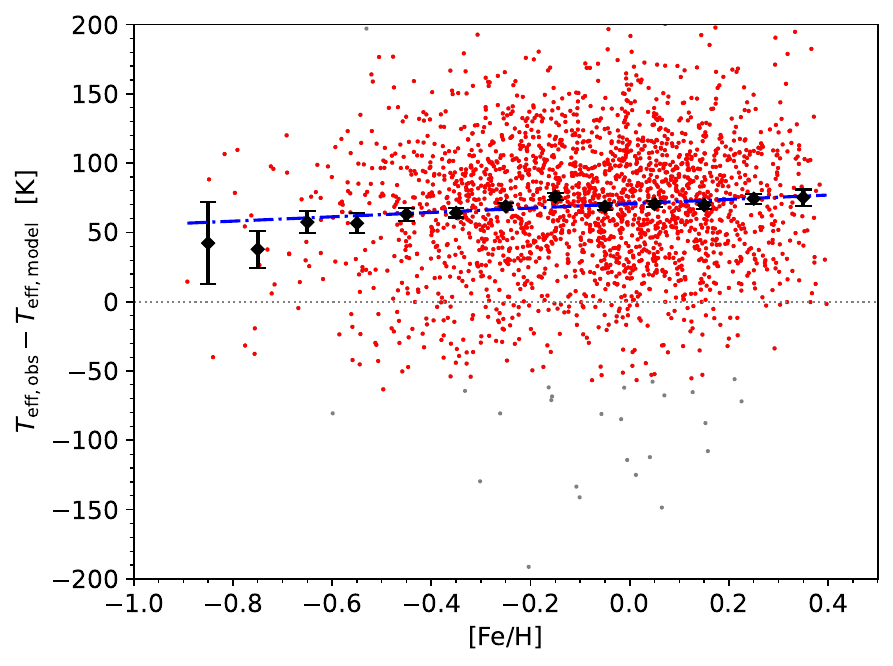}
\includegraphics[width=0.49\textwidth]{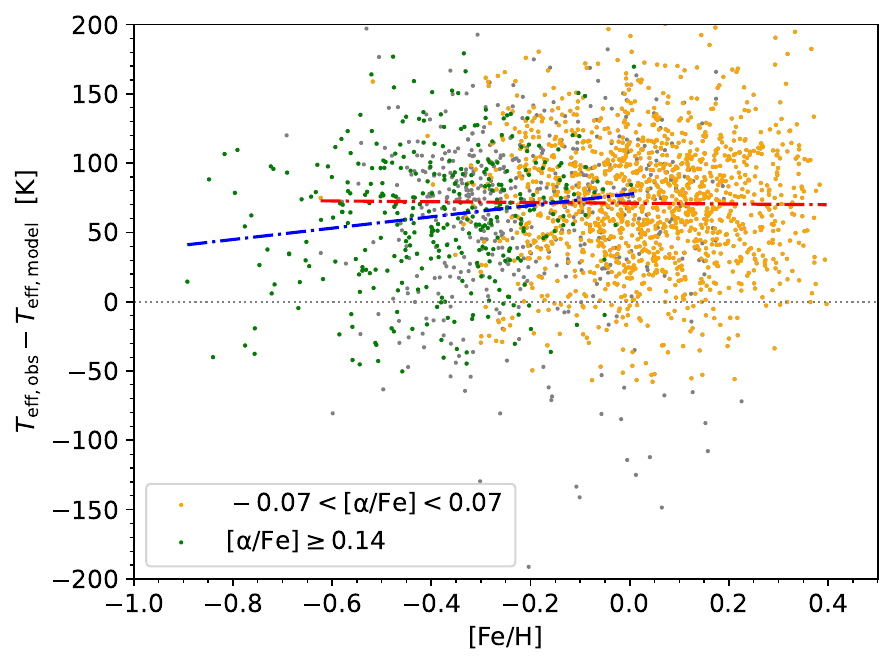}
\caption{Comparison between the modeled effective temperatures and values derived from the IRFM for the red giant sample from \citet{2024ApJ...974...77L}. \textit{Left panel}: Temperature offsets as a function of iron abundance for all stars in the \citet{2024ApJ...974...77L} sample, together with a linear fit and statistical properties in each [Fe/H] bin. \textit{Right panel}: $\alpha$-solar and $\alpha$-rich subsamples and their corresponding linear fits. The color scheme is identical to that used in Fig.~\ref{fig:dTeff}.
\label{fig:dTeff-Li24}
}
\end{figure*}


\begin{table}
\centering
\caption{Results of the linear regression between the temperature mismatch $\Delta T_{\rm eff}$ (observation minus model) and iron abundance for different scenarios.
\label{tb:linear-fit}}
{\begin{tabular*}{\columnwidth}{@{\extracolsep{\fill}}cc}
\toprule[2pt]
	Case 				& Linear fit
  \\
\midrule[1pt]
  	All selected stars (APOKASC3)	& $\Delta T_{\rm eff} = 47.2{\rm [Fe/H]} + 65.8$ K			
  \\
    	$\alpha$-solar group					& $\Delta T_{\rm eff} = -9.4{\rm [Fe/H]} + 72.5$ K
  \\
    $\alpha$-rich group 					& $\Delta T_{\rm eff} = 45.3{\rm [Fe/H]} + 42.2$ K
  \\
  All selected stars (Li24)			& $\Delta T_{\rm eff} = 15.7{\rm [Fe/H]} + 70.7$ K			
  \\
    	$\alpha$-solar group					& $\Delta T_{\rm eff} = -2.7{\rm [Fe/H]} + 71.1$ K
  \\
    $\alpha$-rich group 					& $\Delta T_{\rm eff} = 41.0{\rm [Fe/H]} + 77.6$ K
  \\
\bottomrule[2pt]
\end{tabular*}}
\tablefoot{
The theoretical effective temperatures are compared with the APOKASC3 catalog as well as measurements from the IRFM for the \citet{2024ApJ...974...77L} sample.
}
\end{table}

  Models computed with the 1D-3D coupling method have been validated against observations of the Sun and dwarfs from the \textit{Kepler}-LEGACY samples \citep{2018MNRAS.481L..35J,2020MNRAS.491.1160M,2026ApJ...996...83L}, plus a few red giants \citep{2025MNRAS.540.3400Z}. Here, we perform a systematic validation of \gastag{} evolutionary tracks in the red giant regime by comparing the model-predicted effective temperatures with measurements from the third data release of the APOGEE-\textit{Kepler} joint spectroscopic and asteroseismic catalog (\citealt{2025ApJS..276...69P}, hereinafter APOKASC3). The APOKASC3 catalog contains about 15000 evolved stars, mostly red giants and red clump stars. Their effective temperature and elemental abundances are provided by the APOGEE DR17, while the asteroseismic observables, more specifically the large frequency separation $\Delta\nu$ and the frequency of maximum oscillation power $\nu_{\max}$, were measured from \textit{Kepler}. Stellar masses and radii can therefore be derived from the asteroseismic scaling relations based on the observed $\Delta\nu$, $\nu_{\max}$, and $T_{\rm eff}$ values \citep{2016ApJ...822...15S}:
\begin{equation} \label{eq:SSR}
\begin{aligned} 
\frac{M}{M_{\odot}} &\simeq 
\left(\frac{\nu_{\rm max}}{f_{\nu_{\max}} \nu_{\rm max, \odot}}\right)^3 
\left(\frac{\Delta\nu}{f_{\Delta\nu} \Delta\nu_{\odot}}\right)^{-4} 
\left(\frac{T_{\rm eff}}{T_{\rm eff, \odot}} \right)^{3/2},
\\
\frac{R}{R_{\odot}} &\simeq 
\left(\frac{\nu_{\rm max}}{f_{\nu_{\max}} \nu_{\rm max, \odot}}\right) 
\left(\frac{\Delta\nu}{f_{\Delta\nu} \Delta\nu_{\odot}}\right)^{-2} 
\left(\frac{T_{\rm eff}}{T_{\rm eff, \odot}} \right)^{1/2}.
\end{aligned}
\end{equation}
Considerable efforts have been made to quantify the uncertainty of this empirical relationship. Among others, \citet{2017ApJ...844..102H} and \citet{2018MNRAS.476.1931S} demonstrate that for dwarfs, subgiants, and giants with radii below $10 R_{\odot}$, radii estimated from the original scaling relations ($f_{\Delta\nu} = 1$, $f_{\nu_{\max}} = 1$, \citealt{1986ApJ...306L..37U,1991ApJ...368..599B}) agree with those derived from \textit{Gaia} parallaxes and photometric data to within 5\%. However, asteroseismic masses obtained from the original scaling relations are about 15\% higher than dynamical masses of eclipsing binaries \citep{2018MNRAS.476.3729B}. Corrections to the $\Delta\nu$ scaling relation through the $f_{\Delta\nu}$ factor can be provided via stellar structural models (see, e.g., \citealt{2011ApJ...743..161W,2016ApJ...822...15S,2017MNRAS.467.1433R}). Applying the $f_{\Delta\nu}$-corrected scaling relations to stars in the APOKASC sample, \citet{2019ApJ...885..166Z} and \citet{2025ApJ...979..135A} find that the seismic radii align closely with \textit{Gaia} radii to within about 2\% for low-luminosity giants. The disagreement is more pronounced for high-luminosity giants with $R \gtrsim 30 R_{\odot}$. It is worth noting that a calibration of $f_{\nu_{\max}}$ is included in the APOKASC3 catalog, such that the asteroseismic and \textit{Gaia} radii are consistent for stars up to $\approx 50 R_{\odot}$.
  
  \citet{2017ApJ...840...17T} compared theoretical $T_{\rm eff}$ given by YREC and PARSEC grids of stellar evolution models with spectroscopic values from an earlier version of the APOKASC catalog \citep{2014ApJS..215...19P}. For both model grids, they discovered that the temperature mismatches change with metallicity. After demonstrating that neither uncertainties in asteroseismic mass estimations nor spectroscopy can explain the observed trend, \citet{2017ApJ...840...17T} suggested that the metallicity-dependent temperature offset between modeling and observation indicates a positive correlation between $\alpha_{\rm MLT}$ and [Fe/H]. 
  This conclusion was supported by independent studies by \citet{2018ApJ...858...28V} and \citet{2024ApJ...974...77L}, both of whom derived a positive correlation between $\alpha_{\rm MLT}$ and [Fe/H], although the former study was based on dwarfs and subgiants, whereas the latter adopted red giants from the APOKASC2 catalog \citep{2018ApJS..239...32P} in addition to several eclipsing binaries. On the other hand, no proportionality between the mixing-length parameter and metallicity was found in $\alpha_{\rm MLT}$ calibrated from the \stagger{}-grid (\citealt{2015A&A...573A..89M}; see Fig.~6 of \citealt{2024ApJ...974...77L}). The underlying reason for this apparent tension between the $\alpha_{\rm MLT}$ values determined empirically from observational data and those determined from 3D simulations remains unclear. 
  
  As the surface properties of stars modeled with the 1D-3D coupling approach are insensitive to the value of $\alpha_{\rm MLT}$, we are in an ideal position to reexamine this issue based on the \gastag{} evolutionary tracks and the latest APOKASC3 sample. Will the systematic offset between the theoretical and measured effective temperature remain present? If yes, does the temperature mismatch depend on metallicity?
  
  From the APOKASC3 sample, we selected red giant stars based on the evolutionary state provided by the catalog and included only targets with asteroseismic $\log (g/{\rm [cm/s^2]})$ between $2$ and $3.4$. The range for iron abundance is $\rm -1 \leq [Fe/H] \leq 0.4$, selected based on the values given by APOGEE DR17. In addition, stars with estimated asteroseismic masses beyond the coverage of the \gastag{} tracks were excluded. The mass range of the \gastag{} tracks depends on initial iron abundance, as listed in Table \ref{tb:mass-range}. From these criteria we selected a sample of about 3600 stars for further investigation. Their location in the Kiel diagram and measured $\alpha$-element-to-iron ratio are shown in Fig.~\ref{fig:sample}.
  
  For each star in our sample, the modeled $T_{\rm eff}$ was obtained by interpolating evolutionary tracks at the observed $\log g$, stellar mass, and surface [M/H]. The procedure is detailed in Flowchart \ref{chart:Teff}. The effective temperatures predicted by the \gastag{} tracks were compared with spectroscopic measurements from APOGEE. For all stars in our sample, their temperature discrepancies are shown with the observed iron abundances in the \textit{left panel} of Fig.~\ref{fig:dTeff}, together with a linear fit (dash-dotted blue line). Linear regression was performed by excluding the outliers (gray dots in the \textit{left panel} of Fig.~\ref{fig:dTeff}), defined based on the median (med) of the residual of the temperature mismatch $\Delta T_{\rm eff}$: 
\begin{equation}
\left\vert (\Delta T_{\rm eff}  - \Delta T_{\rm eff,fit}) - {\rm med}(\Delta T_{\rm eff}  - \Delta T_{\rm eff,fit}) \right\vert > 3\sigma.
\end{equation}
The deviation $\sigma = 1.4826 \times {\rm MAD}$ was calculated from the median absolute deviation (MAD) of the residual. A systematic offset is clearly identified for red giants with $\rm [Fe/H] \gtrsim -0.5$, where stellar models underestimate $T_{\rm eff}$. For metal-poor stars whose ${\rm [Fe/H]} < -0.5$, the temperature differences scatter slightly above zero. Our finding that theoretical effective temperatures are cooler than observations around solar metallicity and the temperature mismatch decreases with [Fe/H] qualitatively agrees with the result of \citet[see their Fig.~3]{2017ApJ...840...17T}. Quantitatively, the slope of the linear regression, 47.2 K/dex (see also Table \ref{tb:linear-fit}), is much smaller than the values obtained from the fitting of \citet{2017ApJ...840...17T}.

  Meanwhile, comparing APOKASC red giants with their stellar evolution models, \citet{2018A&A...612A..68S} find that disagreements between APOKASC and modeled $T_{\rm eff}$ disappear when restricting the sample to stars with $\alpha$-element abundances close to solar-scaled values. Motivated by this result, we selected two subsets from our sample: a group with $\alpha$-element-to-iron ratio close to the solar value, defined as $\rm -0.07 < [\alpha/Fe] < 0.07$ following \citet{2018A&A...612A..68S}, and a second group of $\alpha$-rich stars whose $\rm [\alpha / Fe] \geq 0.14$. Stars in the $\alpha$-solar and $\alpha$-rich groups are highlighted in orange and green, respectively, in the \textit{right panel} of Fig.~\ref{fig:dTeff}, and linear fits were performed for each group with identical method as for the full sample.
  
  The linear fit for the $\alpha$-solar ($\rm -0.07 < [\alpha/Fe] < 0.07$) stars reveals an approximately 70 K temperature offset independent of metallicity, whereas a systematic offset is not observed by \citet{2018A&A...612A..68S}. Atmosphere boundary condition is the most likely cause of the difference between our work and \citet{2018A&A...612A..68S}. The latter employed a temperature-optical depth, i.e.,~$T(\tau)$, relation extracted from the \citet[VAL-C]{1981ApJS...45..635V} model atmosphere, which is known to produce evolutionary tracks that are approximately 60-70 K warmer along the RGB at solar metallicity than those obtained with the 1D-3D coupling method (cf.~\citealt{2025MNRAS.540.3400Z} Fig.~3).
  
  For the $\alpha$-rich group, the fitted temperature offset exhibits a clear dependence on [Fe/H], with larger discrepancies at $\rm [Fe/H] \gtrsim -0.5$. We note that \gastag{} tracks with $\rm [Fe/H]_i \geq -0.5$ were computed assuming solar-scaled abundances without $\alpha$-enhancement, i.e.,~$\rm [\alpha/Fe] = 0$ (\textit{right panel} of Fig.~\ref{fig:sample}), implying that models are compared to observations with the same overall metallicity but different $\alpha$-element abundances.
  As demonstrated by \citet[their Fig.~1]{2018MNRAS.476..496F}, evolutionary tracks computed using identical $X, Y, Z$ values but different $\alpha$-element abundances yield different effective temperatures in the RGB, with $\alpha$-enhanced tracks (by $\sim 0.4$ dex) being about 50 K warmer in $T_{\rm eff}$ than their solar-scaled counterparts. Had our evolutionary tracks been computed with $\alpha$-element abundances closer to the measured values for stars in the $\alpha$-rich group, we would have expected a reduced mismatch between observations and models at $\rm [Fe/H] \geq -0.5$, and consequently a flatter linear fit for the $\alpha$-rich group. However, given the current status of the \stagger{}-grid, the 1D-3D coupled models with $\alpha$-enhancements at $\rm [Fe/H] \geq -0.5$ are not self-consistent. Additional $\alpha$-enhanced 3D model atmospheres will be useful for both abundance determinations and the 1D-3D coupling approach. 
  
  It is worth emphasizing that the relationship between temperature offset and metallicity is affected by observational uncertainties. As pointed out by \citet{2024ApJ...974...77L} and demonstrated in Fig.~\ref{fig:dTeff-IRFM-APO}, effective temperatures determined from the infrared flux method (IRFM) scale of \citet{2021MNRAS.507.2684C} are higher than APOGEE values for metal-poor stars while generally lower for metal-rich ones. To investigate whether observational uncertainties impact the trend between $\Delta T_{\rm eff}$ and [Fe/H], we repeated the comparison using temperatures measured from IRFM for the red giant sample used in \citet{2024ApJ...974...77L}. The differences in measured and modeled effective temperatures are shown against the iron abundances in Fig.~\ref{fig:dTeff-Li24}, together with the linear fit to the full \citet{2024ApJ...974...77L} sample as well as two subgroups defined according to $\rm [\alpha/Fe]$ (see also Table \ref{tb:linear-fit}). 
  Comparing with temperatures derived from IRFM, we find a marginal correlation between $\Delta T_{\rm eff}$ and [Fe/H] of only 16 K per dex. A constant temperature offset of approximately 70 K remains for the $\alpha$-solar group, as seen from the \textit{right panel} of Fig.~\ref{fig:dTeff-Li24}.

  This nearly constant offset may be partially explained by observational uncertainties in the effective temperature scale. Uncertainties in the absolute calibration of photometric systems affect the IRFM $T_{\rm eff}$ scale, at the 1\% level \citep{2010A&A...512A..54C}. Uncertainties in interferometric angular diameters also limit the accuracy with which $T_{\rm eff}$ scales can be tested or calibrated \citep[e.g.,][]{c14}, particularly due to potential systematics in poorly resolved stars. State-of-the-art interferometric measurements now reach an accuracy of about 1–2\% \citep[e.g.,][]{2020MNRAS.493.2377R,2020A&A...640A..25K}. Overall, \citet{2022ApJ...927...31T} conclude that the current fundamental accuracy of the temperature scale is approximately 2\%.

  Nevertheless, the comparison with two sets of observations, combined with the fact that \gastag{} isochrones predict slightly cooler RGB for M67 (discussed in Sect.~\ref{sec:M67}), indicates that evolution calculations based on the 1D-3D coupling method likely underestimate the effective temperature and luminosity for giants at near-solar metallicity. Since stellar properties given by the 1D-3D coupled models are insensitive to $\alpha_{\rm MLT}$ and $\mean{\rm 3D}$ models are arguably among the most realistic choices for outer boundary conditions, it is difficult to locate the source of this error. Uncertainties in low-temperature opacities employed in the construction of 3D model atmospheres may be partly responsible for the temperature mismatch. From the modeling perspective, tracks computed with the 1D-3D coupling approach adopting a different grid of 3D model atmospheres will offer useful insights. 
  
  In short, comparing the effective temperature of APOKASC3 red giants with predictions from \gastag{} tracks that adopt realistic outer boundary conditions and are insensitive to $\alpha_{\rm MLT}$, we find systematic mismatches that correlate moderately or marginally with iron abundance, depending on the adopted observational data. This qualitatively agrees with results of \citet{2017ApJ...840...17T}, whereas the [Fe/H] dependence quantified in this work is weaker. For stars with nearly solar-scaled chemical compositions, $T_{\rm eff}$ given by \gastag{} tracks are about 70 K cooler than observations across the [Fe/H] range. A trend between temperature mismatch and iron abundance is identified for the $\alpha$-rich group, but we expect a weaker metallicity dependence provided correct $\alpha$-element abundances are employed in the modeling.

\section{Validating \gastag{} isochrones against observations of star clusters} \label{sec:CMD}
  
  Stars in a cluster are assumed to have formed nearly coevally from the same star-forming region, and therefore to share a common age and similar initial chemical composition\footnote{This assumption has been challenged by evidence of multiple stellar populations identified in many old clusters.}. These characteristics make star clusters a suitable testbed for validating stellar isochrones.
  In this section, synthetic CMD computed from theoretical isochrones and bolometric corrections are compared with observational data for four clusters spanning a range of metallicities. For each cluster, synthetic CMDs at the observed metallicity that best reproduce the observations are selected by visual inspection. This simple isochrone fitting is sufficient for the present study, as our goal is to examine and validate the \gastag{} isochrone instead of determining ages for clusters.
  
\subsection{M67 (NGC 2682)} \label{sec:M67}

\begin{figure}
\includegraphics[width=\columnwidth]{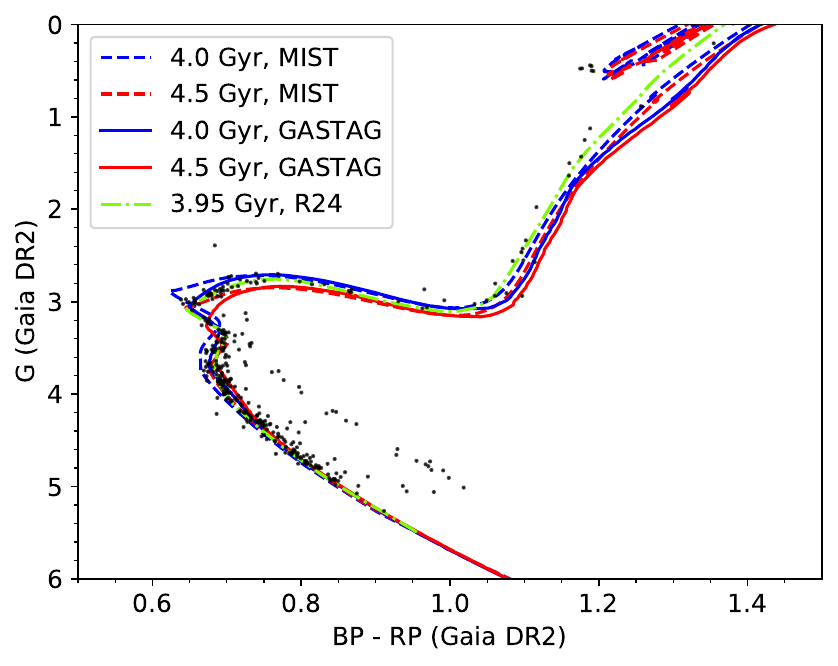}
\caption{CMDs for M67 in \textit{Gaia} photometry. The observed \textit{Gaia} color indices and absolute magnitudes (black dots) are adopted from \citet{2024MNRAS.532.2860R}, where binaries were excluded and differential reddening was corrected (following their Sect.~2.2). No additional manual data selection was applied for this figure. Two solar metallicity \gastag{} isochrones (ages of 4.0 and 4.5 Gyr) are shown as solid blue and red curves, respectively, alongside solar metallicity MIST \citep{2016ApJ...823..102C} isochrones at the same ages. The best-fitting 3.95 Gyr isochrone to M67 from \citet[see their Fig.~7]{2024MNRAS.532.2860R} is also included.
\label{fig:comp-M67}}
\end{figure}

  M67 is an open cluster with approximately solar metallicity \citep{1991AJ....102.1070H,2000A&A...360..499T,2019A&A...627A.117L} and an age close to the solar value ($\sim 4$ Gyr, \citealt{2004PASP..116..997V,litd++24-m67,2024MNRAS.532.2860R}). As a nearby solar-analog cluster, it has been extensively studied from both observational and theoretical perspectives, making it a benchmark system for testing stellar evolution models, including constraints on convective core overshoot and element diffusion (\citealt{2016ApJ...823..102C,2006ApJS..162..375V,2019ApJ...874...97S}, among others).

  We compare solar-metallicity \gastag{} isochrones with the CMD from \textit{Gaia} DR2 \citep{2018A&A...616A...1G} in Fig.~\ref{fig:comp-M67}. The adopted observational data are identical to those used in \citet{2024MNRAS.532.2860R}, where reddening is corrected as described in Sect.~2.2.3 of \citet{2024MNRAS.532.2860R} and observed apparent magnitudes are converted to absolute magnitudes using a distance modulus\footnote{Defined as the difference between apparent and absolute magnitudes.} of 9.614 \citep[Sect.~2.2.4]{2024MNRAS.532.2860R}.
  The 4 Gyr isochrone, shown as the solid blue line in Fig.~\ref{fig:comp-M67}, fits the main-sequence turn-off stars and subgiants excellently but predicts a cooler RGB than observed. We estimate effective temperatures for the observed giants with \textit{Gaia} G-band absolute magnitudes between 0.8 and 2.8 using the color-magnitude relations of \citet{2021MNRAS.507.2684C}\footnote{Code available at \url{https://github.com/casaluca/colte}}, and compare them with the corresponding isochrone $T_{\rm eff}$ at an identical G-band magnitude. For our best-fitting isochrone, the temperature differences are typically 50--100 K. This echoes the conclusion in Sect.~\ref{sec:track-APOGEE} that the 1D-3D coupled models underestimate $T_{\rm eff}$ for giants around solar metallicity. A similarly cooler theoretical $T_{\rm eff}$ along the RGB of M67 is also present in the MIST isochrones, as indicated by the dashed lines in Fig.~\ref{fig:comp-M67} (see also \citealt{2016ApJ...823..102C,2018ApJ...863...65C}).
  
  Nevertheless, by employing $T(\tau)$ relations and a varying $\alpha_{\rm MLT}$ calibrated from the \citet{2013ApJ...769...18T} grid of 3D models, with a scaling factor for $\alpha_{\rm MLT}$ adjusted to reproduce the CMD of M67, \citet{2024MNRAS.532.2860R} constructed isochrones that match the observed CMD remarkably well.
  The method based on 3D-calibrated $T(\tau)$ relations and $\alpha_{\rm MLT}$ used in \citet{2024MNRAS.532.2860R} is an alternative approach to incorporating constraints from 3D model atmospheres into stellar evolution calculations, implemented in \garstec{} as well \citep{2018MNRAS.478.5650M}. The resulting $1 M_{\odot}$, solar metallicity evolutionary track computed with tabulated $T(\tau)$ relations and $\alpha_{\rm MLT}$ calibrated from the \stagger{}-grid was compared with that obtained using the 1D-3D coupling method in \citet[their Fig.~3]{2025MNRAS.540.3400Z}. The two tracks agree well along the RGB. 
  This comparison suggests that the cooler RGB predicted by the \gastag{} tracks and isochrones is unlikely to arise from the particular method used here to improve stellar evolutionary models with 3D simulations.
  On the other hand, the choice of low-temperature opacity in evolution calculations has a notable impact on the luminosity scale of the RGB, as demonstrated in Fig.~B2 of \citet{2024MNRAS.532.2860R}. A more plausible explanation for the temperature mismatch along the RGB may therefore lie in the opacity used in 3D simulations. Verifying this hypothesis would require applying an alternative grid of 3D model atmospheres computed with opacity sources different from those used in the \stagger{}-grid.

\subsection{NGC 6791}

\begin{figure}
\includegraphics[width=\columnwidth]{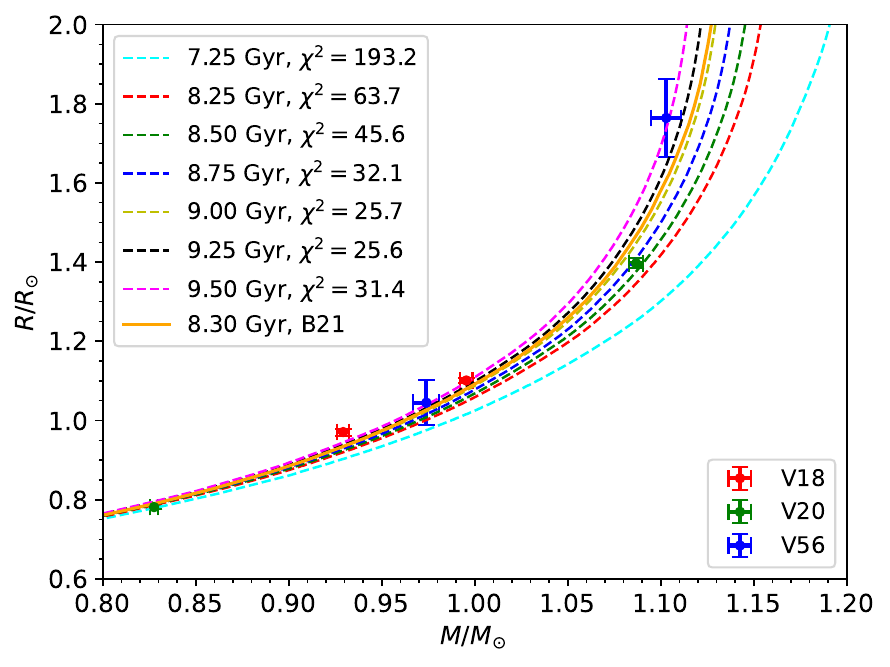}
\caption{Mass-radius relations predicted by the theoretical isochrones compared with the dynamical masses and radii of the three binary systems in NGC 6791 containing main-sequence and subgiant stars \citep{2011A&A...525A...2B,2012A&A...543A.106B,2021A&A...649A.178B}. \gastag{} isochrones at 7.25 Gyr (see text) and from 8.25 to 9.5 Gyr in 0.25 Gyr increments are shown as dashed lines. All assume solar-scaled abundance with $\rm [Fe/H]_i = 0.3$ and an initial helium mass fraction $Y_{\rm i} = 0.274$. The $\chi^2$ values in the legend are calculated from the orthogonal distances between the isochrones and the measurements, accounting for uncertainties in both mass and radius.
The solid gold line denotes the best-fitting isochrone from \citet{2021A&A...649A.178B}, computed with $\rm [Fe/H]_{i} = 0.35$ and $Y_{\rm i} = 0.3$.
\label{fig:comp-6791-bin}}
\end{figure}
\begin{figure*}
\includegraphics[width=0.99\textwidth]{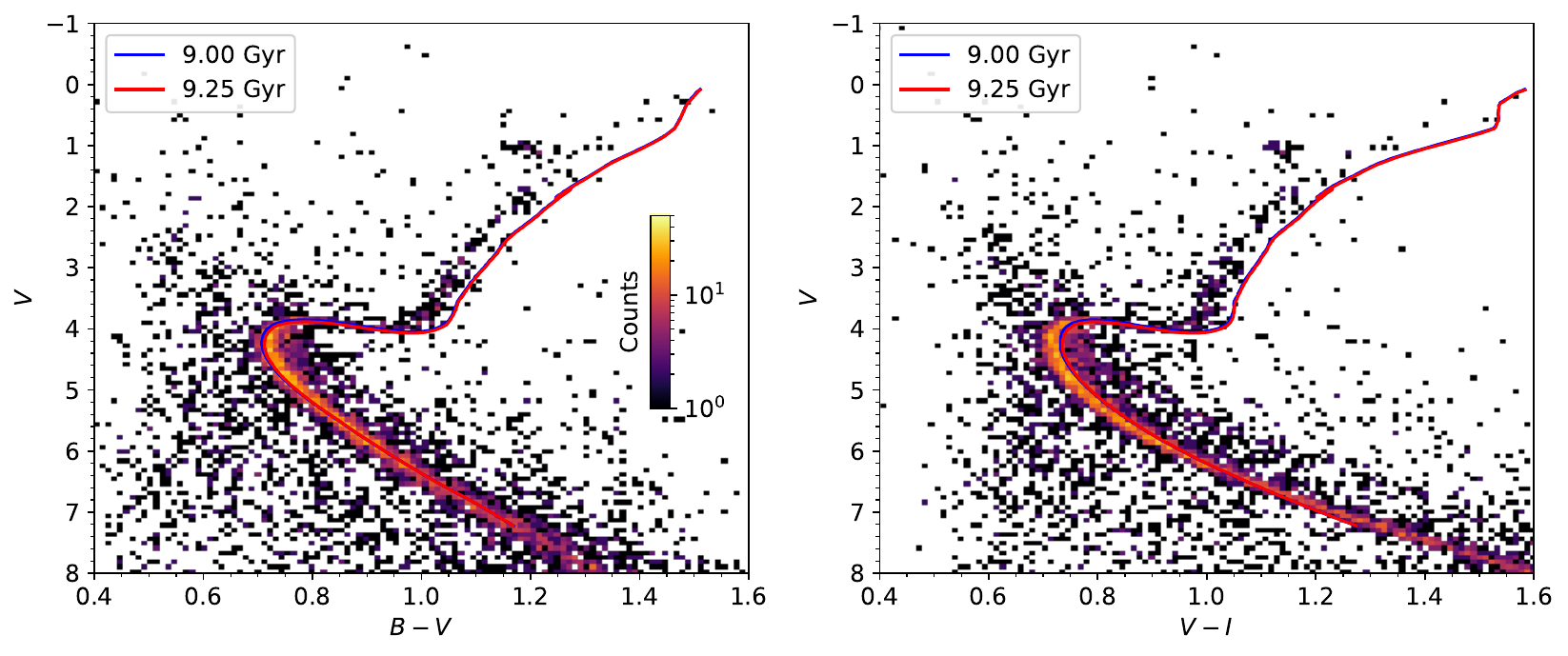}
\caption{CMDs of NGC 6791 in the $UBVRI$ photometry system obtained with the Nordic Optical Telescope, along with two synthetic diagrams at 9 and 9.25 Gyr from our isochrone and bolometric correction calculations. Differential reddening is corrected, as described in \citet{2012A&A...543A.106B}. A uniform reddening of 0.155 mag and a distance modulus of 13.5 mag are applied to the observational data to enable direct comparison with theoretical predictions.
\label{fig:comp-6791-cmd}}
\end{figure*}

  NGC 6791 is among the oldest and most metal-rich open clusters, with age estimates ranging from 6 to 12 Gyr (see \citealt{2014ApJ...786...10W} Table 1 and references therein) and a metal abundance about twice the solar value ($\rm [Fe/H] = 0.31$, \citealt{2018ApJ...867...34V}). Located within the \textit{Kepler} field of view, NGC 6791 is among the most well-studied clusters in asteroseismology. Asteroseismic parameters have been measured for more than 100 oscillating red giants and clump stars (\citealt{2011A&A...530A.100H,2011ApJ...739...13S,2012ApJ...757..190C}, among others). Combining asteroseismic with photometric data, \citet{2012MNRAS.419.2077M} determined the average mass of red giants in NGC 6791 to be approximately $1.23 M_{\odot}$, about $0.09 M_{\odot}$ greater than red clump stars, providing direct evidence for mass-loss along the RGB. Using both key asteroseismic parameters $\Delta\nu$ and $\nu_{\max}$ and photometric magnitudes, \citet{2014ApJ...786...10W} derived a true distance modulus of 13.09 mag for NGC 6791. In addition, several binary systems in this cluster have been carefully analyzed in a series of studies by \citet{2011A&A...525A...2B,2012A&A...543A.106B,2021A&A...649A.178B}. The measured dynamical masses and radii, in combination with the observed CMD, provide strong constraints on the cluster age, which \citet{2021A&A...649A.178B} determined to be 8.3 Gyr.
  
  The isochrone that best represents NGC 6791 should match both the dynamical masses and radii of binaries in the cluster and the observed CMDs. Several selected $\rm [Fe/H] = 0.3$, solar-scaled abundances theoretical isochrones are depicted in Fig.~\ref{fig:comp-6791-bin}, together with the dynamical masses and radii of the three binary systems analyzed by \citet{2011A&A...525A...2B,2012A&A...543A.106B,2021A&A...649A.178B}, as well as their best-fitting isochrone \citep[their Fig.~6]{2021A&A...649A.178B}. 
  The \gastag{} isochrones with ages of 9 and 9.25 Gyr both provide good fits to the observed mass-radius relation. Synthetic CMDs at these two ages are compared with observational data in $B$, $V$, and $I$ filters obtained from the Nordic Optical Telescope \citep{2003PASP..115..413S} in Fig.~\ref{fig:comp-6791-cmd}, where differential reddening is corrected as described in \citet{2012A&A...543A.106B}. The uniform reddening $E(B-V)$, defined as the difference between observed and intrinsic $B - V$ color for all stars, and the distance modulus are adjusted within the observed ranges (cf.~\citealt{2014ApJ...786...10W} Table 1) so that the theoretical and observed main-sequence turn-off regions overlap, yielding $E(B-V) = 0.155$ mag and a distance modulus $(m - \mathcal{M})_V$ of 13.5 mag.
  
  Similar to the M67 case, the synthetic and observed CMDs agree well in the main-sequence and subgiant phase, whereas the \gastag{} isochrones underestimate the color index in the RGB. At an absolute $V$-band magnitude of $\mathcal{M}_V = 3$, the difference between the 9 Gyr theoretical colors and the mean observed values corresponds to temperature offsets of roughly 100 K for the $B - V$ color and 115 K for $V - I$. 
  In terms of masses of NGC 6791 red giants, the mean stellar mass in the magnitude interval $1 < \mathcal{M}_V < 3$ is $1.15 M_{\odot}$ for the 9 Gyr isochrone, which is $0.08 M_{\odot}$ lower than the average RGB mass derived by \citet{2012MNRAS.419.2077M} using asteroseismic scaling relations. This discrepancy far exceeds their reported systematic uncertainty of $0.02 M_{\odot}$. Lowering the age to 7.25 Gyr would bring our predicted mass into agreement with \citet{2012MNRAS.419.2077M}, whereas the mass-radius relation given by the 7.25 Gyr isochrone clearly fails to reproduce dynamical masses and radii of the binaries (Fig.~\ref{fig:comp-6791-bin}).
  In contrast, the mean mass predicted by the \gastag{} isochrone is fully consistent with the average mass of NGC 6791 low-luminosity RGB sample in the APOKASC3 catalog (see Table 4 of \citealt{2025ApJ...979..135A}). The seismic masses in APOKASC3 were determined from asteroseismic scaling relations calibrated to yield consistent radii with those obtained from \textit{Gaia}.
  Therefore, the disagreement between our isochrone mass and the mean RGB mass from \citet{2012MNRAS.419.2077M} likely arises from the calibration method for the asteroseismic scaling relations in their study.
  
  The distance modulus and reddening adopted in the isochrone fitting, when converted to the true distance modulus, also agree well with the measurement by \citet{2014ApJ...786...10W} based on asteroseismic and photometric data. Assuming identical interstellar extinction as in \citet{2014ApJ...786...10W}, i.e.,~$A_V = 3.1 E(B-V)$, the 13.5 distance modulus obtained from our fitting leads to a true distance modulus $(m - \mathcal{M})_0 \approx 13.02$, consistent with their reported value $13.09 \pm 0.1$ mag. This agreement provides independent evidence that the \gastag{} isochrone fits the main-sequence, turn-off, and subgiant region of NGC 6791 well.
  
  The above comparisons indicate that our isochrones faithfully reproduce most of the observed properties of this metal-rich cluster. The only drawback is that the predicted RGB color is too red, which corresponds to temperature mismatches on the order of 100 K. Again, this result echoes the conclusion in Sect.~\ref{sec:track-APOGEE} (Figs.~\ref{fig:dTeff} and \ref{fig:dTeff-Li24}). Taken together, these findings reveal that the temperature discrepancy for $\rm [Fe/H] \gtrsim 0$ red giants is truly mainly due to deficiencies in modeling rather than systematic errors in the observational data.

\subsection{47 Tuc (NGC 104)}

\begin{figure*}
\includegraphics[width=0.99\textwidth]{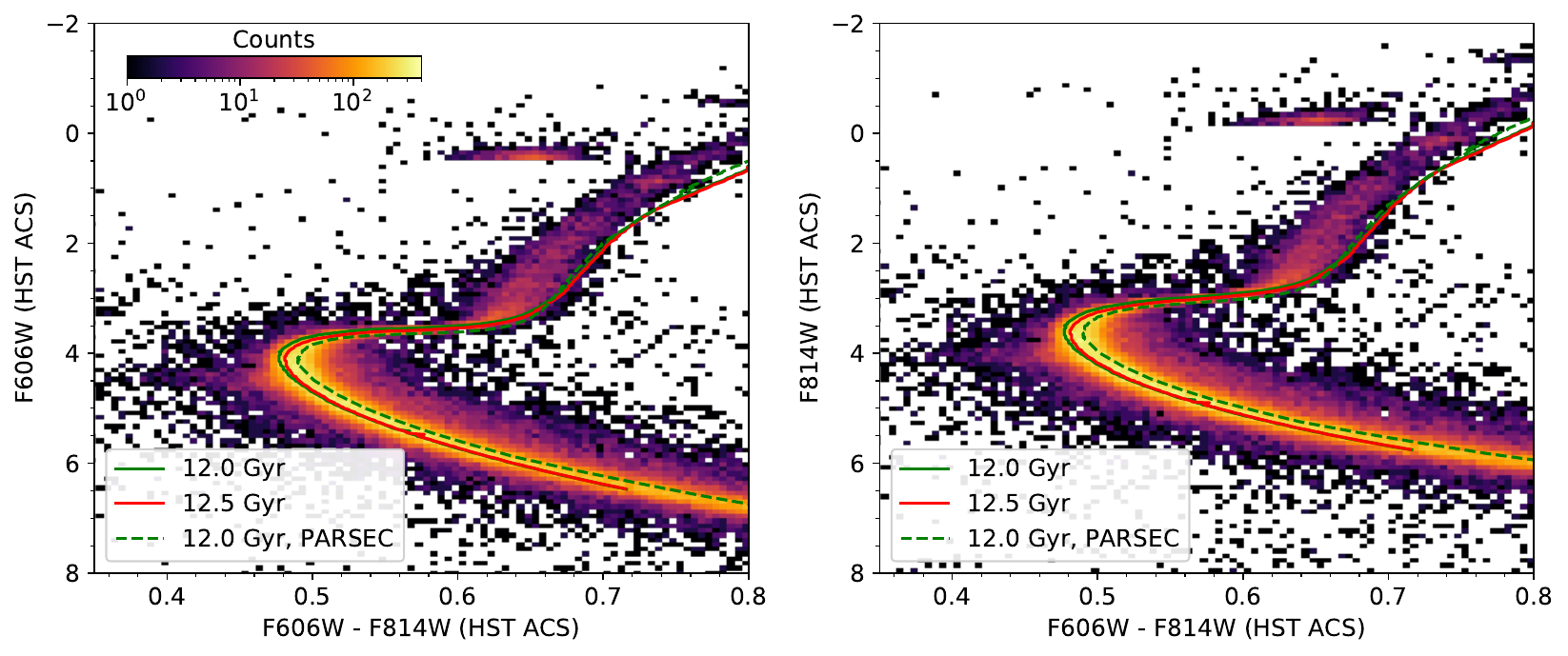}
\caption{Comparison of theoretical isochrones and synthetic colors with the observed CMDs of 47 Tuc from the HST ACS survey. A 0.035 mag correction for uniform reddening and a distance modulus of 13.3 mag are applied to the HST data to facilitate comparison with the theoretical predictions. The \gastag{} isochrones adopt an initial helium mass fraction $Y_{\rm i} = 0.250$ and initial metal abundances $\rm [Fe/H]_i = -0.75$, $\rm [\alpha/Fe]_i = 0.2$. The PARSEC 12 Gyr isochrone, reproduced from Fig.~5 of \citet{2018MNRAS.476..496F}, is shown as the dashed green lines for comparison. 
\label{fig:47Tuc}}
\end{figure*}

  47 Tuc is an old globular cluster with an age of approximately 11--12.5 Gyr \citep{2025ApJ...987...52Y}. It has a mean iron abundance of $\rm [Fe/H] = -0.79 \pm 0.09$ and an $\alpha$-element-to-iron ratio $\rm [\alpha/Fe] \sim 0.3$ \citep{2014ApJ...780...94C}. The CMD of 47 Tuc exhibits a broadened main-sequence and a spread in color along the RGB. The widely accepted interpretation is that multiple stellar populations are present in this cluster. Detailed abundance analysis indicates that the chemical composition of the first population is generally consistent with halo stars at the same [Fe/H]. In contrast, the second population is enhanced in nitrogen and sodium while depleted in carbon and oxygen \citep{2012ApJ...744...58M}. It is likely also enriched in helium \citep{2010MNRAS.408..999D}. As a bright and nearby cluster, 47 Tuc is a suitable target for validating isochrones computed with enhanced $\alpha$-element abundances (e.g., \citealt{2014ApJ...794...72V,2018MNRAS.476..496F}).
  
  The \gastag{} isochrones with $\rm [Fe/H] = -0.75$, $\rm [\alpha/Fe] = 0.2$ are compared with calibrated photometric data from the Hubble Space Telescope (HST) ACS Survey of Galactic Globular Clusters \citep{2007AJ....133.1658S,2008AJ....135.2055A}. Figure~\ref{fig:47Tuc} presents the observed CMD for two HST bandpasses, corrected for a 0.035 mag uniform reddening and a distance modulus of 13.3 mag that converts apparent magnitudes to absolute values, along with the best-fitting isochrones. The adopted reddening and distance modulus values were chosen within the observed ranges (see \citealt{2025ApJ...987...52Y} and references therein) too ensure that the observed and synthetic main-sequence turn-off regions overlap.
  
  Our 12.5 Gyr isochrone provides a good match to the observational data in the main-sequence and subgiant regions, but appears systematically too red along the RGB. At magnitude 2.5 in the F606W filter, the temperature offset between \gastag{} prediction and the center of the heat map ridge is approximately 100 K. The underestimation of the color index, and hence effective temperature, along the RGB of 47 Tuc is a known issue from Dartmouth Stellar Evolution Program and PARSEC isochrones with similar iron abundance, $\alpha$-enhancement and helium mass fraction (cf.~\citealt{2009IAUS..258..171D} Fig.~1, \citealt{2018MNRAS.476..496F} Fig.~5, and the dashed green lines in Fig.~\ref{fig:47Tuc}). 
   
  Because \gastag{} was constructed using a method that greatly reduces uncertainties associated with surface boundary conditions and mixing-length parameters, the difference between modeling and observation can be more confidently attributed to other factors, such as the presence of multiple populations in this cluster or uncertainties in low-temperature opacities. 
  The underlying chemical composition of our isochrones shown in Fig.~\ref{fig:47Tuc}, i.e.,~$\rm [Fe/H] = -0.75$, $\rm [\alpha/Fe] = 0.2$, and $Y = 0.250$, is a reasonable representation of the first generation population of this cluster. However, the second generation is depleted in C and O relative to typical Population II stars and is likely enriched in He. All factors, especially the lower oxygen abundance and higher helium mass fraction, lead to hotter evolutionary tracks at fixed [Fe/H] (see \citealt{2012ApJ...755...15V} Fig.~7 and \citealt{2018MNRAS.476..496F} Fig.~5). A convincing isochrone fit for the second generation would therefore require evolution calculations with helium and CNO abundances tailored to the measured values, which is beyond the scope of this work (but see \citealt{2022MNRAS.509.4208V} for progress in this direction).
  Nevertheless, the presence of multiple populations is unlikely to be the dominant source of the observed color offset. Even when restricting the comparison to first-generation stars alone (Ulloa Solis et al., in prep.), the isochrones remain systematically too cool.

\subsection{M55 (NGC 6809)}

\begin{figure*}
\includegraphics[width=0.99\textwidth]{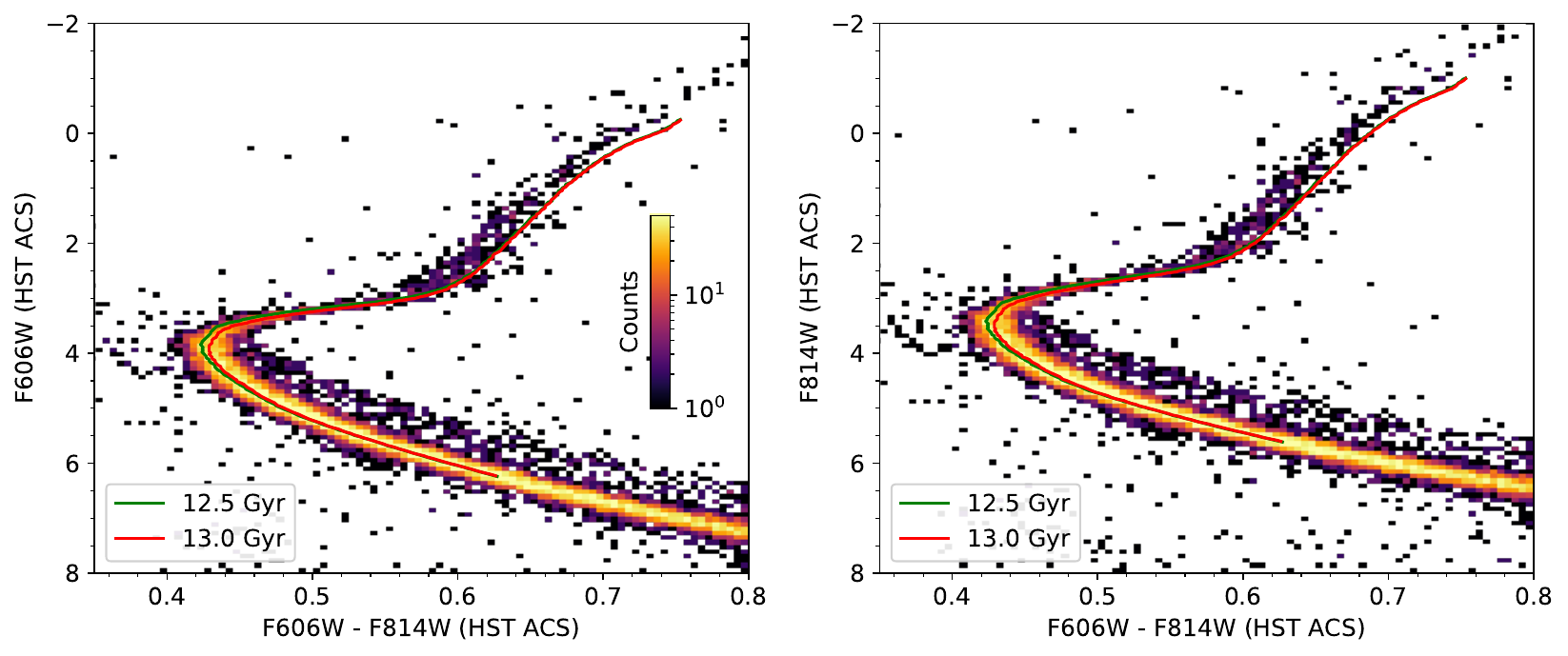}
\caption{Comparison between the \gastag{} synthetic CMDs and observational data from the HST ACS survey for the M55 globular cluster. A 0.11 mag correction for uniform reddening and a true distance modulus of 13.65 mag are applied to the observational data to facilitate comparison with the synthetic color and absolute magnitudes. The \gastag{} isochrones adopt an initial helium mass fraction $Y_{\rm i} = 0.2464$ and initial metal abundances $\rm [Fe/H]_i = -1.9$, $\rm [\alpha/Fe]_i = 0.4$.
\label{fig:M55}}
\end{figure*}

  M55 is an old, metal-poor globular cluster with an estimated age around 12.2 Gyr \citep{2025ApJ...987...52Y}. It has a mean iron abundance $\rm [Fe/H] \simeq -2$, and $\alpha$ and iron-peak element abundances consistent with halo field stars ($\rm [\alpha/Fe] \simeq 0.4$, \citealt{2019MNRAS.483.1674R}). The low metallicity and relative proximity of M55 make it a suitable target to validate our isochrones. The \gastag{} isochrones computed with initial abundances $\rm [Fe/H]_i = -1.9$, $\rm [\alpha/Fe]_i = 0.4$ were compared with CMDs obtained from HST data, as demonstrated in Fig.~\ref{fig:M55}. The best-fitting isochrone has an age of 13 Gyr. 
  In the low-metallicity regime, the agreement between the models and observations is good along the main sequence and the subgiant branch, whereas the \gastag{} isochrone continues to predict slightly redder colors along the RGB. At F606W magnitude 1.5, the difference in $\rm F606W-F814W$ color between the median observed value and the prediction from our 13 Gyr isochrone is approximately 0.016 mag, translating to a temperature offset of roughly 70~K. The underestimation of effective temperatures along the RGB by \gastag{} models is present in all clusters considered.

\section{Summary and conclusions}
 
  We presented the \gastag{} stellar evolutionary tracks and isochrones for low-mass stars, constructed from coupled 1D \garstec{} stellar interior and 3D \stagger{} stellar atmosphere simulations. These tracks and isochrones incorporate one of the most realistic treatments of stellar near-surface layers currently available. They span a metallicity range from $\rm [Fe/H] = -1.9$ to $0.4$. We provided evolutionary tracks and stellar structural models from the zero-age main-sequence to red giants up to at least $100 L_{\odot}$. Each structural model is accompanied by theoretical oscillation frequencies of radial modes for asteroseismic applications.
  Isochrones with ages greater than $\sim 2$ Gyr are available, making them suitable for studies of globular clusters and old open clusters. For each isochrone, bolometric corrections for a large number of photometric systems and bandpasses were computed using the Synthetic Stellar Photometry Package validated and tested by \citet{2014MNRAS.444..392C,2018MNRAS.475.5023C,2018MNRAS.479L.102C}, enabling direct and robust comparison with observed CMDs.

  Our evolution calculations in the RGB were evaluated by comparing the model-predicted $T_{\rm eff}$ with spectroscopic measurements from the APOKASC3 catalog, as well as temperatures determined through the IRFM. Discrepancies between the models and observations are unambiguously present for $\rm [Fe/H] > -0.5$ stars, where our models systematically underestimate the effective temperatures for giants. Moreover, the temperature mismatch $\Delta T_{\rm eff}$ increases with metallicity, in qualitative agreement with the conclusion of \citet{2017ApJ...840...17T}. Quantitatively, the correlation between $\Delta T_{\rm eff}$ and $\rm [Fe/H]$ is moderate or marginal, depending on the adopted measured temperature scale -- a linear regression of the temperature offset against iron abundance yields a slope of approximately 47 $\rm K/dex$ when comparing with APOKASC3 temperatures and 16 $\rm K/dex$ when using IRFM temperatures. For both observational datasets, we find a much weaker metallicity dependence than that reported by \citet{2017ApJ...840...17T}.
  
  The \gastag{} synthetic CMDs, obtained from theoretical isochrones and calculations of bolometric corrections, were validated against observed CMDs for four well-studied star clusters spanning ages from $\sim 4$ to $\sim 13$ Gyr and metallicities from $\rm [Fe/H] = -1.9$ to 0.3. In all cases, our best-fitting isochrones successfully reproduce the main-sequence, turn-off, and subgiant regions of the observed CMDs, and the estimated cluster ages are consistent with the corresponding age ranges reported in previous studies. For the open cluster NGC 6791, which has a large sample of red giants with asteroseismic measurements, the mean red giant mass derived from \gastag{} isochrones agrees well with the asteroseismic masses provided by the APOKASC3 catalog. The true distance modulus of NGC 6791 estimated from our isochrone fitting is also consistent with an independent distance determination based on asteroseismic and photometric data.
  
  However, along the RGB, our isochrones predict colors that are systematically too red for all star clusters considered. The discrepancies in color indices translate to temperature underestimations on the order of 100 K for M67, NGC 6791, and 47 Tuc, and about 70 K for M55. For 47 Tuc, the chemical composition adopted in our calculations is representative of the first generation but cannot account for the second generation, which is depleted in carbon and oxygen and likely enhanced in helium relative to the first generation.
  For M67 and NGC 6791, we conclude that the disagreement between the models and observations arises from deficiencies in the stellar models rather than from systematics in observational data, given that \gastag{} tracks also underestimate $T_{\rm eff}$ for solar-metallicity and metal-rich giants (Sect.~\ref{sec:track-APOGEE}). Since the \gastag{} tracks and isochrones were computed with highly realistic surface boundary conditions, are insensitive to the choice of $\alpha_{\rm MLT}$, and are based on the 1D-3D coupling approach that has been extensively tested and validated \citep{2018MNRAS.481L..35J,2020MNRAS.491.1160M,2025MNRAS.540.3400Z,2026ApJ...996...83L}, the error can be more confidently attributed to other ingredients in the model, such as the adopted helium mass fractions or uncertainties in low-temperature opacities. Previous studies have demonstrated that both factors influence the predicted effective temperatures along the RGB \citep{2018MNRAS.476..496F,2024MNRAS.532.2860R}.
  
  Another important source of uncertainty involves the choice of stellar evolution code. \citet{2020A&A...635A.164S} carefully compared the predicted fundamental properties of low-mass stars across nine widely used stellar evolutionary codes, including \garstec{}. With identical input physics, the effective temperatures predicted by the various stellar evolution codes differ by 30--40 K in the RGB phase for all stellar masses considered. This discrepancy arises from the numerical details of the codes and thus represents minimal systematical uncertainties in evolutionary modeling.

  Our tests and validations suggest several avenues for further improvement and investigation. First, implementing the 1D-3D coupling scheme in other stellar evolutionary codes and/or using an alternative grid of 3D model atmospheres -- computed with different opacity sources -- could provide valuable insight into the temperature discrepancies along the RGB.
  Second, evolutionary tracks and isochrones based on varying helium mass fractions and non-solar-scaled chemical compositions are necessary to better characterize metal-poor stars with different $\alpha$-element-to-iron ratios, as well as different populations in globular clusters. Both aspects are crucial for advancing our understanding of the chemical evolution of the Galaxy. In the context of the 1D-3D coupling approach, this requires additional 3D models with varying helium mass fractions and $\rm [\alpha/Fe]$ values (the latter being the most relevant between $\rm -1 \leq [Fe/H] \leq 0$) to ensure consistent coupling with 1D stellar interior models. Extending the coverage of 3D model atmospheres in chemical composition space not only benefits the 1D-3D coupling method but will also be useful for detailed 3D abundance analyses for stars with non-solar-scaled compositions. Both improvements contribute to a more precise age determination of stars and clusters.

\begin{acknowledgements}
  The authors are grateful to Jakob R{\o}rsted and Andreas J{\o}rgensen for their foundational work and early development of the 1D-3D coupling approach. We thank Claudia Reyes for providing observational data for M67, Xiaoting Fu for providing the PARSEC isochrone for 47 Tuc, and Karsten Brogaard for valuable comments on the color-magnitude diagrams of NGC 6791. Achim Weiss' help and suggestions regarding the details of \garstec{} are greatly appreciated. We also thank J{\o}rgen Christensen-Dalsgaard for his help on the \textsc{adipls} code. YZ thanks the hospitality of Tao Wu and the Yunnan Observatories during his visit. 
  YZ acknowledges support from the European Union's Horizon 2020 research and innovation program under the Marie Skłodowska-Curie grant agreement No.~101150921.
This research was supported by computational resources provided by the Australian Government through the National Computational Infrastructure (NCI) under the National Computational Merit Allocation Scheme and the ANU Merit Allocation Scheme (project y89).
\end{acknowledgements}

\section*{Data availability}

The \gastag{} evolutionary tracks and isochrones are available on Zenodo: \url{https://zenodo.org/records/20416554} and the NIRD Research Data Archive: \url{https://data.archive.sigma2.no/dataset/gastag}. The associated DOI is: \url{https://doi.org/10.11582/2026.h8444rio}


\bibliographystyle{aa} 
\bibliography{References.bib}

@String{mnras = "Mon.\ Not.\ Roy.\ Astr.\ Soc."}

@String{aap = "Astron.\ Astrophys."}

@String{apj =    "Astrophys.\ J."}

@String{apjs =    "Astrophys.\ J. Suppl. Ser."}

@String{aj = "Astron.\ J."}

@String{arxiv =        "http://www.arxiv.org/abs/"}

@ARTICLE{c14,
       author = {{Casagrande}, L. and {Portinari}, L. and {Glass}, I.~S. and {Laney}, D. and {Silva Aguirre}, V. and {Datson}, J. and {Andersen}, J. and {Nordstr{\"o}m}, B. and {Holmberg}, J. and {Flynn}, C. and {Asplund}, M.},
        title = "{Towards stellar effective temperatures and diameters at 1 per cent accuracy for future surveys}",
      journal = {\mnras},
         year = 2014,
        month = apr,
       volume = {439},
       number = {2},
        pages = {2060-2073},
          doi = {10.1093/mnras/stu089},
archivePrefix = {arXiv},
       eprint = {1401.3754},
 primaryClass = {astro-ph.SR},
       adsurl = {https://ui.adsabs.harvard.edu/abs/2014MNRAS.439.2060C}
}

@article{dotter2016,
       author = {{Dotter}, Aaron},
        title = "{MESA Isochrones and Stellar Tracks (MIST) 0: Methods for the Construction of Stellar Isochrones}",
      journal = {\apjs},
         year = 2016,
        month = jan,
       volume = {222},
       number = {1},
          eid = {8},
        pages = {8},
          doi = {10.3847/0067-0049/222/1/8},
archivePrefix = {arXiv},
       eprint = {1601.05144},
 primaryClass = {astro-ph.SR},
       adsurl = {https://ui.adsabs.harvard.edu/abs/2016ApJS..222....8D}
}

@ARTICLE{1958ZA.....46..108B,
       author = {{B{\"o}hm-Vitense}, E.},
        title = "{{\"U}ber die Wasserstoffkonvektionszone in Sternen verschiedener Effektivtemperaturen und Leuchtkr{\"a}fte. Mit 5 Textabbildungen}",
      journal = {Zeitschrift f\"ur Astrophysik},
         year = 1958,
        month = jan,
       volume = {46},
        pages = {108},
       adsurl = {https://ui.adsabs.harvard.edu/abs/1958ZA.....46..108B}
}

@ARTICLE{1965ApJ...142..841H,
       author = {{Henyey}, Louis and {Vardya}, M.~S. and {Bodenheimer}, Peter},
        title = "{Studies in Stellar Evolution. III. The Calculation of Model Envelopes.}",
      journal = {\apj},
         year = 1965,
        month = Oct,
       volume = 142,
        pages = {841},
          doi = {10.1086/148357},
       adsurl = {https://ui.adsabs.harvard.edu/abs/1965ApJ...142..841H}
}

@ARTICLE{1981ApJS...45..635V,
       author = {{Vernazza}, J.~E. and {Avrett}, E.~H. and {Loeser}, R.},
        title = "{Structure of the solar chromosphere. III. Models of the EUV brightness components of the quiet sun.}",
      journal = {\apjs},
         year = 1981,
        month = apr,
       volume = {45},
        pages = {635-725},
          doi = {10.1086/190731},
       adsurl = {https://ui.adsabs.harvard.edu/abs/1981ApJS...45..635V}
}

@ARTICLE{1986ApJ...306L..37U,
       author = {{Ulrich}, R.~K.},
        title = "{Determination of Stellar Ages from Asteroseismology}",
      journal = {\apj},
         year = "1986",
        month = "Jul",
       volume = {306},
        pages = {L37},
          doi = {10.1086/184700},
       adsurl = {https://ui.adsabs.harvard.edu/abs/1986ApJ...306L..37U}
}

@ARTICLE{1986A&A...160..116K,
       author = {{Kuhfuss}, R.},
        title = "{A model for time-dependent turbulent convection}",
      journal = {\aap},
         year = 1986,
        month = may,
       volume = {160},
       number = {1},
        pages = {116-120},
       adsurl = {https://ui.adsabs.harvard.edu/abs/1986A&A...160..116K}
}

@ARTICLE{1988ApJ...331..815M,
   author = {{Mihalas}, D. and {Dappen}, W. and {Hummer}, D.~G.},
    title = "{The equation of state for stellar envelopes. II - Algorithm and selected results}",
  journal = {\apj},
     year = 1988,
    month = aug,
   volume = 331,
    pages = {815-825},
      doi = {10.1086/166601},
   adsurl = {http://adsabs.harvard.edu/abs/1988ApJ...331..815M}
}

@ARTICLE{1988Natur.336..634C,
       author = {{Christensen-Dalsgaard}, J. and {Dappen}, W. and {Lebreton}, Y.},
        title = "{Solar oscillation frequencies and the equation of state}",
      journal = {\nat},
         year = 1988,
        month = dec,
       volume = {336},
       number = {6200},
        pages = {634-638},
          doi = {10.1038/336634a0},
       adsurl = {https://ui.adsabs.harvard.edu/abs/1988Natur.336..634C}
}

@ARTICLE{1990A&A...239..443S,
       author = {{Steffen}, M.},
        title = "{A simple method for monotonic interpolation in one dimension.}",
      journal = {\aap},
         year = 1990,
        month = nov,
       volume = {239},
        pages = {443-450},
       adsurl = {https://ui.adsabs.harvard.edu/abs/1990A&A...239..443S}
}

@ARTICLE{1991AJ....102.1070H,
       author = {{Hobbs}, L.~M. and {Thorburn}, J.~A.},
        title = "{On the Metallicity and the Turnoff Temperature of M67}",
      journal = {\aj},
         year = 1991,
        month = sep,
       volume = {102},
        pages = {1070},
          doi = {10.1086/115933},
       adsurl = {https://ui.adsabs.harvard.edu/abs/1991AJ....102.1070H}
}

@ARTICLE{1991ApJ...368..599B,
   author = {{Brown}, T.~M. and {Gilliland}, R.~L. and {Noyes}, R.~W. and 
	{Ramsey}, L.~W.},
    title = "{Detection of possible p-mode oscillations on Procyon}",
  journal = {\apj},
     year = 1991,
    month = feb,
   volume = 368,
    pages = {599-609},
      doi = {10.1086/169725},
   adsurl = {http://adsabs.harvard.edu/abs/1991ApJ...368..599B}
}

@ARTICLE{1993ApJ...412..752I,
       author = {{Iglesias}, Carlos A. and {Rogers}, Forrest J.},
        title = "{Radiative Opacities for Carbon- and Oxygen-rich Mixtures}",
      journal = {\apj},
         year = "1993",
        month = "Aug",
       volume = {412},
        pages = {752},
          doi = {10.1086/172958},
       adsurl = {https://ui.adsabs.harvard.edu/abs/1993ApJ...412..752I}
}

@ARTICLE{1994ApJ...421..828T,
       author = {{Thoul}, Anne A. and {Bahcall}, John N. and {Loeb}, Abraham},
        title = "{Element Diffusion in the Solar Interior}",
      journal = {\apj},
         year = 1994,
        month = feb,
       volume = {421},
        pages = {828},
          doi = {10.1086/173695},
archivePrefix = {arXiv},
       eprint = {astro-ph/9304005},
 primaryClass = {astro-ph},
       adsurl = {https://ui.adsabs.harvard.edu/abs/1994ApJ...421..828T}
}

@ARTICLE{1996ApJ...464..943I,
       author = {{Iglesias}, Carlos A. and {Rogers}, Forrest J.},
        title = "{Updated Opal Opacities}",
      journal = {\apj},
         year = "1996",
        month = "Jun",
       volume = {464},
        pages = {943},
          doi = {10.1086/177381},
       adsurl = {https://ui.adsabs.harvard.edu/abs/1996ApJ...464..943I}
}

@ARTICLE{1996ApJ...456..902R,
       author = {{Rogers}, Forrest J. and {Swenson}, Fritz J. and {Iglesias}, Carlos A.},
        title = "{OPAL Equation-of-State Tables for Astrophysical Applications}",
      journal = {\apj},
         year = 1996,
        month = jan,
       volume = {456},
        pages = {902},
          doi = {10.1086/176705},
       adsurl = {https://ui.adsabs.harvard.edu/abs/1996ApJ...456..902R}
}

@ARTICLE{1996A&A...313..497F,
       author = {{Freytag}, B. and {Ludwig}, H. -G. and {Steffen}, M.},
        title = "{Hydrodynamical models of stellar convection. The role of overshoot in DA white dwarfs, A-type stars, and the Sun.}",
      journal = {\aap},
         year = 1996,
        month = sep,
       volume = {313},
        pages = {497-516},
       adsurl = {https://ui.adsabs.harvard.edu/abs/1996A&A...313..497F}
}

@ARTICLE{1996Sci...272.1286C,
   author = {{Christensen-Dalsgaard}, J. and {Dappen}, W. and {Ajukov}, S.~V. and 
	{Anderson}, E.~R. and {Antia}, H.~M. and {Basu}, S. and {Baturin}, V.~A. and 
	{Berthomieu}, G. and {Chaboyer}, B. and {Chitre}, S.~M. and 
	{Cox}, A.~N. and {Demarque}, P. and {Donatowicz}, J. and {Dziembowski}, W.~A. and 
	{Gabriel}, M. and {Gough}, D.~O. and {Guenther}, D.~B. and {Guzik}, J.~A. and 
	{Harvey}, J.~W. and {Hill}, F. and {Houdek}, G. and {Iglesias}, C.~A. and 
	{Kosovichev}, A.~G. and {Leibacher}, J.~W. and {Morel}, P. and 
	{Proffitt}, C.~R. and {Provost}, J. and {Reiter}, J. and {Rhodes}, Jr., E.~J. and 
	{Rogers}, F.~J. and {Roxburgh}, I.~W. and {Thompson}, M.~J. and 
	{Ulrich}, R.~K.},
    title = "{The Current State of Solar Modeling}",
  journal = {Science},
     year = 1996,
    month = may,
   volume = 272,
    pages = {1286-1292},
      doi = {10.1126/science.272.5266.1286},
   adsurl = {http://adsabs.harvard.edu/abs/1996Sci...272.1286C}
}

@ARTICLE{1999A&A...351..247V,
       author = {{Varenne}, O. and {Monier}, R.},
        title = "{Chemical abundances of A and F-type stars: the Hyades open cluster}",
      journal = {\aap},
         year = "1999",
        month = "Nov",
       volume = {351},
        pages = {247-266},
       adsurl = {https://ui.adsabs.harvard.edu/abs/1999A&A...351..247V}
}

@ARTICLE{1999NuPhA.656....3A,
       author = {{Angulo}, C. and {Arnould}, M. and {Rayet}, M. and {Descouvemont}, P. and {Baye}, D. and {Leclercq-Willain}, C. and {Coc}, A. and {Barhoumi}, S. and {Aguer}, P. and {Rolfs}, C. and {Kunz}, R. and {Hammer}, J.~W. and {Mayer}, A. and {Paradellis}, T. and {Kossionides}, S. and {Chronidou}, C. and {Spyrou}, K. and {degl'Innocenti}, S. and {Fiorentini}, G. and {Ricci}, B. and {Zavatarelli}, S. and {Providencia}, C. and {Wolters}, H. and {Soares}, J. and {Grama}, C. and {Rahighi}, J. and {Shotter}, A. and {Lamehi Rachti}, M.},
        title = "{A compilation of charged-particle induced thermonuclear reaction rates}",
      journal = {\nphysa},
         year = 1999,
        month = aug,
       volume = {656},
       number = {1},
        pages = {3-183},
          doi = {10.1016/S0375-9474(99)00030-5},
       adsurl = {https://ui.adsabs.harvard.edu/abs/1999NuPhA.656....3A}
}

@ARTICLE{2000A&A...360..499T,
       author = {{Tautvai{\v{s}}iene}, G. and {Edvardsson}, B. and {Tuominen}, I. and {Ilyin}, I.},
        title = "{Chemical composition of evolved stars in the open cluster M 67}",
      journal = {\aap},
         year = 2000,
        month = aug,
       volume = {360},
        pages = {499-508},
          doi = {10.48550/arXiv.astro-ph/0006001},
archivePrefix = {arXiv},
       eprint = {astro-ph/0006001},
 primaryClass = {astro-ph},
       adsurl = {https://ui.adsabs.harvard.edu/abs/2000A&A...360..499T}
}

@ARTICLE{2003PASP..115..413S,
       author = {{Stetson}, Peter B. and {Bruntt}, Hans and {Grundahl}, Frank},
        title = "{Homogeneous Photometry. III. A Star Catalog for the Open Cluster NGC 6791}",
      journal = {\pasp},
         year = 2003,
        month = apr,
       volume = {115},
       number = {806},
        pages = {413-447},
          doi = {10.1086/368337},
       adsurl = {https://ui.adsabs.harvard.edu/abs/2003PASP..115..413S}
}

@ARTICLE{2004PASP..116..997V,
       author = {{VandenBerg}, Don A. and {Stetson}, P.~B.},
        title = "{On the Old Open Clusters M67 and NGC 188: Convective Core Overshooting, Color-Temperature Relations, Distances, and Ages}",
      journal = {\pasp},
         year = 2004,
        month = nov,
       volume = {116},
       number = {825},
        pages = {997-1011},
          doi = {10.1086/426340},
       adsurl = {https://ui.adsabs.harvard.edu/abs/2004PASP..116..997V}
}

@ARTICLE{2004Irwin...feos1,
   author = {{Irwin}, A. W.},
    title = "{The FreeEOS Code for Calculating the Equation of State for Stellar Interiors I: An Improved EFF-Style Approximation for the Fermi-Dirac Integrals}",
  journal = {unpublished, available at \url{http://freeeos.sourceforge.net/eff_fit.pdf}},
     year = 2004,
}

@ARTICLE{2005ApJ...623..585F,
       author = {{Ferguson}, Jason W. and {Alexander}, David R. and {Allard}, France and {Barman}, Travis and {Bodnarik}, Julia G. and {Hauschildt}, Peter H. and {Heffner-Wong}, Amanda and {Tamanai}, Akemi},
        title = "{Low-Temperature Opacities}",
      journal = {\apj},
         year = 2005,
        month = apr,
       volume = {623},
       number = {1},
        pages = {585-596},
          doi = {10.1086/428642},
archivePrefix = {arXiv},
       eprint = {astro-ph/0502045},
 primaryClass = {astro-ph},
       adsurl = {https://ui.adsabs.harvard.edu/abs/2005ApJ...623..585F}
}

@ARTICLE{2006ApJS..162..375V,
       author = {{VandenBerg}, Don A. and {Bergbusch}, Peter A. and {Dowler}, Patrick D.},
        title = "{The Victoria-Regina Stellar Models: Evolutionary Tracks and Isochrones for a Wide Range in Mass and Metallicity that Allow for Empirically Constrained Amounts of Convective Core Overshooting}",
      journal = {\apjs},
         year = 2006,
        month = feb,
       volume = {162},
       number = {2},
        pages = {375-387},
          doi = {10.1086/498451},
archivePrefix = {arXiv},
       eprint = {astro-ph/0510784},
 primaryClass = {astro-ph},
       adsurl = {https://ui.adsabs.harvard.edu/abs/2006ApJS..162..375V}
}

@ARTICLE{2007ApJ...669.1167B,
       author = {{Barnes}, Sydney A.},
        title = "{Ages for Illustrative Field Stars Using Gyrochronology: Viability, Limitations, and Errors}",
      journal = {\apj},
         year = 2007,
        month = nov,
       volume = {669},
       number = {2},
        pages = {1167-1189},
          doi = {10.1086/519295},
archivePrefix = {arXiv},
       eprint = {0704.3068},
 primaryClass = {astro-ph},
       adsurl = {https://ui.adsabs.harvard.edu/abs/2007ApJ...669.1167B}
}

@ARTICLE{2007AJ....133.1658S,
       author = {{Sarajedini}, Ata and {Bedin}, Luigi R. and {Chaboyer}, Brian and {Dotter}, Aaron and {Siegel}, Michael and {Anderson}, Jay and {Aparicio}, Antonio and {King}, Ivan and {Majewski}, Steven and {Mar{\'\i}n-Franch}, A. and {Piotto}, Giampaolo and {Reid}, I. Neill and {Rosenberg}, Alfred},
        title = "{The ACS Survey of Galactic Globular Clusters. I. Overview and Clusters without Previous Hubble Space Telescope Photometry}",
      journal = {\aj},
         year = 2007,
        month = apr,
       volume = {133},
       number = {4},
        pages = {1658-1672},
          doi = {10.1086/511979},
archivePrefix = {arXiv},
       eprint = {astro-ph/0612598},
 primaryClass = {astro-ph},
       adsurl = {https://ui.adsabs.harvard.edu/abs/2007AJ....133.1658S}
}

@ARTICLE{2008PhST..133a4003P,
       author = {{Plez}, B.},
        title = "{MARCS model atmospheres}",
      journal = {Physica Scripta Volume T},
         year = 2008,
        month = dec,
       volume = {133},
          eid = {014003},
        pages = {014003},
          doi = {10.1088/0031-8949/2008/T133/014003},
archivePrefix = {arXiv},
       eprint = {0810.2375},
 primaryClass = {astro-ph},
       adsurl = {https://ui.adsabs.harvard.edu/abs/2008PhST..133a4003P}
}

@ARTICLE{2008Ap&SS.316..113C,
   author = {{Christensen-Dalsgaard}, J.},
    title = "{ADIPLS{\mdash}the Aarhus adiabatic oscillation package}",
  journal = {\apss},
archivePrefix = "arXiv",
   eprint = {0710.3106},
     year = 2008,
    month = aug,
   volume = 316,
    pages = {113-120},
      doi = {10.1007/s10509-007-9689-z},
   adsurl = {http://adsabs.harvard.edu/abs/2008Ap&SS.316..113C}
}

@ARTICLE{2008Ap&SS.316...99W,
   author = {{Weiss}, A. and {Schlattl}, H.},
    title = "{GARSTEC{\mdash}the Garching Stellar Evolution Code. The direct descendant of the legendary Kippenhahn code}",
  journal = {\apss},
     year = 2008,
    month = aug,
   volume = 316,
    pages = {99-106},
      doi = {10.1007/s10509-007-9606-5},
   adsurl = {http://adsabs.harvard.edu/abs/2008Ap&SS.316...99W}
}

@ARTICLE{2008AJ....135.2055A,
       author = {{Anderson}, Jay and {Sarajedini}, Ata and {Bedin}, Luigi R. and {King}, Ivan R. and {Piotto}, Giampaolo and {Reid}, I. Neill and {Siegel}, Michael and {Majewski}, Steven R. and {Paust}, Nathaniel E.~Q. and {Aparicio}, Antonio and {Milone}, Antonino P. and {Chaboyer}, Brian and {Rosenberg}, Alfred},
        title = "{The Acs Survey of Globular Clusters. V. Generating a Comprehensive Star Catalog for each Cluster}",
      journal = {\aj},
         year = 2008,
        month = jun,
       volume = {135},
       number = {6},
        pages = {2055-2073},
          doi = {10.1088/0004-6256/135/6/2055},
archivePrefix = {arXiv},
       eprint = {0804.2025},
 primaryClass = {astro-ph},
       adsurl = {https://ui.adsabs.harvard.edu/abs/2008AJ....135.2055A}
}

@ARTICLE{2008A&A...486..951G,
       author = {{Gustafsson}, B. and {Edvardsson}, B. and {Eriksson}, K. and
         {J{\o}rgensen}, U.~G. and {Nordlund}, {\r{A}}. and {Plez}, B.},
        title = "{A grid of MARCS model atmospheres for late-type stars. I. Methods and general properties}",
      journal = {\aap},
         year = "2008",
        month = "Aug",
       volume = {486},
       number = {3},
        pages = {951-970},
          doi = {10.1051/0004-6361:200809724},
archivePrefix = {arXiv},
       eprint = {0805.0554},
 primaryClass = {astro-ph},
       adsurl = {https://ui.adsabs.harvard.edu/abs/2008A&A...486..951G}
}

@ARTICLE{2009LRSP....6....2N,
   author = {{Nordlund}, {\AA}. and {Stein}, R.~F. and {Asplund}, M.},
    title = "{Solar Surface Convection}",
  journal = {Living Reviews in Solar Physics},
     year = 2009,
    month = apr,
   volume = 6,
      eid = {2},
    pages = {2},
      doi = {10.12942/lrsp-2009-2},
   adsurl = {http://adsabs.harvard.edu/abs/2009LRSP....6....2N}
}

@ARTICLE{2009ARA&A..47..481A,
   author = {{Asplund}, M. and {Grevesse}, N. and {Sauval}, A.~J. and {Scott}, P.
	},
    title = "{The Chemical Composition of the Sun}",
  journal = {\araa},
archivePrefix = "arXiv",
   eprint = {0909.0948},
 primaryClass = "astro-ph.SR",
     year = 2009,
    month = sep,
   volume = 47,
    pages = {481-522},
      doi = {10.1146/annurev.astro.46.060407.145222},
   adsurl = {http://adsabs.harvard.edu/abs/2009ARA&A..47..481A}
}

@INPROCEEDINGS{2009IAUS..258..171D,
       author = {{Dotter}, Aaron and {Kaluzny}, Janusz and {Thompson}, Ian B.},
        title = "{Globular cluster ages from main sequence fitting and detached, eclipsing binaries: The case of 47 Tuc}",
    booktitle = {The Ages of Stars},
         year = 2009,
       editor = {{Mamajek}, Eric E. and {Soderblom}, David R. and {Wyse}, Rosemary F.~G.},
       series = {IAU Symposium},
       volume = {258},
        month = jun,
        pages = {171-176},
          doi = {10.1017/S1743921309031822},
       adsurl = {https://ui.adsabs.harvard.edu/abs/2009IAUS..258..171D}
}

@ARTICLE{2010ApJ...718.1378M,
       author = {{Magic}, Z. and {Serenelli}, A. and {Weiss}, A. and {Chaboyer}, B.},
        title = "{On Using the Color-Magnitude Diagram Morphology of M67 to Test Solar Abundances}",
      journal = {\apj},
         year = 2010,
        month = aug,
       volume = {718},
       number = {2},
        pages = {1378-1387},
          doi = {10.1088/0004-637X/718/2/1378},
archivePrefix = {arXiv},
       eprint = {1004.3308},
 primaryClass = {astro-ph.SR},
       adsurl = {https://ui.adsabs.harvard.edu/abs/2010ApJ...718.1378M}
}

@ARTICLE{2010ApJS..189..240C,
       author = {{Cyburt}, Richard H. and {Amthor}, A. Matthew and {Ferguson}, Ryan and
         {Meisel}, Zach and {Smith}, Karl and {Warren}, Scott and {Heger}, Alexand
        er and {Hoffman}, R.~D. and {Rauscher}, Thomas and {Sakharuk}, Alexand
        er and {Schatz}, Hendrik and {Thielemann}, F.~K. and {Wiescher}, Michael},
        title = "{The JINA REACLIB Database: Its Recent Updates and Impact on Type-I X-ray Bursts}",
      journal = {\apjs},
         year = "2010",
        month = "Jul",
       volume = {189},
       number = {1},
        pages = {240-252},
          doi = {10.1088/0067-0049/189/1/240},
       adsurl = {https://ui.adsabs.harvard.edu/abs/2010ApJS..189..240C}
}

@ARTICLE{2010ARA&A..48..581S,
       author = {{Soderblom}, David R.},
        title = "{The Ages of Stars}",
      journal = {\araa},
         year = 2010,
        month = sep,
       volume = {48},
        pages = {581-629},
          doi = {10.1146/annurev-astro-081309-130806},
archivePrefix = {arXiv},
       eprint = {1003.6074},
 primaryClass = {astro-ph.SR},
       adsurl = {https://ui.adsabs.harvard.edu/abs/2010ARA&A..48..581S}
}

@ARTICLE{2010MNRAS.408..999D,
       author = {{di Criscienzo}, M. and {Ventura}, P. and {D'Antona}, F. and {Milone}, A. and {Piotto}, G.},
        title = "{The helium spread in the globular cluster 47 Tuc}",
      journal = {\mnras},
         year = 2010,
        month = oct,
       volume = {408},
       number = {2},
        pages = {999-1005},
          doi = {10.1111/j.1365-2966.2010.17168.x},
archivePrefix = {arXiv},
       eprint = {1006.2024},
 primaryClass = {astro-ph.SR},
       adsurl = {https://ui.adsabs.harvard.edu/abs/2010MNRAS.408..999D}
}

@ARTICLE{2010A&A...512A..54C,
       author = {{Casagrande}, L. and {Ram{\'\i}rez}, I. and {Mel{\'e}ndez}, J. and {Bessell}, M. and {Asplund}, M.},
        title = "{An absolutely calibrated T$_{eff}$ scale from the infrared flux method. Dwarfs and subgiants}",
      journal = {\aap},
         year = 2010,
        month = mar,
       volume = {512},
          eid = {A54},
        pages = {A54},
          doi = {10.1051/0004-6361/200913204},
archivePrefix = {arXiv},
       eprint = {1001.3142},
 primaryClass = {astro-ph.SR},
       adsurl = {https://ui.adsabs.harvard.edu/abs/2010A&A...512A..54C}
}

@ARTICLE{2010A&A...517A..49H,
   author = {{Hayek}, W. and {Asplund}, M. and {Carlsson}, M. and {Trampedach}, R. and 
	{Collet}, R. and {Gudiksen}, B.~V. and {Hansteen}, V.~H. and 
	{Leenaarts}, J.},
    title = "{Radiative transfer with scattering for domain-decomposed 3D MHD simulations of cool stellar atmospheres. Numerical methods and application to the quiet, non-magnetic, surface of a solar-type star}",
  journal = {\aap},
     year = 2010,
    month = jul,
   volume = 517,
      eid = {A49},
    pages = {A49},
      doi = {10.1051/0004-6361/201014210},
   adsurl = {http://adsabs.harvard.edu/abs/2010A&A...517A..49H}
}

@ARTICLE{2010A&A...523A..71G,
       author = {{Gebran}, M. and {Vick}, M. and {Monier}, R. and {Fossati}, L.},
        title = "{Chemical composition of A and F dwarfs members of the Hyades open cluster}",
      journal = {\aap},
         year = 2010,
        month = nov,
       volume = {523},
          eid = {A71},
        pages = {A71},
          doi = {10.1051/0004-6361/200913273},
archivePrefix = {arXiv},
       eprint = {1006.5284},
 primaryClass = {astro-ph.SR},
       adsurl = {https://ui.adsabs.harvard.edu/abs/2010A&A...523A..71G}
}

@ARTICLE{2011ApJ...743..161W,
   author = {{White}, T.~R. and {Bedding}, T.~R. and {Stello}, D. and {Christensen-Dalsgaard}, J. and 
	{Huber}, D. and {Kjeldsen}, H.},
    title = "{Calculating Asteroseismic Diagrams for Solar-like Oscillations}",
  journal = {\apj},
archivePrefix = "arXiv",
   eprint = {1109.3455},
 primaryClass = "astro-ph.SR",
     year = 2011,
    month = dec,
   volume = 743,
      eid = {161},
    pages = {161},
      doi = {10.1088/0004-637X/743/2/161},
   adsurl = {http://adsabs.harvard.edu/abs/2011ApJ...743..161W}
}

@ARTICLE{2011ApJ...739...13S,
       author = {{Stello}, Dennis and {Meibom}, S{\o}ren and {Gilliland}, Ronald L. and {Grundahl}, Frank and {Hekker}, Saskia and {Mosser}, Beno{\^\i}t and {Kallinger}, Thomas and {Mathur}, Savita and {Garc{\'\i}a}, Rafael A. and {Huber}, Daniel and {Basu}, Sarbani and {Bedding}, Timothy R. and {Brogaard}, Karsten and {Chaplin}, William J. and {Elsworth}, Yvonne P. and {Molenda-{\.Z}akowicz}, Joanna and {Szab{\'o}}, Robert and {Still}, Martin and {Jenkins}, Jon M. and {Christensen-Dalsgaard}, J{\o}rgen and {Kjeldsen}, Hans and {Serenelli}, Aldo M. and {Wohler}, Bill},
        title = "{An Asteroseismic Membership Study of the Red Giants in Three Open Clusters Observed by Kepler: NGC 6791, NGC 6819, and NGC 6811}",
      journal = {\apj},
         year = 2011,
        month = sep,
       volume = {739},
       number = {1},
          eid = {13},
        pages = {13},
          doi = {10.1088/0004-637X/739/1/13},
archivePrefix = {arXiv},
       eprint = {1107.1234},
 primaryClass = {astro-ph.SR},
       adsurl = {https://ui.adsabs.harvard.edu/abs/2011ApJ...739...13S}
}

@ARTICLE{2011A&A...529A.158H,
       author = {{Hayek}, W. and {Asplund}, M. and {Collet}, R. and {Nordlund}, {\r{A}}.},
        title = "{3D LTE spectral line formation with scattering in red giant stars}",
      journal = {\aap},
         year = 2011,
        month = may,
       volume = {529},
          eid = {A158},
        pages = {A158},
          doi = {10.1051/0004-6361/201015782},
archivePrefix = {arXiv},
       eprint = {1108.3366},
 primaryClass = {astro-ph.SR},
       adsurl = {https://ui.adsabs.harvard.edu/abs/2011A&A...529A.158H}
}

@ARTICLE{2011A&A...530A.100H,
       author = {{Hekker}, S. and {Basu}, S. and {Stello}, D. and {Kallinger}, T. and {Grundahl}, F. and {Mathur}, S. and {Garc{\'\i}a}, R.~A. and {Mosser}, B. and {Huber}, D. and {Bedding}, T.~R. and {Szab{\'o}}, R. and {De Ridder}, J. and {Chaplin}, W.~J. and {Elsworth}, Y. and {Hale}, S.~J. and {Christensen-Dalsgaard}, J. and {Gilliland}, R.~L. and {Still}, M. and {McCauliff}, S. and {Quintana}, E.~V.},
        title = "{Asteroseismic inferences on red giants in open clusters NGC 6791, NGC 6819, and NGC 6811 using Kepler}",
      journal = {\aap},
         year = 2011,
        month = jun,
       volume = {530},
          eid = {A100},
        pages = {A100},
          doi = {10.1051/0004-6361/201016303},
archivePrefix = {arXiv},
       eprint = {1104.4393},
 primaryClass = {astro-ph.SR},
       adsurl = {https://ui.adsabs.harvard.edu/abs/2011A&A...530A.100H}
}

@ARTICLE{2011A&A...525A...2B,
       author = {{Brogaard}, K. and {Bruntt}, H. and {Grundahl}, F. and {Clausen}, J.~V. and {Frandsen}, S. and {Vandenberg}, D.~A. and {Bedin}, L.~R.},
        title = "{Age and helium content of the open cluster NGC 6791 from multiple eclipsing binary members. I. Measurements, methods, and first results}",
      journal = {\aap},
         year = 2011,
        month = jan,
       volume = {525},
          eid = {A2},
        pages = {A2},
          doi = {10.1051/0004-6361/201015503},
archivePrefix = {arXiv},
       eprint = {1009.5537},
 primaryClass = {astro-ph.SR},
       adsurl = {https://ui.adsabs.harvard.edu/abs/2011A&A...525A...2B}
}

@ARTICLE{2012A&A...543A.106B,
       author = {{Brogaard}, K. and {VandenBerg}, D.~A. and {Bruntt}, H. and {Grundahl}, F. and {Frandsen}, S. and {Bedin}, L.~R. and {Milone}, A.~P. and {Dotter}, A. and {Feiden}, G.~A. and {Stetson}, P.~B. and {Sandquist}, E. and {Miglio}, A. and {Stello}, D. and {Jessen-Hansen}, J.},
        title = "{Age and helium content of the open cluster NGC 6791 from multiple eclipsing binary members. II. Age dependencies and new insights}",
      journal = {\aap},
         year = 2012,
        month = jul,
       volume = {543},
          eid = {A106},
        pages = {A106},
          doi = {10.1051/0004-6361/201219196},
archivePrefix = {arXiv},
       eprint = {1205.4071},
 primaryClass = {astro-ph.SR},
       adsurl = {https://ui.adsabs.harvard.edu/abs/2012A&A...543A.106B}
}

@ARTICLE{2012ApJ...744...58M,
       author = {{Milone}, A.~P. and {Piotto}, G. and {Bedin}, L.~R. and {King}, I.~R. and {Anderson}, J. and {Marino}, A.~F. and {Bellini}, A. and {Gratton}, R. and {Renzini}, A. and {Stetson}, P.~B. and {Cassisi}, S. and {Aparicio}, A. and {Bragaglia}, A. and {Carretta}, E. and {D'Antona}, F. and {Di Criscienzo}, M. and {Lucatello}, S. and {Monelli}, M. and {Pietrinferni}, A.},
        title = "{Multiple Stellar Populations in 47 Tucanae}",
      journal = {\apj},
         year = 2012,
        month = jan,
       volume = {744},
       number = {1},
          eid = {58},
        pages = {58},
          doi = {10.1088/0004-637X/744/1/5810.1086/141918},
archivePrefix = {arXiv},
       eprint = {1109.0900},
 primaryClass = {astro-ph.SR},
       adsurl = {https://ui.adsabs.harvard.edu/abs/2012ApJ...744...58M}
}

@ARTICLE{2012ApJ...755...15V,
       author = {{VandenBerg}, Don A. and {Bergbusch}, Peter A. and {Dotter}, Aaron and {Ferguson}, Jason W. and {Michaud}, Georges and {Richer}, Jacques and {Proffitt}, Charles R.},
        title = "{Models for Metal-poor Stars with Enhanced Abundances of C, N, O, Ne, Na, Mg, Si, S, Ca, and Ti, in Turn, at Constant Helium and Iron Abundances}",
      journal = {\apj},
         year = 2012,
        month = aug,
       volume = {755},
       number = {1},
          eid = {15},
        pages = {15},
          doi = {10.1088/0004-637X/755/1/15},
archivePrefix = {arXiv},
       eprint = {1206.1820},
 primaryClass = {astro-ph.SR},
       adsurl = {https://ui.adsabs.harvard.edu/abs/2012ApJ...755...15V}
}

@ARTICLE{2012ApJ...757..190C,
       author = {{Corsaro}, Enrico and {Stello}, Dennis and {Huber}, Daniel and {Bedding}, Timothy R. and {Bonanno}, Alfio and {Brogaard}, Karsten and {Kallinger}, Thomas and {Benomar}, Othman and {White}, Timothy R. and {Mosser}, Benoit and {Basu}, Sarbani and {Chaplin}, William J. and {Christensen-Dalsgaard}, J{\o}rgen and {Elsworth}, Yvonne P. and {Garc{\'\i}a}, Rafael A. and {Hekker}, Saskia and {Kjeldsen}, Hans and {Mathur}, Savita and {Meibom}, S{\o}ren and {Hall}, Jennifer R. and {Ibrahim}, Khadeejah A. and {Klaus}, Todd C.},
        title = "{Asteroseismology of the Open Clusters NGC 6791, NGC 6811, and NGC 6819 from 19 Months of Kepler Photometry}",
      journal = {\apj},
         year = 2012,
        month = oct,
       volume = {757},
       number = {2},
          eid = {190},
        pages = {190},
          doi = {10.1088/0004-637X/757/2/190},
archivePrefix = {arXiv},
       eprint = {1205.4023},
 primaryClass = {astro-ph.SR},
       adsurl = {https://ui.adsabs.harvard.edu/abs/2012ApJ...757..190C}
}

@ARTICLE{2012MNRAS.419.2077M,
       author = {{Miglio}, A. and {Brogaard}, K. and {Stello}, D. and {Chaplin}, W.~J. and {D'Antona}, F. and {Montalb{\'a}n}, J. and {Basu}, S. and {Bressan}, A. and {Grundahl}, F. and {Pinsonneault}, M. and {Serenelli}, A.~M. and {Elsworth}, Y. and {Hekker}, S. and {Kallinger}, T. and {Mosser}, B. and {Ventura}, P. and {Bonanno}, A. and {Noels}, A. and {Silva Aguirre}, V. and {Szabo}, R. and {Li}, J. and {McCauliff}, S. and {Middour}, C.~K. and {Kjeldsen}, H.},
        title = "{Asteroseismology of old open clusters with Kepler: direct estimate of the integrated red giant branch mass-loss in NGC 6791 and 6819}",
      journal = {\mnras},
         year = 2012,
        month = jan,
       volume = {419},
       number = {3},
        pages = {2077-2088},
          doi = {10.1111/j.1365-2966.2011.19859.x},
archivePrefix = {arXiv},
       eprint = {1109.4376},
 primaryClass = {astro-ph.SR},
       adsurl = {https://ui.adsabs.harvard.edu/abs/2012MNRAS.419.2077M}
}

@book{kippenhahn2012stellar,
  title={Stellar Structure and Evolution},
  author={Kippenhahn, R. and Weigert, A. and Weiss, A.},
  isbn={9783642303043},
  series={Astronomy and Astrophysics Library},
  year={2012},
  publisher={Springer Berlin Heidelberg}
}

@MISC{2012ascl.soft11002I,
       author = {{Irwin}, Alan W.},
        title = "{FreeEOS: Equation of State for stellar interiors calculations}",
         year = "2012",
        month = "Nov",
          eid = {ascl:1211.002},
        pages = {ascl:1211.002},
archivePrefix = {ascl},
       eprint = {1211.002},
       adsurl = {https://ui.adsabs.harvard.edu/abs/2012ascl.soft11002I}
}

@ARTICLE{2013A&A...554A.118P,
   author = {{Pereira}, T.~M.~D. and {Asplund}, M. and {Collet}, R. and {Thaler}, I. and 
	{Trampedach}, R. and {Leenaarts}, J.},
    title = "{How realistic are solar model atmospheres?}",
  journal = {\aap},
archivePrefix = "arXiv",
   eprint = {1304.4932},
 primaryClass = "astro-ph.SR",
     year = 2013,
    month = jun,
   volume = 554,
      eid = {A118},
    pages = {A118},
      doi = {10.1051/0004-6361/201321227},
   adsurl = {http://adsabs.harvard.edu/abs/2013A&A...554A.118P}
}

@ARTICLE{2013A&A...557A..26M,
       author = {{Magic}, Z. and {Collet}, R. and {Asplund}, M. and {Trampedach}, R. and {Hayek}, W. and {Chiavassa}, A. and {Stein}, R.~F. and {Nordlund}, {\r{A}}.},
        title = "{The Stagger-grid: A grid of 3D stellar atmosphere models. I. Methods and general properties}",
      journal = {\aap},
         year = 2013,
        month = sep,
       volume = {557},
          eid = {A26},
        pages = {A26},
          doi = {10.1051/0004-6361/201321274},
archivePrefix = {arXiv},
       eprint = {1302.2621},
 primaryClass = {astro-ph.SR},
       adsurl = {https://ui.adsabs.harvard.edu/abs/2013A&A...557A..26M}
}

@ARTICLE{2013A&A...559A...4C,
       author = {{Cruz}, M.~A. and {Serenelli}, A. and {Weiss}, A.},
        title = "{S-process in extremely metal-poor, low-mass stars}",
      journal = {\aap},
         year = 2013,
        month = nov,
       volume = {559},
          eid = {A4},
        pages = {A4},
          doi = {10.1051/0004-6361/201219513},
archivePrefix = {arXiv},
       eprint = {1308.2224},
 primaryClass = {astro-ph.SR},
       adsurl = {https://ui.adsabs.harvard.edu/abs/2013A&A...559A...4C}
}

@ARTICLE{2013ApJ...769...18T,
   author = {{Trampedach}, R. and {Asplund}, M. and {Collet}, R. and {Nordlund}, {\AA}. and 
	{Stein}, R.~F.},
    title = "{A Grid of Three-dimensional Stellar Atmosphere Models of Solar Metallicity. I. General Properties, Granulation, and Atmospheric Expansion}",
  journal = {\apj},
archivePrefix = "arXiv",
   eprint = {1303.1780},
 primaryClass = "astro-ph.SR",
     year = 2013,
    month = may,
   volume = 769,
      eid = {18},
    pages = {18},
      doi = {10.1088/0004-637X/769/1/18},
   adsurl = {http://adsabs.harvard.edu/abs/2013ApJ...769...18T}
}

@ARTICLE{2013ARA&A..51..353C,
   author = {{Chaplin}, W.~J. and {Miglio}, A.},
    title = "{Asteroseismology of Solar-Type and Red-Giant Stars}",
  journal = {\araa},
archivePrefix = "arXiv",
   eprint = {1303.1957},
 primaryClass = "astro-ph.SR",
     year = 2013,
    month = aug,
   volume = 51,
    pages = {353-392},
      doi = {10.1146/annurev-astro-082812-140938},
   adsurl = {http://adsabs.harvard.edu/abs/2013ARA&A..51..353C}
}

@ARTICLE{2014MNRAS.444..392C,
       author = {{Casagrande}, L. and {VandenBerg}, Don A.},
        title = "{Synthetic stellar photometry - I. General considerations and new transformations for broad-band systems}",
      journal = {\mnras},
         year = 2014,
        month = oct,
       volume = {444},
       number = {1},
        pages = {392-419},
          doi = {10.1093/mnras/stu1476},
archivePrefix = {arXiv},
       eprint = {1407.6095},
 primaryClass = {astro-ph.SR},
       adsurl = {https://ui.adsabs.harvard.edu/abs/2014MNRAS.444..392C}
}

@ARTICLE{2014ApJ...780...94C,
       author = {{Cordero}, M.~J. and {Pilachowski}, C.~A. and {Johnson}, C.~I. and {McDonald}, I. and {Zijlstra}, A.~A. and {Simmerer}, J.},
        title = "{Detailed Abundances for a Large Sample of Giant Stars in the Globular Cluster 47 Tucanae (NGC 104)}",
      journal = {\apj},
         year = 2014,
        month = jan,
       volume = {780},
       number = {1},
          eid = {94},
        pages = {94},
          doi = {10.1088/0004-637X/780/1/94},
archivePrefix = {arXiv},
       eprint = {1311.1541},
 primaryClass = {astro-ph.SR},
       adsurl = {https://ui.adsabs.harvard.edu/abs/2014ApJ...780...94C}
}

@ARTICLE{2014ApJ...794...72V,
       author = {{VandenBerg}, Don A. and {Bergbusch}, Peter A. and {Ferguson}, Jason W. and {Edvardsson}, Bengt},
        title = "{Isochrones for Old (>5 Gyr) Stars and Stellar Populations. I. Models for -2.4 <= [Fe/H] <=+0.6, 0.25 <= Y <= 0.33, and -0.4 <= [{\ensuremath{\alpha}}/Fe] <=+0.4}",
      journal = {\apj},
         year = 2014,
        month = oct,
       volume = {794},
       number = {1},
          eid = {72},
        pages = {72},
          doi = {10.1088/0004-637X/794/1/72},
archivePrefix = {arXiv},
       eprint = {1409.1283},
 primaryClass = {astro-ph.SR},
       adsurl = {https://ui.adsabs.harvard.edu/abs/2014ApJ...794...72V}
}

@ARTICLE{2014ApJ...786...10W,
       author = {{Wu}, T. and {Li}, Y. and {Hekker}, S.},
        title = "{Asteroseismic Study on Cluster Distance Moduli for Red Giant Branch Stars in NGC 6791 and NGC 6819}",
      journal = {\apj},
         year = 2014,
        month = may,
       volume = {786},
       number = {1},
          eid = {10},
        pages = {10},
          doi = {10.1088/0004-637X/786/1/10},
archivePrefix = {arXiv},
       eprint = {1403.5838},
 primaryClass = {astro-ph.SR},
       adsurl = {https://ui.adsabs.harvard.edu/abs/2014ApJ...786...10W}
}

@ARTICLE{2014ApJS..215...19P,
       author = {{Pinsonneault}, Marc H. and {Elsworth}, Yvonne and {Epstein}, Courtney and {Hekker}, Saskia and {M{\'e}sz{\'a}ros}, Sz. and {Chaplin}, William J. and {Johnson}, Jennifer A. and {Garc{\'\i}a}, Rafael A. and {Holtzman}, Jon and {Mathur}, Savita and {Garc{\'\i}a P{\'e}rez}, Ana and {Silva Aguirre}, Victor and {Girardi}, L{\'e}o and {Basu}, Sarbani and {Shetrone}, Matthew and {Stello}, Dennis and {Allende Prieto}, Carlos and {An}, Deokkeun and {Beck}, Paul and {Beers}, Timothy C. and {Bizyaev}, Dmitry and {Bloemen}, Steven and {Bovy}, Jo and {Cunha}, Katia and {De Ridder}, Joris and {Frinchaboy}, Peter M. and {Garc{\'\i}a-Hern{\'a}ndez}, D.~A. and {Gilliland}, Ronald and {Harding}, Paul and {Hearty}, Fred R. and {Huber}, Daniel and {Ivans}, Inese and {Kallinger}, Thomas and {Majewski}, Steven R. and {Metcalfe}, Travis S. and {Miglio}, Andrea and {Mosser}, Benoit and {Muna}, Demitri and {Nidever}, David L. and {Schneider}, Donald P. and {Serenelli}, Aldo and {Smith}, Verne V. and {Tayar}, Jamie and {Zamora}, Olga and {Zasowski}, Gail},
        title = "{The APOKASC Catalog: An Asteroseismic and Spectroscopic Joint Survey of Targets in the Kepler Fields}",
      journal = {\apjs},
         year = 2014,
        month = dec,
       volume = {215},
       number = {2},
          eid = {19},
        pages = {19},
          doi = {10.1088/0067-0049/215/2/19},
archivePrefix = {arXiv},
       eprint = {1410.2503},
 primaryClass = {astro-ph.SR},
       adsurl = {https://ui.adsabs.harvard.edu/abs/2014ApJS..215...19P}
}

@ARTICLE{2015A&A...573A..89M,
   author = {{Magic}, Z. and {Weiss}, A. and {Asplund}, M.},
    title = "{The Stagger-grid: A grid of 3D stellar atmosphere models. III. The relation to mixing length convection theory}",
  journal = {\aap},
archivePrefix = "arXiv",
   eprint = {1403.1062},
 primaryClass = "astro-ph.SR",
     year = 2015,
    month = jan,
   volume = 573,
      eid = {A89},
    pages = {A89},
      doi = {10.1051/0004-6361/201423760},
   adsurl = {http://adsabs.harvard.edu/abs/2015A&A...573A..89M}
}

@ARTICLE{2015MNRAS.452.2127S,
       author = {{Silva Aguirre}, V. and {Davies}, G.~R. and {Basu}, S. and {Christensen-Dalsgaard}, J. and {Creevey}, O. and {Metcalfe}, T.~S. and {Bedding}, T.~R. and {Casagrande}, L. and {Handberg}, R. and {Lund}, M.~N. and {Nissen}, P.~E. and {Chaplin}, W.~J. and {Huber}, D. and {Serenelli}, A.~M. and {Stello}, D. and {Van Eylen}, V. and {Campante}, T.~L. and {Elsworth}, Y. and {Gilliland}, R.~L. and {Hekker}, S. and {Karoff}, C. and {Kawaler}, S.~D. and {Kjeldsen}, H. and {Lundkvist}, M.~S.},
        title = "{Ages and fundamental properties of Kepler exoplanet host stars from asteroseismology}",
      journal = {\mnras},
         year = 2015,
        month = sep,
       volume = {452},
       number = {2},
        pages = {2127-2148},
          doi = {10.1093/mnras/stv1388},
archivePrefix = {arXiv},
       eprint = {1504.07992},
 primaryClass = {astro-ph.SR},
       adsurl = {https://ui.adsabs.harvard.edu/abs/2015MNRAS.452.2127S}
}

@ARTICLE{2016A&A...589A..93D,
       author = {{Deheuvels}, S. and {Brand{\~a}o}, I. and {Silva Aguirre}, V. and {Ballot}, J. and {Michel}, E. and {Cunha}, M.~S. and {Lebreton}, Y. and {Appourchaux}, T.},
        title = "{Measuring the extent of convective cores in low-mass stars using Kepler data: toward a calibration of core overshooting}",
      journal = {\aap},
         year = 2016,
        month = may,
       volume = {589},
          eid = {A93},
        pages = {A93},
          doi = {10.1051/0004-6361/201527967},
archivePrefix = {arXiv},
       eprint = {1603.02332},
 primaryClass = {astro-ph.SR},
       adsurl = {https://ui.adsabs.harvard.edu/abs/2016A&A...589A..93D}
}

@ARTICLE{2016ApJ...822...15S,
       author = {{Sharma}, Sanjib and {Stello}, Dennis and {Bland-Hawthorn}, Joss and {Huber}, Daniel and {Bedding}, Timothy R.},
        title = "{Stellar Population Synthesis Based Modeling of the Milky Way Using Asteroseismology of 13,000 Kepler Red Giants}",
      journal = {\apj},
         year = 2016,
        month = may,
       volume = {822},
       number = {1},
          eid = {15},
        pages = {15},
          doi = {10.3847/0004-637X/822/1/15},
archivePrefix = {arXiv},
       eprint = {1603.05661},
 primaryClass = {astro-ph.GA},
       adsurl = {https://ui.adsabs.harvard.edu/abs/2016ApJ...822...15S}
}

@ARTICLE{2016ApJ...823..102C,
       author = {{Choi}, Jieun and {Dotter}, Aaron and {Conroy}, Charlie and {Cantiello}, Matteo and {Paxton}, Bill and {Johnson}, Benjamin D.},
        title = "{Mesa Isochrones and Stellar Tracks (MIST). I. Solar-scaled Models}",
      journal = {\apj},
         year = 2016,
        month = jun,
       volume = {823},
       number = {2},
          eid = {102},
        pages = {102},
          doi = {10.3847/0004-637X/823/2/102},
archivePrefix = {arXiv},
       eprint = {1604.08592},
 primaryClass = {astro-ph.SR},
       adsurl = {https://ui.adsabs.harvard.edu/abs/2016ApJ...823..102C}
}

@ARTICLE{2017MNRAS.472.3264J,
   author = {{J{\o}rgensen}, A.~C.~S. and {Weiss}, A. and {Mosumgaard}, J.~R. and 
	{Silva Aguirre}, V. and {Sahlholdt}, C.~L.},
    title = "{Theoretical oscillation frequencies for solar-type dwarfs from stellar models with {\#12296}3D{\#12297}-atmospheres}",
  journal = {\mnras},
archivePrefix = "arXiv",
   eprint = {1709.06332},
 primaryClass = "astro-ph.SR",
     year = 2017,
    month = dec,
   volume = 472,
    pages = {3264-3276},
      doi = {10.1093/mnras/stx2226},
   adsurl = {http://adsabs.harvard.edu/abs/2017MNRAS.472.3264J}
}

@ARTICLE{2017MNRAS.467.1433R,
       author = {{Rodrigues}, Tha{\'\i}se S. and {Bossini}, Diego and {Miglio}, Andrea and {Girardi}, L{\'e}o and {Montalb{\'a}n}, Josefina and {Noels}, Arlette and {Trabucchi}, Michele and {Coelho}, Hugo Rodrigues and {Marigo}, Paola},
        title = "{Determining stellar parameters of asteroseismic targets: going beyond the use of scaling relations}",
      journal = {\mnras},
         year = 2017,
        month = may,
       volume = {467},
       number = {2},
        pages = {1433-1448},
          doi = {10.1093/mnras/stx120},
archivePrefix = {arXiv},
       eprint = {1701.04791},
 primaryClass = {astro-ph.SR},
       adsurl = {https://ui.adsabs.harvard.edu/abs/2017MNRAS.467.1433R}
}

@ARTICLE{2017ApJ...835..172L,
       author = {{Lund}, Mikkel N. and {Silva Aguirre}, V{\'\i}ctor and {Davies}, Guy R. and
         {Chaplin}, William J. and {Christensen-Dalsgaard}, J{\o}rgen and
         {Houdek}, G{\"u}nter and {White}, Timothy R. and {Bedding}, Timothy R. and
         {Ball}, Warrick H. and {Huber}, Daniel},
        title = "{Standing on the Shoulders of Dwarfs: the Kepler Asteroseismic LEGACY Sample. I. Oscillation Mode Parameters}",
      journal = {\apj},
         year = "2017",
        month = "Feb",
       volume = {835},
       number = {2},
          eid = {172},
        pages = {172},
          doi = {10.3847/1538-4357/835/2/172},
archivePrefix = {arXiv},
       eprint = {1612.00436},
 primaryClass = {astro-ph.SR},
       adsurl = {https://ui.adsabs.harvard.edu/abs/2017ApJ...835..172L}
}

@ARTICLE{2017ApJ...835..173S,
       author = {{Silva Aguirre}, V{\'\i}ctor and {Lund}, Mikkel N. and {Antia}, H.~M. and
         {Ball}, Warrick H. and {Basu}, Sarbani and
         {Christensen-Dalsgaard}, J{\o}rgen and {Lebreton}, Yveline and
         {Reese}, Daniel R. and {Verma}, Kuldeep and {Casagrande}, Luca},
        title = "{Standing on the Shoulders of Dwarfs: the Kepler Asteroseismic LEGACY Sample. II.Radii, Masses, and Ages}",
      journal = {\apj},
         year = "2017",
        month = "Feb",
       volume = {835},
       number = {2},
          eid = {173},
        pages = {173},
          doi = {10.3847/1538-4357/835/2/173},
archivePrefix = {arXiv},
       eprint = {1611.08776},
 primaryClass = {astro-ph.SR},
       adsurl = {https://ui.adsabs.harvard.edu/abs/2017ApJ...835..173S}
}

@ARTICLE{2017ApJ...840...99D,
       author = {{Dotter}, Aaron and {Conroy}, Charlie and {Cargile}, Phillip and
         {Asplund}, Martin},
        title = "{The Influence of Atomic Diffusion on Stellar Ages and Chemical Tagging}",
      journal = {\apj},
         year = "2017",
        month = "May",
       volume = {840},
       number = {2},
          eid = {99},
        pages = {99},
          doi = {10.3847/1538-4357/aa6d10},
archivePrefix = {arXiv},
       eprint = {1704.03465},
 primaryClass = {astro-ph.SR},
       adsurl = {https://ui.adsabs.harvard.edu/abs/2017ApJ...840...99D}
}

@ARTICLE{2017ApJ...844..102H,
       author = {{Huber}, Daniel and {Zinn}, Joel and {Bojsen-Hansen}, Mathias and
         {Pinsonneault}, Marc and {Sahlholdt}, Christian and {Serenelli}, Aldo and
         {Silva Aguirre}, Victor and {Stassun}, Keivan and {Stello}, Dennis and
         {Tayar}, Jamie and {Bastien}, Fabienne and {Bedding}, Timothy R. and
         {Buchhave}, Lars A. and {Chaplin}, William J. and {Davies}, Guy R. and
         {Garc{\'\i}a}, Rafael A. and {Latham}, David W. and {Mathur}, Savita and
         {Mosser}, Benoit and {Sharma}, Sanjib},
        title = "{Asteroseismology and Gaia: Testing Scaling Relations Using 2200 Kepler  Stars with TGAS Parallaxes}",
      journal = {\apj},
         year = 2017,
        month = aug,
       volume = {844},
       number = {2},
          eid = {102},
        pages = {102},
          doi = {10.3847/1538-4357/aa75ca},
archivePrefix = {arXiv},
       eprint = {1705.04697},
 primaryClass = {astro-ph.SR},
       adsurl = {https://ui.adsabs.harvard.edu/abs/2017ApJ...844..102H}
}

@ARTICLE{2017ApJ...840...17T,
       author = {{Tayar}, Jamie and {Somers}, Garrett and {Pinsonneault}, Marc H. and {Stello}, Dennis and {Mints}, Alexey and {Johnson}, Jennifer A. and {Zamora}, O. and {Garc{\'\i}a-Hern{\'a}ndez}, D.~A. and {Maraston}, Claudia and {Serenelli}, Aldo and {Allende Prieto}, Carlos and {Bastien}, Fabienne A. and {Basu}, Sarbani and {Bird}, J.~C. and {Cohen}, R.~E. and {Cunha}, Katia and {Elsworth}, Yvonne and {Garc{\'\i}a}, Rafael A. and {Girardi}, Leo and {Hekker}, Saskia and {Holtzman}, Jon and {Huber}, Daniel and {Mathur}, Savita and {M{\'e}sz{\'a}ros}, Szabolcs and {Mosser}, B. and {Shetrone}, Matthew and {Silva Aguirre}, Victor and {Stassun}, Keivan and {Stringfellow}, Guy S. and {Zasowski}, Gail and {Roman-Lopes}, A.},
        title = "{The Correlation between Mixing Length and Metallicity on the Giant Branch: Implications for Ages in the Gaia Era}",
      journal = {\apj},
         year = 2017,
        month = may,
       volume = {840},
       number = {1},
          eid = {17},
        pages = {17},
          doi = {10.3847/1538-4357/aa6a1e},
archivePrefix = {arXiv},
       eprint = {1704.01164},
 primaryClass = {astro-ph.SR},
       adsurl = {https://ui.adsabs.harvard.edu/abs/2017ApJ...840...17T}
}

@article{2018A&A...611A..11C,
  title = {The {{STAGGER}}-Grid: {{A}} Grid of {{3D}} Stellar Atmosphere Models. {{V}}. {{Synthetic}} Stellar Spectra and Broad-Band Photometry},
  shorttitle = {The {{STAGGER}}-Grid},
  author = {Chiavassa, A. and Casagrande, L. and Collet, R. and Magic, Z. and Bigot, L. and Th{\'e}venin, F. and Asplund, M.},
  year = {2018},
  month = mar,
  volume = {611},
  pages = {A11},
  issn = {0004-6361},
  doi = {10.1051/0004-6361/201732147},
  journal = {\aap}
}

@ARTICLE{2018A&A...612A..68S,
       author = {{Salaris}, M. and {Cassisi}, S. and {Schiavon}, R.~P. and {Pietrinferni}, A.},
        title = "{Effective temperatures of red giants in the APOKASC catalogue and the mixing length calibration in stellar models}",
      journal = {\aap},
         year = 2018,
        month = apr,
       volume = {612},
          eid = {A68},
        pages = {A68},
          doi = {10.1051/0004-6361/201732340},
archivePrefix = {arXiv},
       eprint = {1801.09441},
 primaryClass = {astro-ph.SR},
       adsurl = {https://ui.adsabs.harvard.edu/abs/2018A&A...612A..68S}
}

@ARTICLE{2018A&A...616A...1G,
       author = {{Gaia Collaboration} and {Brown}, A.~G.~A. and {Vallenari}, A. and {Prusti}, T. and {de Bruijne}, J.~H.~J. and {Babusiaux}, C. and {Bailer-Jones}, C.~A.~L. and {Biermann}, M. and {Evans}, D.~W. and {Eyer}, L. and {Jansen}, F. and {Jordi}, C. and {Klioner}, S.~A. and {Lammers}, U. and {Lindegren}, L. and {Luri}, X. and {Mignard}, F. and {Panem}, C. and {Pourbaix}, D. and {Randich}, S. and {Sartoretti}, P. and {Siddiqui}, H.~I. and {Soubiran}, C. and {van Leeuwen}, F. and {Walton}, N.~A. and {Arenou}, F. and {Bastian}, U. and {Cropper}, M. and {Drimmel}, R. and {Katz}, D. and {Lattanzi}, M.~G. and {Bakker}, J. and {Cacciari}, C. and {Casta{\~n}eda}, J. and {Chaoul}, L. and {Cheek}, N. and {De Angeli}, F. and {Fabricius}, C. and {Guerra}, R. and {Holl}, B. and {Masana}, E. and {Messineo}, R. and {Mowlavi}, N. and {Nienartowicz}, K. and {Panuzzo}, P. and {Portell}, J. and {Riello}, M. and {Seabroke}, G.~M. and {Tanga}, P. and {Th{\'e}venin}, F. and {Gracia-Abril}, G. and {Comoretto}, G. and {Garcia-Reinaldos}, M. and {Teyssier}, D. and {Altmann}, M. and {Andrae}, R. and {Audard}, M. and {Bellas-Velidis}, I. and {Benson}, K. and {Berthier}, J. and {Blomme}, R. and {Burgess}, P. and {Busso}, G. and {Carry}, B. and {Cellino}, A. and {Clementini}, G. and {Clotet}, M. and {Creevey}, O. and {Davidson}, M. and {De Ridder}, J. and {Delchambre}, L. and {Dell'Oro}, A. and {Ducourant}, C. and {Fern{\'a}ndez-Hern{\'a}ndez}, J. and {Fouesneau}, M. and {Fr{\'e}mat}, Y. and {Galluccio}, L. and {Garc{\'\i}a-Torres}, M. and {Gonz{\'a}lez-N{\'u}{\~n}ez}, J. and {Gonz{\'a}lez-Vidal}, J.~J. and {Gosset}, E. and {Guy}, L.~P. and {Halbwachs}, J.-L. and {Hambly}, N.~C. and {Harrison}, D.~L. and {Hern{\'a}ndez}, J. and {Hestroffer}, D. and {Hodgkin}, S.~T. and {Hutton}, A. and {Jasniewicz}, G. and {Jean-Antoine-Piccolo}, A. and {Jordan}, S. and {Korn}, A.~J. and {Krone-Martins}, A. and {Lanzafame}, A.~C. and {Lebzelter}, T. and {L{\"o}ffler}, W. and {Manteiga}, M. and {Marrese}, P.~M. and {Mart{\'\i}n-Fleitas}, J.~M. and {Moitinho}, A. and {Mora}, A. and {Muinonen}, K. and {Osinde}, J. and {Pancino}, E. and {Pauwels}, T. and {Petit}, J.-M. and {Recio-Blanco}, A. and {Richards}, P.~J. and {Rimoldini}, L. and {Robin}, A.~C. and {Sarro}, L.~M. and {Siopis}, C. and {Smith}, M. and {Sozzetti}, A. and {S{\"u}veges}, M. and {Torra}, J. and {van Reeven}, W. and {Abbas}, U. and {Abreu Aramburu}, A. and {Accart}, S. and {Aerts}, C. and {Altavilla}, G. and {{\'A}lvarez}, M.~A. and {Alvarez}, R. and {Alves}, J. and {Anderson}, R.~I. and {Andrei}, A.~H. and {Anglada Varela}, E. and {Antiche}, E. and {Antoja}, T. and {Arcay}, B. and {Astraatmadja}, T.~L. and {Bach}, N. and {Baker}, S.~G. and {Balaguer-N{\'u}{\~n}ez}, L. and {Balm}, P. and {Barache}, C. and {Barata}, C. and {Barbato}, D. and {Barblan}, F. and {Barklem}, P.~S. and {Barrado}, D. and {Barros}, M. and {Barstow}, M.~A. and {Bartholom{\'e} Mu{\~n}oz}, S. and {Bassilana}, J.-L. and {Becciani}, U. and {Bellazzini}, M. and {Berihuete}, A. and {Bertone}, S. and {Bianchi}, L. and {Bienaym{\'e}}, O. and {Blanco-Cuaresma}, S. and {Boch}, T. and {Boeche}, C. and {Bombrun}, A. and {Borrachero}, R. and {Bossini}, D. and {Bouquillon}, S. and {Bourda}, G. and {Bragaglia}, A. and {Bramante}, L. and {Breddels}, M.~A. and {Bressan}, A. and {Brouillet}, N. and {Br{\"u}semeister}, T. and {Brugaletta}, E. and {Bucciarelli}, B. and {Burlacu}, A. and {Busonero}, D. and {Butkevich}, A.~G. and {Buzzi}, R. and {Caffau}, E. and {Cancelliere}, R. and {Cannizzaro}, G. and {Cantat-Gaudin}, T. and {Carballo}, R. and {Carlucci}, T. and {Carrasco}, J.~M. and {Casamiquela}, L. and {Castellani}, M. and {Castro-Ginard}, A. and {Charlot}, P. and {Chemin}, L. and {Chiavassa}, A. and {Cocozza}, G. and {Costigan}, G. and {Cowell}, S. and {Crifo}, F. and {Crosta}, M. and {Crowley}, C. and {Cuypers}, J. and {Dafonte}, C. and {Damerdji}, Y. and {Dapergolas}, A. and {David}, P. and {David}, M. and {de Laverny}, P. and {De Luise}, F.},
        title = "{Gaia Data Release 2. Summary of the contents and survey properties}",
      journal = {\aap},
         year = 2018,
        month = aug,
       volume = {616},
          eid = {A1},
        pages = {A1},
          doi = {10.1051/0004-6361/201833051},
archivePrefix = {arXiv},
       eprint = {1804.09365},
 primaryClass = {astro-ph.GA},
       adsurl = {https://ui.adsabs.harvard.edu/abs/2018A&A...616A...1G}
}

@ARTICLE{2018ApJ...863...65C,
       author = {{Choi}, Jieun and {Conroy}, Charlie and {Ting}, Yuan-Sen and {Cargile}, Phillip A. and {Dotter}, Aaron and {Johnson}, Benjamin D.},
        title = "{Star Cluster Ages in the Gaia Era}",
      journal = {\apj},
         year = 2018,
        month = aug,
       volume = {863},
       number = {1},
          eid = {65},
        pages = {65},
          doi = {10.3847/1538-4357/aad18c},
archivePrefix = {arXiv},
       eprint = {1807.03789},
 primaryClass = {astro-ph.SR},
       adsurl = {https://ui.adsabs.harvard.edu/abs/2018ApJ...863...65C}
}

@ARTICLE{2018ApJ...858...28V,
       author = {{Viani}, Lucas S. and {Basu}, Sarbani and {Ong J.}, M. Joel and {Bonaca}, Ana and {Chaplin}, William J.},
        title = "{Investigating the Metallicity-Mixing-length Relation}",
      journal = {\apj},
         year = 2018,
        month = may,
       volume = {858},
       number = {1},
          eid = {28},
        pages = {28},
          doi = {10.3847/1538-4357/aab7eb},
archivePrefix = {arXiv},
       eprint = {1803.05924},
 primaryClass = {astro-ph.SR},
       adsurl = {https://ui.adsabs.harvard.edu/abs/2018ApJ...858...28V}
}

@ARTICLE{2018ApJ...859..100C,
       author = {{Claret}, Antonio and {Torres}, Guillermo},
        title = "{The Dependence of Convective Core Overshooting on Stellar Mass: Additional Binary Systems and Improved Calibration}",
      journal = {\apj},
         year = 2018,
        month = jun,
       volume = {859},
       number = {2},
          eid = {100},
        pages = {100},
          doi = {10.3847/1538-4357/aabd35},
archivePrefix = {arXiv},
       eprint = {1804.03148},
 primaryClass = {astro-ph.SR},
       adsurl = {https://ui.adsabs.harvard.edu/abs/2018ApJ...859..100C}
}

@ARTICLE{2018ApJ...867...34V,
       author = {{Villanova}, Sandro and {Carraro}, Giovanni and {Geisler}, Doug and {Monaco}, Lorenzo and {Assmann}, Paulina},
        title = "{NGC 6791: A Probable Bulge Cluster without Multiple Populations}",
      journal = {\apj},
         year = 2018,
        month = nov,
       volume = {867},
       number = {1},
          eid = {34},
        pages = {34},
          doi = {10.3847/1538-4357/aae4e5},
archivePrefix = {arXiv},
       eprint = {1809.09661},
 primaryClass = {astro-ph.GA},
       adsurl = {https://ui.adsabs.harvard.edu/abs/2018ApJ...867...34V}
}

@ARTICLE{2018ApJS..239...32P,
       author = {{Pinsonneault}, Marc H. and {Elsworth}, Yvonne P. and {Tayar}, Jamie and {Serenelli}, Aldo and {Stello}, Dennis and {Zinn}, Joel and {Mathur}, Savita and {Garc{\'\i}a}, Rafael A. and {Johnson}, Jennifer A. and {Hekker}, Saskia and {Huber}, Daniel and {Kallinger}, Thomas and {M{\'e}sz{\'a}ros}, Szabolcs and {Mosser}, Benoit and {Stassun}, Keivan and {Girardi}, L{\'e}o and {Rodrigues}, Tha{\'\i}se S. and {Silva Aguirre}, Victor and {An}, Deokkeun and {Basu}, Sarbani and {Chaplin}, William J. and {Corsaro}, Enrico and {Cunha}, Katia and {Garc{\'\i}a-Hern{\'a}ndez}, D.~A. and {Holtzman}, Jon and {J{\"o}nsson}, Henrik and {Shetrone}, Matthew and {Smith}, Verne V. and {Sobeck}, Jennifer S. and {Stringfellow}, Guy S. and {Zamora}, Olga and {Beers}, Timothy C. and {Fern{\'a}ndez-Trincado}, J.~G. and {Frinchaboy}, Peter M. and {Hearty}, Fred R. and {Nitschelm}, Christian},
        title = "{The Second APOKASC Catalog: The Empirical Approach}",
      journal = {\apjs},
         year = 2018,
        month = dec,
       volume = {239},
       number = {2},
          eid = {32},
        pages = {32},
          doi = {10.3847/1538-4365/aaebfd},
archivePrefix = {arXiv},
       eprint = {1804.09983},
 primaryClass = {astro-ph.SR},
       adsurl = {https://ui.adsabs.harvard.edu/abs/2018ApJS..239...32P}
}

@ARTICLE{2018MNRAS.475.3369C,
   author = {{Collet}, R. and {Nordlund}, {\AA}. and {Asplund}, M. and {Hayek}, W. and 
	{Trampedach}, R.},
    title = "{The benchmark halo giant HD 122563: CNO abundances revisited with three-dimensional hydrodynamic model stellar atmospheres}",
  journal = {\mnras},
archivePrefix = "arXiv",
   eprint = {1712.08099},
 primaryClass = "astro-ph.SR",
     year = 2018,
    month = apr,
   volume = 475,
    pages = {3369-3392},
      doi = {10.1093/mnras/sty002},
   adsurl = {http://adsabs.harvard.edu/abs/2018MNRAS.475.3369C}
}

@ARTICLE{2018MNRAS.478.5650M,
       author = {{Mosumgaard}, Jakob R{\o}rsted and {Ball}, Warrick H. and
         {Silva Aguirre}, V{\'\i}ctor and {Weiss}, Achim and
         {Christensen-Dalsgaard}, J{\o}rgen},
        title = "{Stellar models with calibrated convection and temperature stratification from 3D hydrodynamics simulations}",
      journal = {\mnras},
         year = "2018",
        month = "Aug",
       volume = {478},
       number = {4},
        pages = {5650-5659},
          doi = {10.1093/mnras/sty1442},
archivePrefix = {arXiv},
       eprint = {1806.00020},
 primaryClass = {astro-ph.SR},
       adsurl = {https://ui.adsabs.harvard.edu/abs/2018MNRAS.478.5650M}
}

@ARTICLE{2018MNRAS.481L..35J,
       author = {{J{\o}rgensen}, Andreas Christ S{\o}lvsten and
         {Mosumgaard}, Jakob R{\o}rsted and {Weiss}, Achim and
         {Silva Aguirre}, V{\'\i}ctor and {Christensen-Dalsgaard}, J{\o}rgen},
        title = "{Coupling 1D stellar evolution with 3D-hydrodynamical simulations on the fly - I. A new standard solar model}",
      journal = {\mnras},
         year = 2018,
        month = nov,
       volume = {481},
       number = {1},
        pages = {L35-L39},
          doi = {10.1093/mnrasl/sly152},
archivePrefix = {arXiv},
       eprint = {1808.08886},
 primaryClass = {astro-ph.SR},
       adsurl = {https://ui.adsabs.harvard.edu/abs/2018MNRAS.481L..35J}
}

@ARTICLE{2018MNRAS.476.1931S,
       author = {{Sahlholdt}, C.~L. and {Silva Aguirre}, V. and {Casagrande}, L. and {Mosumgaard}, J.~R. and {Bojsen-Hansen}, M.},
        title = "{Testing asteroseismic radii of dwarfs and subgiants with Kepler and Gaia}",
      journal = {\mnras},
         year = 2018,
        month = may,
       volume = {476},
       number = {2},
        pages = {1931-1941},
          doi = {10.1093/mnras/sty319},
archivePrefix = {arXiv},
       eprint = {1802.01127},
 primaryClass = {astro-ph.SR},
       adsurl = {https://ui.adsabs.harvard.edu/abs/2018MNRAS.476.1931S}
}

@ARTICLE{2018MNRAS.476.3729B,
       author = {{Brogaard}, K. and {Hansen}, C.~J. and {Miglio}, A. and {Slumstrup}, D. and {Frandsen}, S. and {Jessen-Hansen}, J. and {Lund}, M.~N. and {Bossini}, D. and {Thygesen}, A. and {Davies}, G.~R. and {Chaplin}, W.~J. and {Arentoft}, T. and {Bruntt}, H. and {Grundahl}, F. and {Handberg}, R.},
        title = "{Establishing the accuracy of asteroseismic mass and radius estimates of giant stars - I. Three eclipsing systems at [Fe/H] {\ensuremath{\sim}} -0.3 and the need for a large high-precision sample}",
      journal = {\mnras},
         year = 2018,
        month = may,
       volume = {476},
       number = {3},
        pages = {3729-3743},
          doi = {10.1093/mnras/sty268},
archivePrefix = {arXiv},
       eprint = {1801.08167},
 primaryClass = {astro-ph.SR},
       adsurl = {https://ui.adsabs.harvard.edu/abs/2018MNRAS.476.3729B}
}

@ARTICLE{2018MNRAS.475.5023C,
       author = {{Casagrande}, L. and {VandenBerg}, Don A.},
        title = "{Synthetic Stellar Photometry - II. Testing the bolometric flux scale and tables of bolometric corrections for the Hipparcos/Tycho, Pan-STARRS1, SkyMapper, and JWST systems}",
      journal = {\mnras},
         year = 2018,
        month = apr,
       volume = {475},
       number = {4},
        pages = {5023-5040},
          doi = {10.1093/mnras/sty149},
archivePrefix = {arXiv},
       eprint = {1801.05508},
 primaryClass = {astro-ph.SR},
       adsurl = {https://ui.adsabs.harvard.edu/abs/2018MNRAS.475.5023C}
}

@ARTICLE{2018MNRAS.479L.102C,
       author = {{Casagrande}, L. and {VandenBerg}, Don A.},
        title = "{On the use of Gaia magnitudes and new tables of bolometric corrections}",
      journal = {\mnras},
         year = 2018,
        month = sep,
       volume = {479},
       number = {1},
        pages = {L102-L107},
          doi = {10.1093/mnrasl/sly104},
archivePrefix = {arXiv},
       eprint = {1806.01953},
 primaryClass = {astro-ph.SR},
       adsurl = {https://ui.adsabs.harvard.edu/abs/2018MNRAS.479L.102C}
}

@ARTICLE{2018MNRAS.476..496F,
       author = {{Fu}, Xiaoting and {Bressan}, Alessandro and {Marigo}, Paola and {Girardi}, L{\'e}o and {Montalb{\'a}n}, Josefina and {Chen}, Yang and {Nanni}, Ambra},
        title = "{New PARSEC data base of {\ensuremath{\alpha}}-enhanced stellar evolutionary tracks and isochrones - I. Calibration with 47 Tuc (NGC 104) and the improvement on RGB bump}",
      journal = {\mnras},
         year = 2018,
        month = may,
       volume = {476},
       number = {1},
        pages = {496-511},
          doi = {10.1093/mnras/sty235},
archivePrefix = {arXiv},
       eprint = {1801.07137},
 primaryClass = {astro-ph.SR},
       adsurl = {https://ui.adsabs.harvard.edu/abs/2018MNRAS.476..496F}
}

@ARTICLE{2019A&A...627A.117L,
       author = {{Liu}, F. and {Asplund}, M. and {Yong}, D. and {Feltzing}, S. and {Dotter}, A. and {Mel{\'e}ndez}, J. and {Ram{\'\i}rez}, I.},
        title = "{Chemical (in)homogeneity and atomic diffusion in the open cluster M 67}",
      journal = {\aap},
         year = 2019,
        month = jul,
       volume = {627},
          eid = {A117},
        pages = {A117},
          doi = {10.1051/0004-6361/201935306},
archivePrefix = {arXiv},
       eprint = {1902.11008},
 primaryClass = {astro-ph.SR},
       adsurl = {https://ui.adsabs.harvard.edu/abs/2019A&A...627A.117L}
}

@ARTICLE{2019ApJ...885..166Z,
       author = {{Zinn}, Joel C. and {Pinsonneault}, Marc H. and {Huber}, Daniel and
         {Stello}, Dennis and {Stassun}, Keivan and {Serenelli}, Aldo},
        title = "{Testing the Radius Scaling Relation with Gaia DR2 in the Kepler Field}",
      journal = {\apj},
         year = 2019,
        month = nov,
       volume = {885},
       number = {2},
          eid = {166},
        pages = {166},
          doi = {10.3847/1538-4357/ab44a9},
archivePrefix = {arXiv},
       eprint = {1910.00719},
 primaryClass = {astro-ph.SR},
       adsurl = {https://ui.adsabs.harvard.edu/abs/2019ApJ...885..166Z}
}

@ARTICLE{2019ApJ...874...97S,
       author = {{Souto}, Diogo and {Allende Prieto}, C. and {Cunha}, Katia and {Pinsonneault}, Marc and {Smith}, Verne V. and {Garcia-Dias}, R. and {Bovy}, Jo and {Garc{\'\i}a-Hern{\'a}ndez}, D.~A. and {Holtzman}, Jon and {Johnson}, J.~A. and {J{\"o}nsson}, Henrik and {Majewski}, Steve R. and {Shetrone}, Matthew and {Sobeck}, Jennifer and {Zamora}, Olga and {Pan}, Kaike and {Nitschelm}, Christian},
        title = "{Chemical Abundances of Main-sequence, Turnoff, Subgiant, and Red Giant Stars from APOGEE Spectra. II. Atomic Diffusion in M67 Stars}",
      journal = {\apj},
         year = 2019,
        month = mar,
       volume = {874},
       number = {1},
          eid = {97},
        pages = {97},
          doi = {10.3847/1538-4357/ab0b43},
archivePrefix = {arXiv},
       eprint = {1902.10199},
 primaryClass = {astro-ph.SR},
       adsurl = {https://ui.adsabs.harvard.edu/abs/2019ApJ...874...97S}
}

@ARTICLE{2019MNRAS.489.1850V,
       author = {{Verma}, Kuldeep and {Silva Aguirre}, V{\'\i}ctor},
        title = "{Helium settling in F stars: constraining turbulent mixing using observed helium glitch signature}",
      journal = {\mnras},
         year = "2019",
        month = "Oct",
       volume = {489},
       number = {2},
        pages = {1850-1858},
          doi = {10.1093/mnras/stz2272},
       adsurl = {https://ui.adsabs.harvard.edu/abs/2019MNRAS.489.1850V}
}

@ARTICLE{2019MNRAS.483.1674R,
       author = {{Rain}, M.~J. and {Villanova}, S. and {Mun{\~o}z}, C. and {Valenzuela-Calderon}, C.},
        title = "{Chemical evolution of the metal-poor globular cluster NGC 6809}",
      journal = {\mnras},
         year = 2019,
        month = feb,
       volume = {483},
       number = {2},
        pages = {1674-1685},
          doi = {10.1093/mnras/sty3208},
archivePrefix = {arXiv},
       eprint = {1811.12457},
 primaryClass = {astro-ph.SR},
       adsurl = {https://ui.adsabs.harvard.edu/abs/2019MNRAS.483.1674R}
}

@ARTICLE{2019MNRAS.489.1742F,
       author = {{Feuillet}, Diane K. and {Frankel}, Neige and {Lind}, Karin and {Frinchaboy}, Peter M. and {Garc{\'\i}a-Hern{\'a}ndez}, D.~A. and {Lane}, Richard R. and {Nitschelm}, Christian and {Roman-Lopes}, Alexandre},
        title = "{Spatial variations in the Milky Way disc metallicity-age relation}",
      journal = {\mnras},
         year = 2019,
        month = oct,
       volume = {489},
       number = {2},
        pages = {1742-1752},
          doi = {10.1093/mnras/stz2221},
archivePrefix = {arXiv},
       eprint = {1908.02772},
 primaryClass = {astro-ph.GA},
       adsurl = {https://ui.adsabs.harvard.edu/abs/2019MNRAS.489.1742F}
}

@ARTICLE{2020A&A...640A..25K,
       author = {{Karovicova}, I. and {White}, T.~R. and {Nordlander}, T. and {Casagrande}, L. and {Ireland}, M. and {Huber}, D. and {Jofr{\'e}}, P.},
        title = "{Fundamental stellar parameters of benchmark stars from CHARA interferometry. I. Metal-poor stars}",
      journal = {\aap},
         year = 2020,
        month = aug,
       volume = {640},
          eid = {A25},
        pages = {A25},
          doi = {10.1051/0004-6361/202037590},
archivePrefix = {arXiv},
       eprint = {2006.05411},
 primaryClass = {astro-ph.SR},
       adsurl = {https://ui.adsabs.harvard.edu/abs/2020A&A...640A..25K}
}

@ARTICLE{2020A&A...640A..81N,
       author = {{Nissen}, P.~E. and {Christensen-Dalsgaard}, J. and {Mosumgaard}, J.~R. and {Silva Aguirre}, V. and {Spitoni}, E. and {Verma}, K.},
        title = "{High-precision abundances of elements in solar-type stars. Evidence of two distinct sequences in abundance-age relations}",
      journal = {\aap},
         year = 2020,
        month = aug,
       volume = {640},
          eid = {A81},
        pages = {A81},
          doi = {10.1051/0004-6361/202038300},
archivePrefix = {arXiv},
       eprint = {2006.06013},
 primaryClass = {astro-ph.SR},
       adsurl = {https://ui.adsabs.harvard.edu/abs/2020A&A...640A..81N}
}

@ARTICLE{2020A&A...635A.164S,
       author = {{Silva Aguirre}, V. and {Christensen-Dalsgaard}, J. and {Cassisi}, S. and {Miller Bertolami}, M. and {Serenelli}, A. and {Stello}, D. and {Weiss}, A. and {Angelou}, G. and {Jiang}, C. and {Lebreton}, Y. and {Spada}, F. and {Bellinger}, E.~P. and {Deheuvels}, S. and {Ouazzani}, R.~M. and {Pietrinferni}, A. and {Mosumgaard}, J.~R. and {Townsend}, R.~H.~D. and {Battich}, T. and {Bossini}, D. and {Constantino}, T. and {Eggenberger}, P. and {Hekker}, S. and {Mazumdar}, A. and {Miglio}, A. and {Nielsen}, K.~B. and {Salaris}, M.},
        title = "{The Aarhus red giants challenge. I. Stellar structures in the red giant branch phase}",
      journal = {\aap},
         year = 2020,
        month = mar,
       volume = {635},
          eid = {A164},
        pages = {A164},
          doi = {10.1051/0004-6361/201935843},
archivePrefix = {arXiv},
       eprint = {1912.04909},
 primaryClass = {astro-ph.SR},
       adsurl = {https://ui.adsabs.harvard.edu/abs/2020A&A...635A.164S}
}

@ARTICLE{2020A&A...641A...6P,
       author = {{Planck Collaboration} and {Aghanim}, N. and {Akrami}, Y. and {Ashdown}, M. and {Aumont}, J. and {Baccigalupi}, C. and {Ballardini}, M. and {Banday}, A.~J. and {Barreiro}, R.~B. and {Bartolo}, N. and {Basak}, S. and {Battye}, R. and {Benabed}, K. and {Bernard}, J.-P. and {Bersanelli}, M. and {Bielewicz}, P. and {Bock}, J.~J. and {Bond}, J.~R. and {Borrill}, J. and {Bouchet}, F.~R. and {Boulanger}, F. and {Bucher}, M. and {Burigana}, C. and {Butler}, R.~C. and {Calabrese}, E. and {Cardoso}, J.-F. and {Carron}, J. and {Challinor}, A. and {Chiang}, H.~C. and {Chluba}, J. and {Colombo}, L.~P.~L. and {Combet}, C. and {Contreras}, D. and {Crill}, B.~P. and {Cuttaia}, F. and {de Bernardis}, P. and {de Zotti}, G. and {Delabrouille}, J. and {Delouis}, J.-M. and {Di Valentino}, E. and {Diego}, J.~M. and {Dor{\'e}}, O. and {Douspis}, M. and {Ducout}, A. and {Dupac}, X. and {Dusini}, S. and {Efstathiou}, G. and {Elsner}, F. and {En{\ss}lin}, T.~A. and {Eriksen}, H.~K. and {Fantaye}, Y. and {Farhang}, M. and {Fergusson}, J. and {Fernandez-Cobos}, R. and {Finelli}, F. and {Forastieri}, F. and {Frailis}, M. and {Fraisse}, A.~A. and {Franceschi}, E. and {Frolov}, A. and {Galeotta}, S. and {Galli}, S. and {Ganga}, K. and {G{\'e}nova-Santos}, R.~T. and {Gerbino}, M. and {Ghosh}, T. and {Gonz{\'a}lez-Nuevo}, J. and {G{\'o}rski}, K.~M. and {Gratton}, S. and {Gruppuso}, A. and {Gudmundsson}, J.~E. and {Hamann}, J. and {Handley}, W. and {Hansen}, F.~K. and {Herranz}, D. and {Hildebrandt}, S.~R. and {Hivon}, E. and {Huang}, Z. and {Jaffe}, A.~H. and {Jones}, W.~C. and {Karakci}, A. and {Keih{\"a}nen}, E. and {Keskitalo}, R. and {Kiiveri}, K. and {Kim}, J. and {Kisner}, T.~S. and {Knox}, L. and {Krachmalnicoff}, N. and {Kunz}, M. and {Kurki-Suonio}, H. and {Lagache}, G. and {Lamarre}, J.-M. and {Lasenby}, A. and {Lattanzi}, M. and {Lawrence}, C.~R. and {Le Jeune}, M. and {Lemos}, P. and {Lesgourgues}, J. and {Levrier}, F. and {Lewis}, A. and {Liguori}, M. and {Lilje}, P.~B. and {Lilley}, M. and {Lindholm}, V. and {L{\'o}pez-Caniego}, M. and {Lubin}, P.~M. and {Ma}, Y.-Z. and {Mac{\'\i}as-P{\'e}rez}, J.~F. and {Maggio}, G. and {Maino}, D. and {Mandolesi}, N. and {Mangilli}, A. and {Marcos-Caballero}, A. and {Maris}, M. and {Martin}, P.~G. and {Martinelli}, M. and {Mart{\'\i}nez-Gonz{\'a}lez}, E. and {Matarrese}, S. and {Mauri}, N. and {McEwen}, J.~D. and {Meinhold}, P.~R. and {Melchiorri}, A. and {Mennella}, A. and {Migliaccio}, M. and {Millea}, M. and {Mitra}, S. and {Miville-Desch{\^e}nes}, M.-A. and {Molinari}, D. and {Montier}, L. and {Morgante}, G. and {Moss}, A. and {Natoli}, P. and {N{\o}rgaard-Nielsen}, H.~U. and {Pagano}, L. and {Paoletti}, D. and {Partridge}, B. and {Patanchon}, G. and {Peiris}, H.~V. and {Perrotta}, F. and {Pettorino}, V. and {Piacentini}, F. and {Polastri}, L. and {Polenta}, G. and {Puget}, J.-L. and {Rachen}, J.~P. and {Reinecke}, M. and {Remazeilles}, M. and {Renzi}, A. and {Rocha}, G. and {Rosset}, C. and {Roudier}, G. and {Rubi{\~n}o-Mart{\'\i}n}, J.~A. and {Ruiz-Granados}, B. and {Salvati}, L. and {Sandri}, M. and {Savelainen}, M. and {Scott}, D. and {Shellard}, E.~P.~S. and {Sirignano}, C. and {Sirri}, G. and {Spencer}, L.~D. and {Sunyaev}, R. and {Suur-Uski}, A.-S. and {Tauber}, J.~A. and {Tavagnacco}, D. and {Tenti}, M. and {Toffolatti}, L. and {Tomasi}, M. and {Trombetti}, T. and {Valenziano}, L. and {Valiviita}, J. and {Van Tent}, B. and {Vibert}, L. and {Vielva}, P. and {Villa}, F. and {Vittorio}, N. and {Wandelt}, B.~D. and {Wehus}, I.~K. and {White}, M. and {White}, S.~D.~M. and {Zacchei}, A. and {Zonca}, A.},
        title = "{Planck 2018 results. VI. Cosmological parameters}",
      journal = {\aap},
         year = 2020,
        month = sep,
       volume = {641},
          eid = {A6},
        pages = {A6},
          doi = {10.1051/0004-6361/201833910},
archivePrefix = {arXiv},
       eprint = {1807.06209},
 primaryClass = {astro-ph.CO},
       adsurl = {https://ui.adsabs.harvard.edu/abs/2020A&A...641A...6P}
}

@ARTICLE{2020AJ....160..108B,
       author = {{Berger}, Travis A. and {Huber}, Daniel and {Gaidos}, Eric and {van Saders}, Jennifer L. and {Weiss}, Lauren M.},
        title = "{The Gaia-Kepler Stellar Properties Catalog. II. Planet Radius Demographics as a Function of Stellar Mass and Age}",
      journal = {\aj},
         year = 2020,
        month = sep,
       volume = {160},
       number = {3},
          eid = {108},
        pages = {108},
          doi = {10.3847/1538-3881/aba18a},
archivePrefix = {arXiv},
       eprint = {2005.14671},
 primaryClass = {astro-ph.EP},
       adsurl = {https://ui.adsabs.harvard.edu/abs/2020AJ....160..108B}
}

@ARTICLE{2020MNRAS.493.2377R,
       author = {{Rains}, Adam D. and {Ireland}, Michael J. and {White}, Timothy R. and {Casagrande}, Luca and {Karovicova}, I.},
        title = "{Precision angular diameters for 16 southern stars with VLTI/PIONIER}",
      journal = {\mnras},
         year = 2020,
        month = apr,
       volume = {493},
       number = {2},
        pages = {2377-2394},
          doi = {10.1093/mnras/staa282},
archivePrefix = {arXiv},
       eprint = {2004.02343},
 primaryClass = {astro-ph.SR},
       adsurl = {https://ui.adsabs.harvard.edu/abs/2020MNRAS.493.2377R}
}

@ARTICLE{2020MNRAS.491.1160M,
       author = {{Mosumgaard}, Jakob R{\o}rsted and {J{\o}rgensen}, Andreas Christ S{\o}lvsten and {Weiss}, Achim and {Silva Aguirre}, V{\'\i}ctor and {Christensen-Dalsgaard}, J{\o}rgen},
        title = "{Coupling 1D stellar evolution with 3D-hydrodynamical simulations on-the-fly II: stellar evolution and asteroseismic applications}",
      journal = {\mnras},
         year = 2020,
        month = jan,
       volume = {491},
       number = {1},
        pages = {1160-1173},
          doi = {10.1093/mnras/stz2979},
archivePrefix = {arXiv},
       eprint = {1910.10163},
 primaryClass = {astro-ph.SR},
       adsurl = {https://ui.adsabs.harvard.edu/abs/2020MNRAS.491.1160M}
}

@ARTICLE{2020NatAs...4..382C,
       author = {{Chaplin}, William J. and {Serenelli}, Aldo M. and {Miglio}, Andrea and {Morel}, Thierry and {Mackereth}, J. Ted and {Vincenzo}, Fiorenzo and {Kjeldsen}, Hans and {Basu}, Sarbani and {Ball}, Warrick H. and {Stokholm}, Amalie and {Verma}, Kuldeep and {Mosumgaard}, Jakob R{\o}rsted and {Silva Aguirre}, Victor and {Mazumdar}, Anwesh and {Ranadive}, Pritesh and {Antia}, H.~M. and {Lebreton}, Yveline and {Ong}, Joel and {Appourchaux}, Thierry and {Bedding}, Timothy R. and {Christensen-Dalsgaard}, J{\o}rgen and {Creevey}, Orlagh and {Garc{\'\i}a}, Rafael A. and {Handberg}, Rasmus and {Huber}, Daniel and {Kawaler}, Steven D. and {Lund}, Mikkel N. and {Metcalfe}, Travis S. and {Stassun}, Keivan G. and {Bazot}, Mich{\"a}el and {Beck}, Paul G. and {Bell}, Keaton J. and {Bergemann}, Maria and {Buzasi}, Derek L. and {Benomar}, Othman and {Bossini}, Diego and {Bugnet}, Lisa and {Campante}, Tiago L. and {Orhan}, Zeynep {\c{c}}elik and {Corsaro}, Enrico and {Gonz{\'a}lez-Cuesta}, Luc{\'\i}a and {Davies}, Guy R. and {Di Mauro}, Maria Pia and {Egeland}, Ricky and {Elsworth}, Yvonne P. and {Gaulme}, Patrick and {Ghasemi}, Hamed and {Guo}, Zhao and {Hall}, Oliver J. and {Hasanzadeh}, Amir and {Hekker}, Saskia and {Howe}, Rachel and {Jenkins}, Jon M. and {Jim{\'e}nez}, Antonio and {Kiefer}, Ren{\'e} and {Kuszlewicz}, James S. and {Kallinger}, Thomas and {Latham}, David W. and {Lundkvist}, Mia S. and {Mathur}, Savita and {Montalb{\'a}n}, Josefina and {Mosser}, Benoit and {Bed{\'o}n}, Andres Moya and {Nielsen}, Martin Bo and {{\"O}rtel}, Sibel and {Rendle}, Ben M. and {Ricker}, George R. and {Rodrigues}, Tha{\'\i}se S. and {Roxburgh}, Ian W. and {Safari}, Hossein and {Schofield}, Mathew and {Seager}, Sara and {Smalley}, Barry and {Stello}, Dennis and {Szab{\'o}}, R{\'o}bert and {Tayar}, Jamie and {Theme{\ss}l}, Nathalie and {Thomas}, Alexandra E.~L. and {Vanderspek}, Roland K. and {van Rossem}, Walter E. and {Vrard}, Mathieu and {Weiss}, Achim and {White}, Timothy R. and {Winn}, Joshua N. and {Y{\i}ld{\i}z}, Mutlu},
        title = "{Age dating of an early Milky Way merger via asteroseismology of the naked-eye star {\ensuremath{\nu}} Indi}",
      journal = {Nature Astronomy},
         year = 2020,
        month = jan,
       volume = {4},
        pages = {382-389},
          doi = {10.1038/s41550-019-0975-9},
archivePrefix = {arXiv},
       eprint = {2001.04653},
 primaryClass = {astro-ph.GA},
       adsurl = {https://ui.adsabs.harvard.edu/abs/2020NatAs...4..382C}
}

@ARTICLE{2021A&A...650A..58M,
       author = {{Mombarg}, J.~S.~G. and {Van Reeth}, T. and {Aerts}, C.},
        title = "{Constraining stellar evolution theory with asteroseismology of {\ensuremath{\gamma}} Doradus stars using deep learning. Stellar masses, ages, and core-boundary mixing}",
      journal = {\aap},
         year = 2021,
        month = jun,
       volume = {650},
          eid = {A58},
        pages = {A58},
          doi = {10.1051/0004-6361/202039543},
archivePrefix = {arXiv},
       eprint = {2103.13394},
 primaryClass = {astro-ph.SR},
       adsurl = {https://ui.adsabs.harvard.edu/abs/2021A&A...650A..58M}
}

@ARTICLE{2021A&A...646A.133H,
       author = {{Higl}, J. and {M{\"u}ller}, E. and {Weiss}, A.},
        title = "{Calibrating core overshooting parameters with two-dimensional hydrodynamical simulations}",
      journal = {\aap},
         year = 2021,
        month = feb,
       volume = {646},
          eid = {A133},
        pages = {A133},
          doi = {10.1051/0004-6361/202039532},
archivePrefix = {arXiv},
       eprint = {2012.05262},
 primaryClass = {astro-ph.SR},
       adsurl = {https://ui.adsabs.harvard.edu/abs/2021A&A...646A.133H}
}

@ARTICLE{2021A&A...649A.178B,
       author = {{Brogaard}, K. and {Grundahl}, F. and {Sandquist}, E.~L. and {Slumstrup}, D. and {Jensen}, M.~L. and {Thomsen}, J.~B. and {J{\o}rgensen}, J.~H. and {Larsen}, J.~R. and {Bj{\o}rn}, S.~T. and {S{\o}rensen}, C.~T.~G. and {Bruntt}, H. and {Arentoft}, T. and {Frandsen}, S. and {Jessen-Hansen}, J. and {Orosz}, J.~A. and {Mathieu}, R. and {Geller}, A. and {Ryde}, N. and {Stello}, D. and {Meibom}, S. and {Platais}, I.},
        title = "{Age and helium content of the open cluster NGC 6791 from multiple eclipsing binary members. III. Constraints from a subgiant}",
      journal = {\aap},
         year = 2021,
        month = may,
       volume = {649},
          eid = {A178},
        pages = {A178},
          doi = {10.1051/0004-6361/202140911},
archivePrefix = {arXiv},
       eprint = {2104.14330},
 primaryClass = {astro-ph.SR},
       adsurl = {https://ui.adsabs.harvard.edu/abs/2021A&A...649A.178B}
}

@ARTICLE{2021MNRAS.503...13Z,
       author = {{Zhou}, Yixiao and {Nordlander}, Thomas and {Casagrande}, Luca and {Joyce}, Meridith and {Li}, Yaguang and {Amarsi}, Anish M. and {Reggiani}, Henrique and {Asplund}, Martin},
        title = "{The relationship between photometric and spectroscopic oscillation amplitudes from 3D stellar atmosphere simulations}",
      journal = {\mnras},
         year = 2021,
        month = may,
       volume = {503},
       number = {1},
        pages = {13-27},
          doi = {10.1093/mnras/stab337},
archivePrefix = {arXiv},
       eprint = {2102.02135},
 primaryClass = {astro-ph.SR},
       adsurl = {https://ui.adsabs.harvard.edu/abs/2021MNRAS.503...13Z}
}

@ARTICLE{2021MNRAS.500.4277J,
       author = {{J{\o}rgensen}, Andreas Christ S{\o}lvsten and {Montalb{\'a}n}, Josefina and {Angelou}, George C. and {Miglio}, Andrea and {Weiss}, Achim and {Scuflaire}, Richard and {Noels}, Arlette and {Mosumgaard}, Jakob R{\o}rsted and {Silva Aguirre}, V{\'\i}ctor},
        title = "{On the impact of the structural surface effect on global stellar properties and asteroseismic analyses}",
      journal = {\mnras},
         year = 2021,
        month = jan,
       volume = {500},
       number = {4},
        pages = {4277-4295},
          doi = {10.1093/mnras/staa3476},
archivePrefix = {arXiv},
       eprint = {2009.11251},
 primaryClass = {astro-ph.SR},
       adsurl = {https://ui.adsabs.harvard.edu/abs/2021MNRAS.500.4277J}
}

@ARTICLE{2021MNRAS.507.2684C,
       author = {{Casagrande}, Luca and {Lin}, Jane and {Rains}, Adam D. and {Liu}, Fan and {Buder}, Sven and {Horner}, Jonathan and {Asplund}, Martin and {Lewis}, Geraint F. and {Martell}, Sarah L. and {Nordlander}, Thomas and {Stello}, Dennis and {Ting}, Yuan-Sen and {Wittenmyer}, Robert A. and {Bland-Hawthorn}, Joss and {Casey}, Andrew R. and {De Silva}, Gayandhi M. and {D'Orazi}, Valentina and {Freeman}, Ken C. and {Hayden}, Michael R. and {Kos}, Janez and {Lind}, Karin and {Schlesinger}, Katharine J. and {Sharma}, Sanjib and {Simpson}, Jeffrey D. and {Zucker}, Daniel B. and {Zwitter}, Toma{\v{z}}},
        title = "{The GALAH survey: effective temperature calibration from the InfraRed Flux Method in the Gaia system}",
      journal = {\mnras},
         year = 2021,
        month = oct,
       volume = {507},
       number = {2},
        pages = {2684-2696},
          doi = {10.1093/mnras/stab2304},
archivePrefix = {arXiv},
       eprint = {2011.02517},
 primaryClass = {astro-ph.SR},
       adsurl = {https://ui.adsabs.harvard.edu/abs/2021MNRAS.507.2684C}
}

@ARTICLE{2022ApJ...927...31T,
       author = {{Tayar}, Jamie and {Claytor}, Zachary R. and {Huber}, Daniel and {van Saders}, Jennifer},
        title = "{A Guide to Realistic Uncertainties on the Fundamental Properties of Solar-type Exoplanet Host Stars}",
      journal = {\apj},
         year = 2022,
        month = mar,
       volume = {927},
       number = {1},
          eid = {31},
        pages = {31},
          doi = {10.3847/1538-4357/ac4bbc},
archivePrefix = {arXiv},
       eprint = {2012.07957},
 primaryClass = {astro-ph.EP},
       adsurl = {https://ui.adsabs.harvard.edu/abs/2022ApJ...927...31T}
}

@ARTICLE{2022MNRAS.509.4208V,
       author = {{VandenBerg}, Don A. and {Casagrande}, Luca and {Edvardsson}, Bengt},
        title = "{Models for metal-poor stars with different initial abundances of C, N, O, Mg, and Si. II. Application to the colour-magnitude diagrams of the globular clusters 47 Tuc, NGC 6362, M 5, M 3, M 55, and M92}",
      journal = {\mnras},
         year = 2022,
        month = jan,
       volume = {509},
       number = {3},
        pages = {4208-4228},
          doi = {10.1093/mnras/stab2998},
archivePrefix = {arXiv},
       eprint = {2111.02623},
 primaryClass = {astro-ph.SR},
       adsurl = {https://ui.adsabs.harvard.edu/abs/2022MNRAS.509.4208V}
}

@ARTICLE{2022MNRAS.514.2527B,
       author = {{Borre}, Camilla C. and {Aguirre B{\o}rsen-Koch}, V{\'\i}ctor and {Helmi}, Amina and {Koppelman}, Helmer H. and {Nielsen}, Martin B. and {R{\o}rsted}, Jakob L. and {Stello}, Dennis and {Stokholm}, Amalie and {Winther}, Mark L. and {Davies}, Guy R. and {Hon}, Marc and {Kruijssen}, J.~M. Diederik and {Laporte}, Chervin F.~P. and {Reyes}, Claudia and {Yu}, Jie},
        title = "{Age determination of galaxy merger remnant stars using asteroseismology}",
      journal = {\mnras},
         year = 2022,
        month = aug,
       volume = {514},
       number = {2},
        pages = {2527-2544},
          doi = {10.1093/mnras/stac1498},
archivePrefix = {arXiv},
       eprint = {2111.01669},
 primaryClass = {astro-ph.GA},
       adsurl = {https://ui.adsabs.harvard.edu/abs/2022MNRAS.514.2527B}
}

@ARTICLE{2022Natur.603..599X,
       author = {{Xiang}, Maosheng and {Rix}, Hans-Walter},
        title = "{A time-resolved picture of our Milky Way's early formation history}",
      journal = {\nat},
         year = 2022,
        month = mar,
       volume = {603},
       number = {7902},
        pages = {599-603},
          doi = {10.1038/s41586-022-04496-5},
archivePrefix = {arXiv},
       eprint = {2203.12110},
 primaryClass = {astro-ph.GA},
       adsurl = {https://ui.adsabs.harvard.edu/abs/2022Natur.603..599X}
}

@ARTICLE{2023AJ....166...18Y,
       author = {{Ying}, Jiaqi Martin and {Chaboyer}, Brian and {Boudreaux}, Emily M. and {Slaughter}, Catherine and {Boylan-Kolchin}, Michael and {Weisz}, Daniel},
        title = "{The Absolute Age of M92}",
      journal = {\aj},
         year = 2023,
        month = jul,
       volume = {166},
       number = {1},
          eid = {18},
        pages = {18},
          doi = {10.3847/1538-3881/acd9b1},
archivePrefix = {arXiv},
       eprint = {2306.02180},
 primaryClass = {astro-ph.SR},
       adsurl = {https://ui.adsabs.harvard.edu/abs/2023AJ....166...18Y}
}

@ARTICLE{2023A&A...677A..98Z,
       author = {{Zhou}, Yixiao and {Amarsi}, Anish M. and {Aguirre B{\o}rsen-Koch}, Victor and {Karlsmose}, Klara G. and {Collet}, Remo and {Nordlander}, Thomas},
        title = "{3D Stagger model atmospheres with FreeEOS. I. Exploring the impact of microphysics on the Sun}",
      journal = {\aap},
         year = 2023,
        month = sep,
       volume = {677},
          eid = {A98},
        pages = {A98},
          doi = {10.1051/0004-6361/202346398},
archivePrefix = {arXiv},
       eprint = {2307.05403},
 primaryClass = {astro-ph.SR},
       adsurl = {https://ui.adsabs.harvard.edu/abs/2023A&A...677A..98Z}
}

@ARTICLE{2023MNRAS.525.1416W,
       author = {{Winther}, Mark Lykke and {Aguirre B{\o}rsen-Koch}, V{\'\i}ctor and {R{\o}rsted}, Jakob Lysgaard and {Stokholm}, Amalie and {Verma}, Kuldeep},
        title = "{Did Kepler-444 have a long-lived convective core?}",
      journal = {\mnras},
         year = 2023,
        month = oct,
       volume = {525},
       number = {1},
        pages = {1416-1430},
          doi = {10.1093/mnras/stad1802},
archivePrefix = {arXiv},
       eprint = {2306.08430},
 primaryClass = {astro-ph.SR},
       adsurl = {https://ui.adsabs.harvard.edu/abs/2023MNRAS.525.1416W}
}

@ARTICLE{2024ApJ...970...24S,
       author = {{Stein}, Robert F. and {Nordlund}, {\r{A}}ke and {Collet}, Remo and {Trampedach}, Regner},
        title = "{The Stagger Code for Accurate and Efficient, Radiation-coupled Magnetohydrodynamic Simulations}",
      journal = {\apj},
         year = 2024,
        month = jul,
       volume = {970},
       number = {1},
          eid = {24},
        pages = {24},
          doi = {10.3847/1538-4357/ad4706},
archivePrefix = {arXiv},
       eprint = {2405.02483},
 primaryClass = {astro-ph.IM},
       adsurl = {https://ui.adsabs.harvard.edu/abs/2024ApJ...970...24S}
}



 

\end{document}